\documentclass[aps,a4paper,11pt,nofootinbib,preprintnumbers]{revtex4}
\usepackage{amsmath}
\usepackage{graphicx,amssymb,bm,latexsym,color,hyperref}
\makeatletter
\@addtoreset{equation}{section}

\makeatother
\newcommand{\be}{\begin{equation}}
\newcommand{\ee}{\end{equation}}
\newcommand{\bea}{\begin{eqnarray}}
\newcommand{\eea}{\end{eqnarray}}

   \newcommand{\pslash}{p \hspace{-0.15cm}/}
   
   \newcommand{\sgn}{\mathop{\mathrm{sgn}}}

   \definecolor{comment}{rgb}{0.9,0,0}
   
   \newcommand{\Eqref}[1]{Eq.~\eqref{#1}}

\begin{document}

\begin{titlepage}

\title{Multi-meson Yukawa interactions at criticality}

\author{Gian Paolo Vacca}
\email{vacca@bo.infn.it}
\affiliation{INFN, Sezione di Bologna, via Irnerio 46, I-40126 Bologna}

\author{Luca Zambelli}
\email{luca.zambelli@uni-jena.de}
\affiliation{Theoretisch-Physikalisches Institut, Friedrich-Schiller-Universit{\"a}t Jena, D-07743 Jena, Germany}

\pacs{}

\begin{abstract}
The critical behavior of a relativistic $\mathbb{Z}_2$-symmetric Yukawa model at zero temperature
and density is discussed for
a continuous number of fermion degrees of freedom and of spacetime dimensions,
with emphasis on the role played by multi-meson exchange in the Yukawa sector.
We argue that this should be generically taken into account in studies based on the functional renormalization group,
either in four-dimensional high-energy models or in lower-dimensional condensed-matter systems.
By means of the latter method, we describe the generation of multi-critical models in less then three dimensions,
both at infinite and finite number of flavors.
We also provide different estimates of the critical exponents of the chiral Ising universality class in three dimensions
for various field contents, from a couple of massless Dirac fermions down to the supersymmetric theory with a single Majorana spinor.
\end{abstract} 

\maketitle

\end{titlepage}
\newpage
\setcounter{page}{2}

\section{Introduction}
\label{sec:intro}
In this paper we will study the renormalization group (RG) flow of a simple Yukawa model 
describing relativistic fermions interacting through the exchange of scalar fluctuations. 
We will discuss some of its critical properties in a continuum of spacetime dimensions $2<d\leq4$,
dedicating most of the analysis to the $d=3$ case.
The class of models we want to consider is described by the following generic 
bare Lagrangian 
\be
{\cal L}=\frac{1}{2}\, \partial^\mu\phi\partial_\mu\phi+V(\phi)+
\bar{\psi}\gamma^\mu i \partial_\mu\psi+i\,H(\phi)\,\bar{\psi}\,\psi \ .
\ee 
where we have $N_f$ copies of fermions, whose representation will be kept general in the following,
and one real scalar field.
The requirement of power-counting renormalizability would
further restrict the interactions inside the potentials $V$ and $H$ (and would generically require
the inclusion of derivative interactions too) but we are not
going to impose such conditions, since we are interested in describing the possible conformal
models in this family, even if strongly coupled.
In case the potentials $V$ and $H$ are even and odd respectively,
the system is characterized by a chiral $\mathbb{Z}_2$ symmetry,
besides the U($N_f$) symmetry.
For this reason, the model with bare potentials
\be
V(\phi)=\frac{\bar{m}^2}{2}\phi^2+\frac{\bar{\lambda}_2}{2}\phi^4\ , \quad
H(\phi)=\bar{y} \phi
\ee
is often called Gross-Neveu-Yukawa model, since it shares these symmetries
with the purely fermionic Gross-Neveu model~\cite{Gross:1974jv}
and can be obtained from it by means of a Hubbard-Stratonovich transformation.

Even for more general bare Lagrangians that are not related by any bosonization technique,
the Yukawa models and chiral fermionic models remain deeply connected.
The three dimensional Gross-Neveu model shows a second order quantum phase transition 
that separates the phase with preserved chiral symmetry from the one where this 
is spontaneously broken and a chiral condensate of fermions appears.
The latter can be effectively described as a scalar degree of freedom, 
therefore this transition can be unveiled also as a dynamical
effect in interacting scalar-spinor systems.
Indeed, it is found that the critical properties of the Gross-Neveu model in $2<d<4$ dimensions 
are compatible with the ones of the Yukawa model, thus indicating that the
two are in the same universality class, which for generic but non-vanishing
flavor number is also called the chiral Ising universality class.
In both parameterizations, this is described by a non-Gau\ss ian fixed point (FP)
of the RG flow. As a consequence, non-perturbative tools are best suited for the
investigation of its properties, and for the extraction of key quantities such as the corresponding
critical exponents.
Indeed several methods have been applied to this problem, including $\epsilon$-expansions~\cite{Rosenstein:1988dj,Rosenstein:1993zf,ZinnJustin:1991yn,Gracey:1990,Vasiliev:1997sk},
large $N_f$ expansions~\cite{ZinnJustin:1991yn,Gracey:1993kb,Gracey:1993kc},
lattice simulations~\cite{Hands:1992be,Karkkainen:1993ef,Focht:1995bw,Chandrasekharan:2013aya,Wang:2014cbw}
and functional RG equations~\cite{Rosa:2000ju,Hofling:2002hj,Braun:2010tt,Sonoda:2011qd,Janssen:2014gea,Borchardt:2015rxa}.

These critical properties have great physical relevance for the description of several systems.
In condensed matter, three dimensional relativistic fermionic systems, 
such as QED${}_3$ and the Thirring model, play the role of building blocks 
for theories of high-$T_C$ superconductivity~\cite{Dorey:1990sz},
and for the description of electrons in graphene~\cite{Semenoff:1984dq}.
Understanding the phase diagram and critical properties of these models 
at variable $N_f$ represents pretty much the same challenge as the one 
posed by the Gross-Neveu and Yukawa model, and one can even address them in
a unified picture~\cite{Gies:2010st}.	
Even the simple Yukawa model discussed in this work can find applications to extremely 
nontrivial phenomena in condensed matter.
For the case of two massless Dirac fermions, its quantum critical
phase transition in $d=3$ might be a close relative of the
putative transition between the semi-metallic and the Mott-insulating phases of electrons
in graphene~\cite{Janssen:2014gea}.
For a single Dirac field instead, it is considered to be in the same universality class of spinless fermions 
on the honeycomb lattice with repulsive nearest neighbors interactions~\cite{Wang:2014cbw}.

For a single Majorana spinor, it is a precious example of a three dimensional
model showing emergent supersymmetry. Indeed, it is known that in this case
the critical theory not only enjoys ${\cal N}=1$ supersymmetry,
but also possesses only one relevant component,
which means that by tuning a single macroscopic parameter one can discriminate between
two distinct phases with preserved or spontaneously broken supersymmetry~\cite{Sonoda:2011qd,Grover:2013rc}.
On these grounds, a potential experimental realization of supersymmetry was
proposed in~\cite{Grover:2013rc}, at the boundary of topological superconductors.
A similar phenomenon occurs for Yukawa systems with complex scalars and spinors,
which have been argued to give rise to an emergent ${\cal N}=2$ 
supersymmetry~\cite{Balents:2008vd}.

The phase diagram of Gross-Neveu and Yukawa models has been analysed in $d<4$
also for a better understanding of their $d\to4$ limit.
Clearly, nonperturbative phenomena in the latter case can have many applications in particle physics.
These range from the chiral phase transition in QCD~\cite{Nambu:1961tp},
where these models serve as simplified versions of quark-meson models~\cite{Jungnickel:1995fp},
to the Higgs sector of the standard model~\cite{Gies:2009hq,Gies:2009sv},
and to toy-models of composite-Higgs extensions~\cite{Nambu:1988}.

In the present work we will analyze a more general truncation scheme for the
functional renormalization group (FRG) study of these systems, showing under
which conditions this brings important improvements
in the results obtained by means of the latter nonperturbative method.
Such a truncation scheme amounts to allowing for a generic potential
$H(\phi)$, that essentially describes  vertices with two fermions and 
an arbitrary number of scalars. This kind of interactions have been
neglected in the FRG studies of fermionic models for a long time.
Only recently they have been discussed in other works considering more complicated
models and different but related questions.
For example, in~\cite{Zanusso:2009bs} the flow equations for this Yukawa system coupled
to quantum gravity were derived, but only the linear coupling $H(\phi)=\bar{y} \phi$
was considered in explicit studies of these equations.
Most prominently, in~\cite{Pawlowski:2014zaa} the effect of higher Yukawa couplings
 on the chiral phase structure of QCD at finite temperature and chemical potential
was analyzed by means of an effective quark-meson model.
It was observed, within polynomial truncations of a Yukawa potential $H(\phi)$, 
that higher order quark-meson interactions are quantitatively
important in the description of the chiral transition.

A similar but different study will be performed here, for the present $\mathbb{Z}_2$-symmetric
Yukawa model, in lower dimensionality and for a generic number of flavors.
We will confine ourselves to the study of the zero-temperature system at criticality, 
looking for scaling solutions for various $d$ and $N_f$, and comparing the results obtained with different methods.
In Sect.~\ref{sec:largeXf} we start with the leading order of the $1/N_f$-expansion,
reproducing known results in three dimensions, and generalizing them to multi-critical theories
below three dimensions. Technical details regarding this analysis are sketched in App.~\ref{sec:largeXf_appendix}.
In Sect.~\ref{sec:LPA_Xf1_genericd} we turn to a finite number of fermions and, 
by neglecting the wave function renormalization of the fields, we observe how critical Yukawa theories arise 
while continuously lowering the dimensionality towards two. To this end, we consider the
FP equations for the two generic functions $V(\phi)$ and $H(\phi)$, and solve them numerically
without resorting to any truncation.
In Sect.~\ref{sec:d=3_numerical_FP}, still neglecting the wave function renormalizations,
we adopt a different strategy for the numerical integration of the FP equations, and compute the
global FP potentials in three dimensions, for
various flavor numbers. For the case of a single Majorana spinor, we also apply these numerical methods to the computation of  
the critical exponents and perturbations.
In Sect.~\ref{sec:polynomial} we discuss polynomial truncations, showing how these can
give results in satisfactory agreement with the global numerical analysis. As a consequence,
we use them for a self-consistent inclusion of the wave function renormalizations, and produce 
estimates of the critical exponents in three dimensions and for various number of fermions,
which we compare with some of the existing literature.
Finally, in Sect.~\ref{sec:d=4} we address the $d\to 4$ limit at low number of fermions,
and in Sect.~\ref{sec:conclusions} we draw a summary of our results.
Yet, to introduce our work, we need to provide the reader with the definition of the
approximations involved in the computation of the flow equations, and with the resulting
beta-functions. This is the object of the next Section and of App.~\ref{sec:regulators_and_thresholds}.

\section{The RG flow of a simple Yukawa model with multi-meson-exchange.}
\label{sec:model_and_flow}


The functional renormalization group (FRG) is a representation of quantum dynamics
based on Wilson's idea of floating cutoff $k$. 
In this work we will adopt its formulation in terms of a scale-dependent 1PI effective action, 
called average effective action~\cite{Wetterich:1992yh}.
For a given system, the form of this action is determined by the field content $\Phi$
 and by the symmetry properties, as well as by an initial condition (bare action) 
and boundary conditions for the integration of the flow equation
\begin{equation}\label{flowequation}
\dot{\Gamma}_k[\Phi]
        =\frac{1}{2}\mathrm{STr}\!\left[\left(\Gamma^{(2)}_k[\Phi]+R_k\right)^{-1}\dot{R}_k\right]\, .
\end{equation}
Here $(\Gamma^{(2)}_k[\Phi]+R_k)^{-1}$ represents the matrix of regularized propagators,
while $R_k$ is a momentum-dependent mass-like regulator. Since the dot stands for differentiation
with respect to the RG time $t=\log k$, this flow equation comprehends the infinite set of beta
functions for the infinitely many allowed interactions inside $\Gamma_k$.
Extracting them amounts to projecting both sides of the equation on each separate
interaction functional. In practical computations, one drops infinitely many operators, 
thus performing a nonperturbative approximation called truncation of the theory space.
To this end, several systematic strategies are available and appropriate in different circumstances,
such as the vertex expansion or the derivate expansion. For reviews see~\cite{ReviewRG}.

In this work we will consider the following truncation:
\be\label{eq:truncation}
\Gamma_k \!\left[\phi,\psi,\bar{\psi} \right] \!=\! \int\!\!
{\rm d}^d x \left(\frac{1}{2}Z_{\phi,k}\, \partial^\mu\phi\partial_\mu\phi+V_k(\phi)+
Z_{\psi,k} \bar{\psi}\gamma^\mu i \partial_\mu\psi+i\,H_k(\phi)\,\bar{\psi}\,\psi \right)\, .
\ee
Here $\phi$ is a real scalar field, while $\psi$ denotes $N_f$ copies of a spinor field with $d_\gamma$ real 
components. The latter parameter is related to the symmetries of the system and plays therefore a crucial role
in pure fermionic as well as in fermion-boson models.
Yet, as long as we truncate the theory space to the ansatz of~\Eqref{eq:truncation},
focusing on the mechanism of $\mathbb{Z}_2$-symmetry breaking, we can simply deal
 with the total number of real Grassmannian degrees
of freedom $X_f=d_\gamma N_f$, considering it as an arbitrary real number.
As soon as $X_f\geq2$ the truncation above is missing purely fermionic derivative-free interactions,
that are indeed symmetry-sensitive and that would contribute to the leading (zeroth) order of the derivative expansion.
Furthermore, it is also missing field-dependent contributions to the wave functions renormalizations
$Z_{\phi}$ and $Z_{\psi}$, which would appear in the next-to-leading (first) order of the derivative expansion.
In the following we will call the ansatz of~\Eqref{eq:truncation}
a local potential approximation (LPA) for this simple Yukawa model,
whenever the wave functions renormalizations are neglected ($Z_{\phi,k}=Z_{\psi,k}=1$), 
and therefore the fields have no anomalous dimensions $\eta_{\phi,\psi}=-\partial_t\log Z_{\phi,\psi}$.
The inclusion of the latter will be named LPA${}^\prime$.
Our justification for the choice of this truncation is in the exhaustive 
evidence that similar ans\"atze give a good description of the existence and
properties of conformal models in $2<d\leq4$ for linear systems with scalar
degrees of freedom~\cite{ReviewRG}.

Projection of the Wetterich equation on the truncation of~\Eqref{eq:truncation}
yields the running of the corresponding parameters.
Since we are interested in reproducing conformal models, that correspond to scaling solutions of
the RG flow, it is useful to consider rescaled amplitudes
\begin{equation*}
\phi\longrightarrow\frac{k^{(d-2)/2}}{Z_\phi^{1/2}}\phi\ ,\quad  \quad \psi\longrightarrow\frac{k^{(d-1)/2}}{Z_\psi^{1/2}}\psi
\end{equation*}
since the new dimensionless renormalized field would then be constant at criticality.
As a consequence we will focus on the potentials for these fields
\begin{equation*}
v_k(\phi)=k^{-d}V_k\bigg(\frac{Z_\phi^{1/2}\phi}{k^{(d-2)/2}}\bigg)\ ,\quad  \quad h_k(\phi)=\frac{k^{-1}}{Z_\psi}H_k\bigg(\frac{Z_\phi^{1/2}\phi}{k^{(d-2)/2}}\bigg)\ .
\end{equation*}
In this new set of variables the flow equations read
  \be\label{eq:floweq_v}
 \dot{v}=-d v +\frac{d-2+\eta _{\phi }}{2} \phi\, v'+2 v_d
 \left\{ 
l_0^{(\mathrm{B})d}(v'')-X_f l_0^{(\mathrm{F})d}(h^2)
 \right\}
  \ee
 \be\label{eq:floweq_h}
 \dot{h}=h \left(\eta _{\psi }-1\right)+\frac{d-2+\eta _{\phi }}{2} \phi\,  h' +
2 v_d\left\{ 2 h (h')^2 l_{1,1}^{(\mathrm{FB})d}(h^2,v'')  - h'' l_1^{(\mathrm{B})d}(v'') \right\}
 \ee
 \be\label{eq:floweq_etaphi}
 \eta_\phi=\frac{4 v_d}{d} \left\{(v^{(3)})^2\  m_{4}^{(\mathrm{B})d}(v'')+
2X_f  (h')^2\left[ m_{4}^{(\mathrm{F})d}(h^2)-h^2 m_{2}^{(\mathrm{F})d}(h^2) \right]\right\}_{\phi_0}
 \ee
 \be\label{eq:floweq_etapsi}
 \eta_\psi=\frac{8 v_d}{d} \left\{(h')^2 m_{1,2}^{(\mathrm{FB})d}(h^2,v'') \right\}_{\phi_0}
 \ee
where $v_d=(2^{d+1}\pi^{d/2} \Gamma(d/2))^{-1}$, the threshold functions $l^{(F/B)d}$ and $m^{(F/B)d}$ on the right hand side denote regulator-dependent 
contributions from loops containing fermionic or bosonic propagators,
and the equations for the anomalous dimensions are to be evaluated at the minimum $\phi_0$ of
the scalar potential. 
Their definition can be found in App.~\ref{sec:regulators_and_thresholds},
together with the explicit form they take for the linear regulator, which is our 
choice in this work since it allows for a simple analytic computation of such integrals.
For this linear regulator the flow equations of the two potentials, read
 \bea
 \dot{v}&=& -d v +\frac{d-2+\eta _{\phi }}{2} \phi\, v'+C_d
 \left(  
 \frac{1-\frac{\eta _{\phi}}{d+2}}{1+v''}
 -X_f  \frac{1-\frac{\eta _{\psi }}{d+1}} {1+h^2}
 \right)\label{LPAeqv}\\
 \dot{h}&=&h \left(\eta _{\psi }-1\right)+\frac{d-2+\eta _{\phi }}{2} \phi\,  h' \nonumber\\
 &{}&+C_d 
 \left[  
 2  h \left(h'\right)^2 \left(\frac{1-\frac{\eta _{\psi }}{d+1}}{\left(1+h^2\right)^2
   \left(1+v''\right)}+\frac{1-\frac{\eta _{\phi }}{d+2}}{\left(1+h^2\right)
   \left(1+v''\right)^2}\right)-\frac{h'' \left(1-\frac{\eta _{\phi
   }}{d+2}\right)}{\left(1+v''\right)^2}
   \right]\label{LPAeqh}
 \eea
 where we have denoted for convenience $C_d=4v_d/d$.

A simple way of facilitating the stability of the vacuum is the requirement of $\mathbb{Z}_2$ symmetry, i.e. invariance over $\phi\to -\phi$.
For standard Yukawa system, with a linear bare Yukawa interaction $H(\phi)=y \phi$,
this requires a discrete chiral symmetry $\psi \to i \psi$ and $\bar{\psi}\to i \bar{\psi}$. 
A generalization of local interactions with such a symmetry then requires an odd $H(\phi)$.
There is also the possibility to let the spinors unchanged under the transformation, which would require an even function $H(\phi)$.

The goal of this work is to construct global FP solutions of the flow equations compatible with the symmetry conditions, and to study the properties of the RG flow in their neighborhood.
The FPs, which describe scaling solutions, are computed by solving the coupled system of two ordinary differential equations  $\dot{v}=0$ and $\dot{h}=0$ or, in some cases, from the equivalent system for the quantities ($v$, $y=h^2$).
The dependence of such scaling solutions on the two parameters $d$ and $X_f$  is one of the main themes discussed in the literature as well as in the present work. 
Regarding the former, we will assume $2<d\leq4$ and qualitatively discuss how the number of critical models
varies with $d$, but we will especially concentrate on the properties of the $d=3$ system.
For the latter, we restrict ourselves to non-negative number
of degrees of freedom, and we start from the two simple limiting cases one can address.
The simplest is $X_f\!\to\!0$. In this case, the fermion sector remains nontrivial,
see Eqs.~(\ref{eq:floweq_h},\ref{eq:floweq_etapsi}), 
but is not allowed to influence the scalar dynamics, which is therefore identical to the fermion-free model, see Eqs.~(\ref{eq:floweq_v},\ref{eq:floweq_etaphi}).
Hence, as far as criticality is concerned, 
we expect to observe the same pattern of FPs that can be observed without fermions,
with the same critical exponents in the scalar sector, even if at generically nonvanishing values of the Yukawa couplings.
The second limit which brings radical simplifications is $X_f\to\infty$, and it is discussed in the next Section.

\section{Leading order large $-\, X_f$ expansion.}
\label{sec:largeXf}

Large-$N_f$ methods are a traditional and successful way to analyze the strongly
coupled domain of the three dimensional Gross-Neveu model, 
which is renormalizable at any order in a $1/N_f$-expansion~\cite{ZinnJustin:1991yn,Gracey:1993kb,Gracey:1993kc}.
As a consequence, any other nonperturbative method is challenged to reproduce known results in this
limit. For this reason, before moving to the finite-$X_f$ results provided by the FRG,
let us start with discussing the behavior of this simple Yukawa model 
with many fermionic degrees of freedom, within the basic parameterization of its dynamics provided
by~\Eqref{eq:truncation}, in a continuous set of dimensions $2<d<4$.
This FRG analysis, for the case of a linear Yukawa function, has already been performed in~\cite{Braun:2010tt}.
Our results can be considered as an extension of it, to include a generic function $h(\phi)$.
As we will see, the main advantage that this brings at large-$X_f$ is the possibility to describe also multi-critical
models in $d<3$.

In this Section let us replace $v$ with $X_f \,  v$,
as well as $\eta_\phi$ with $X_f \,  \eta_\phi$,
and look at the leading order in $1/X_f$.
The first simplification is the fact that only canonical scaling terms and pure fermion loops survive.
Therefore the flow equations at this order reduce to
  \bea
 \dot{v}&=&-d v +\frac{d-2+\eta _{\phi }}{2} \phi\, v'
-2 v_d  l_0^{(\mathrm{F})d}(h^2)\\
 \dot{h}&=&h \left(\eta _{\psi }-1\right)+\frac{d-2+\eta _{\phi }}{2} \phi\,  h' \\
 \eta_\phi&=&\frac{4 v_d}{d}  (h')^2
\left[ m_{4}^{(\mathrm{F})d}(h^2)-2h^2 m_{2}^{(\mathrm{F})d}(h^2) \right]\\
 \eta_\psi&=&0\ .
 \eea
Let us draw some general considerations about the FP solutions,
by postponing the task of consistently solving the flow equation for $\eta_\phi$.
The equation for $h$ is almost regulator-independent
(apart for the value of $\eta_\phi$)
and the solution is a simple power
\be
h(\phi)=c_h\, \phi^{2/(d-2+\eta_\phi)}\ .
\ee
This is real only if the exponent is rational and with an odd denominator.
Furthermore it is smooth only if the exponent is a positive integer.
The FP solution for $v$ is instead regulator dependent.
Adopting the linear regulator, in $2<d<4$ it reads
\be
v(\phi)=c_v\, \phi^{2d/(d-2+\eta_\phi)}-\frac{4v_d}{d^2}\, {}_2F_1\!
\left(1,-\frac{d}{2};1-\frac{d}{2};-h(\phi)^2\right)\ .
\ee
The function ${}_2F_1\!\left(1,-\frac{d}{2};1-\frac{d}{2},-x\right)$, which actually can be reduced 
to a Hurwitz-Lerch function $-\frac{d}{2}\Phi\left(-x,1,-\frac{d}{2}\right)$,
has a logarithmic singularity at $x=-1$,
therefore the condition that $h(\phi)$ be real entails that this singularity is
always avoided, and that the potential is globally defined.
On the other hand, the smoothness of $v$ is not for granted.
Since
\be\label{eq:hyperFsmallx}
{}_2F_1\left(1,-\frac{d}{2};1-\frac{d}{2};-x\right)=1-\frac{d}{d-2}x-\frac{d}{4-d}x^2+O(x^3)
\ee
and since this function is always convex, the leading $\phi$-dependence of $v$ at its minimum,
i.e. at the origin, is provided by $h^2(\phi)$ itself.
Hence, the latter must be a smooth function, because we
want the couplings associated to the derivatives of the potential at
the minimum to be well defined at the FP.
The same reasoning, if applied to the Yukawa couplings, 
leads to the requirement that $h(\phi)$ be
smooth at the origin.
This translates into a quantization condition
on the dimensionality of the scalar field
\be\label{eq:prequantizedetaphi}
\frac{d-2+\eta_\phi}{2}=\frac{1}{n}\ , \quad n\in\mathbb N
\ee
which is a consequence of the large-$X_f$ limit.

We find it helpful, for the interpretation of this relation,
to consider a similar condition at the purely scalar FPs,
with trivial Yukawa interaction.
With this we mean the limit $X_f\to\infty$
followed by $c_h\to0$, which is clearly not
the same as the fermion-free model; yet,
by consistency, this limit should describe 
the classical properties of the latter model.
Indeed, if $c_h=0$ the only condition left is that
the homogeneous part of the FP scalar potential
be smooth and stable, that is
\be
\frac{d-2+\eta_\phi}{2}=\frac{d}{2n}\ , \quad n\in\mathbb N\, .
\ee
The meaning of this constraint is well known.
By neglecting the quantum corrections, hence setting $\eta_\phi=0$,
one would deduce that the smooth bounded solutions $v(\phi)=c_v \, \phi^{2n}$ 
are allowed only in
\be
d_n=\frac{2n}{n-1}=2+\frac{2}{n-1}\ ,\quad n\in\mathbb N\ .
\ee
This is the usual tree-level counting according to which
the interaction $\phi^{2n}$ is marginal in $d_n$ and becomes relevant for $d<d_n$.
From the quantum point of view, these dimensions are the corresponding upper 
critical dimensions for multi-critical universality classes.
For any $n$, below $d_n$ a new FP with nontrivial $\eta_\phi$
branches from the Gau\ss ian FP and survives for $2\leq d<d_n$~\cite{Felder,Codello:2012sc}.
In the purely scalar model,
this is already visible within a simple LPA of the FRG,
where it is indeed possible to unveil and describe some
properties of these universality classes
in a whole continuum of dimensions $2<d<d_m$.
In the leading order of the large-$X_f$ expansion,
the fact that quantum effects allow for these FPs at any $2\leq d<d_n$
remains invisible. This is because in the LPA one sets $\eta_\phi=0$,
and in the LPA${}^\prime$ the $c_h\to0$ limit again forces a vanishing anomalous dimension.
This simply signals that the two limits $X_f\to\infty$ and $h(\phi)\to 0$ do not commute.

A similar analysis can be applied to the Yukawa system.
Namely, if one forces classical scaling and
sets $\eta_\phi=0$, the large-$X_f$ limit constrains
$d$ to the critical values
\be\label{eq:uppercritdim}
d_n=2+\frac{2}{n}\ , \quad n\in\mathbb N
\ee
that are exactly the dimensions at which the
interaction terms  $\phi^n\,\bar{\psi}\,\psi$
become marginal. Notice that they coincide with
the critical dimensions of an even scalar potential, and that
by selecting odd or even functions
$h(\phi)$ one can reduce the number of critical dimensions 
for $h$ by a factor of two.
As soon as anomalous scaling is allowed,
the large-$X_f$ limit tells us that the
nontrivial FPs can indeed exist
for $d<d_n$, and quantizes the
corresponding anomalous dimensions
\be\label{eq:quantizedetaphi}
\eta_\phi=\frac{2}{n}+2-d=d_n-d\ ,\quad n\in\mathbb{N}\ .
\ee
Notice that they get smaller and smaller, the closer
$d$ is to the upper critical dimension $d_n$.
As a consequence, the value of $X_f$ at which one expects a breakdown of the LPA with $\eta_\phi=0$
must be a decreasing function of $(d_n-d)$.
Unfortunately, the latter is maximum for the very interesting $n=1$ scaling solution,
which includes the $d=3$ Gross-Neveu universality class.
However, even in this case, for small enough $X_f$
we have no a-priori reason to discard the use of the LPA for a first
study of the critical Yukawa models.
On the other hand, for the $n=1$ scaling solution the LPA${}^\prime$ is able to reproduce~\Eqref{eq:quantizedetaphi}
and therefore provides a consistent picture of this critical model for any $X_f$, see App.~\ref{sec:largeXf_appendix}.
This is not the case for the $n>1$ multi-critical models, whose nontrivial scaling properties 
require larger truncations of the FRG.

Before going on and discussing the finite-$X_f$ results, let's comment
on the universal critical exponents that one should approach in a 
large-$X_f$ limit, since they provide an important reference point
for the finite-$X_f$ investigations.
The eigenvalue problem for the linearized flow in vicinity of the
large-$X_f$ FPs is solved in App.~\ref{sec:largeXf_appendix}, both in
the LPA and in the LPA${}^\prime$.
The result is that one can safely split the problem into two classes of perturbations.
The former have $\delta h(\phi)=0$ and $\delta v(\phi)=\delta c_v \phi^M$, where we 
required the potential to be smooth, thus quantizing the corresponding 
critical exponents to the values
\be
\theta_{M}=d-M\left(\frac{d-2+\eta_\phi}{2}\right)=d-\frac{M}{n}\ , \quad M\in\mathbb N 
\ee
i.e. the dimensionality
of the couplings in front of $\delta v(\phi)$.
The latter have $\delta h(\phi)=\delta c_h \phi^N$, a nontrivial $\delta v(\phi)$,
and again
\be
\theta_{N}=1-N\left(\frac{d-2+\eta_\phi}{2}\right)=1-\frac{N}{n}\ , \quad N\in\mathbb N 
\ee
where we used~\Eqref{eq:prequantizedetaphi} as before.
As a consequence,  the large-$X_f$ exponents are independent of $c_h$ and $c_v$.
They are Gau\ss ian in the sense that they are directly linked to the dimensionality of the fields by naive dimensional counting,
but the latter dimensionality, as far as the scalar is concerned, is deeply non-Gau\ss ian and
actually independent of $d$.

As usual one can observe a hierarchy among FPs with different $n$.
For example, let us restrict ourselves to the slice of theory space parameterized by the couplings inside $h(\phi)$ only.
Then, for a FP labelled by the integer $n$, there are $n$ relevant
operators, namely $\phi^0,\dots,\phi^{n-1}$, and one marginal operator, $\phi^n$ itself.
Within the LPA, the latter can be exactly marginal  since it corresponds to
shifts in $c_h$. For the $n=1$ FP, the LPA${}^\prime$ is enough to change this
conclusion, since the flow equation for $\eta_\phi$ provides a condition that fixes the FP value of
$c_h$. For $n>1$, higher truncations are needed.
Thus, the $\bar{n}$-th FP can provide UV completion for theories
approaching the $n$-th FP in the IR, only if $n<\bar{n}$.
The detailed study of the global flows among these FPs is in principle a straightforward
task in the large-$X_f$ approximation, but it is out of the purposes of the present work.
We confine ourselves to sketching some properties of the FP potentials and of the
linearized perturbations in vicinity of the FPs, which can be found in App.~\ref{sec:largeXf_appendix},
together with some comments on how these nontrivial critical theories disappear in $d=4$.

\section{LPA at finite $X_f$ and generic $d$. Some features from numerical investigations.}
\label{sec:LPA_Xf1_genericd}

In the previous Section we described how the large-$X_f$
expansion supports the expectation that, as the number of dimensions is lowered from
$d=4$ towards $d=2$, across the upper critical dimensions of \Eqref{eq:uppercritdim},
new universality classes become accessible in the theory space of Yukawa models. 
In this Section we are going to present evidence that
this happens also at finite $X_f$. Here and in the rest of this work,
we restrict our analysis to the subset of theory space
which enjoys a conventional $\mathbb{Z}_2$-symmetry, such that $v$ is even and $h$ 
is odd. Furthermore we adopt the LPA and neglect the flow equations 
for the wave function renormalization of the fields. 
As it was argued in the previous Section, as well as in App.~\ref{sec:largeXf_appendix}
with more details, one cannot expect this approximation to perform well for any $n$
and $X_f$. Therefore the following studies should be understood as a first step
towards a proper description of these universality classes. 
Only the $d=3$ chiral Ising universality class will be later analyzed also in the
LPA${}^\prime$, by resorting to polynomial truncations of the potentials, see 
Sect.~\ref{sec:polynomial}.

Since we look for odd Yukawa potentials, we can restrict the list of the operators that become relevant at the corresponding
critical dimensions:
\bea\label{eq:Z2_relevant_dimensions}
\phi^{2n}\, &:& \qquad  d_c^{v}(n\ge2)=\frac{2n}{n-1}=4,3, \frac{8}{3},\frac{5}{2},\frac{12}{5} \cdots \nonumber\\
\phi^{2n+1}\bar{\psi}\psi\, &:& \qquad d_c^{h}(n\ge0)=\frac{4(n+1)}{2n+1}=4, \frac{8}{3},\frac{12}{5} \cdots 
\eea
In order to reveal the new universality classes appearing below these dimensions, we follow the strategy developed in~\cite{Morris:1994ki},
that was already successfully applied to the purely scalar model in continuous dimensions~\cite{Codello:2012sc}. 
This consists in solving the FP condition, which is a Cauchy problem involving a system of two coupled second order ODEs, 
by a numerical shooting method, i.e. varying the initial conditions in a space of parameters which is  two dimensional,
since two of the four boundary conditions are fixed by the symmetry requirements ($v'(0)=0$ and $h(0)=0$).
For the potential $v$ we choose as parameter $\sigma=v''(0)$, relating it to $v(0)$ using the differential equation. For $h$ we use $h_1=h'(0)$.
Trying to numerically solve the non linear differential equations with generic initial conditions,
 one typically encounters a singularity at some value of $\phi_c(\sigma,h_1)$ where the algorithm stops.
Such value increases in a steep way close to the initial conditions which correspond to a global solution, even if the numerical errors mask partially this behavior. 
As a consequence, in our case a three-dimensional plot for $\phi_c(\sigma,h_1)$ is very useful to gain a first understanding of the positions of the possible FPs.

\begin{figure}[!t]
\begin{center}
\includegraphics[width=0.3\textwidth]{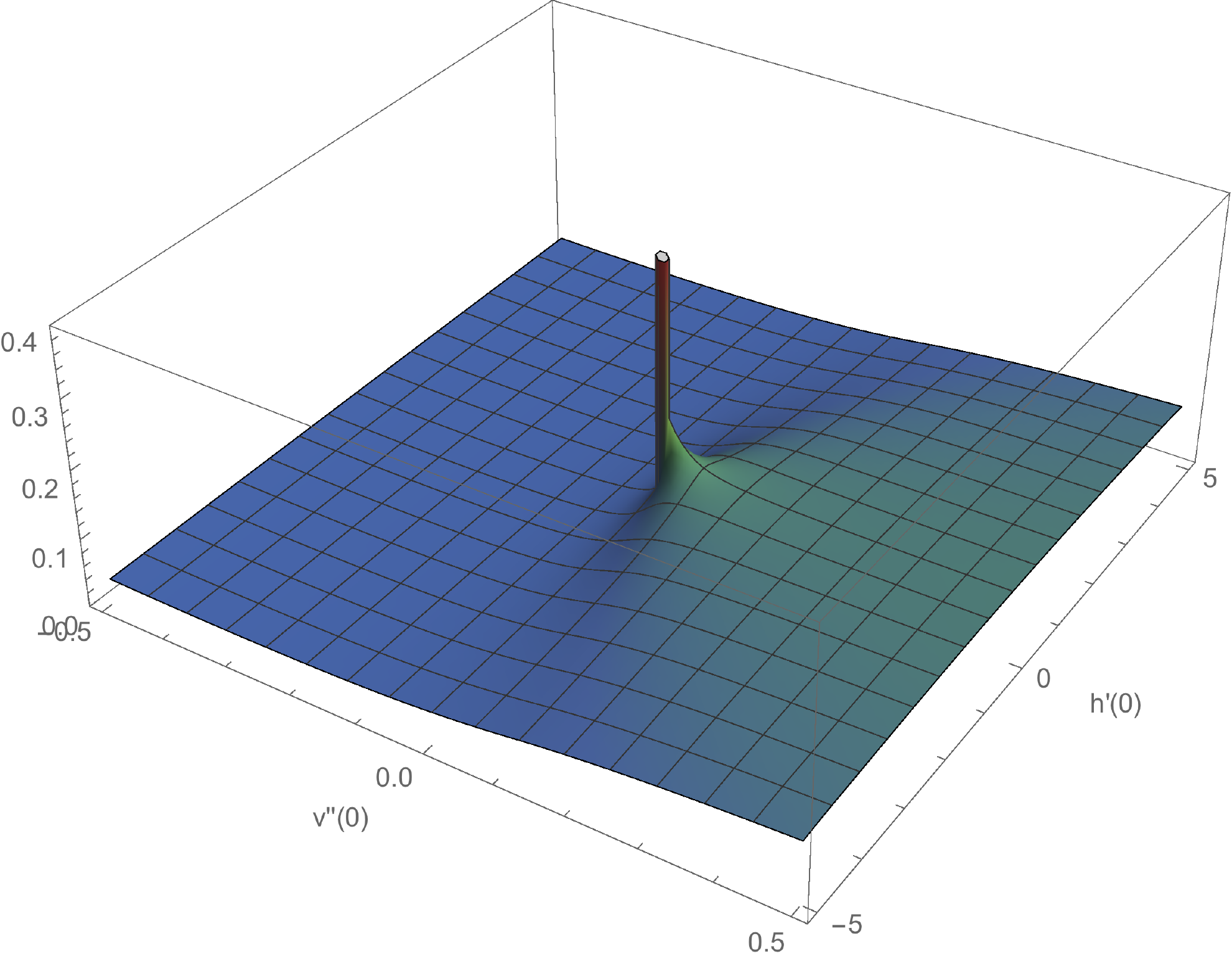}
\includegraphics[width=0.3\textwidth]{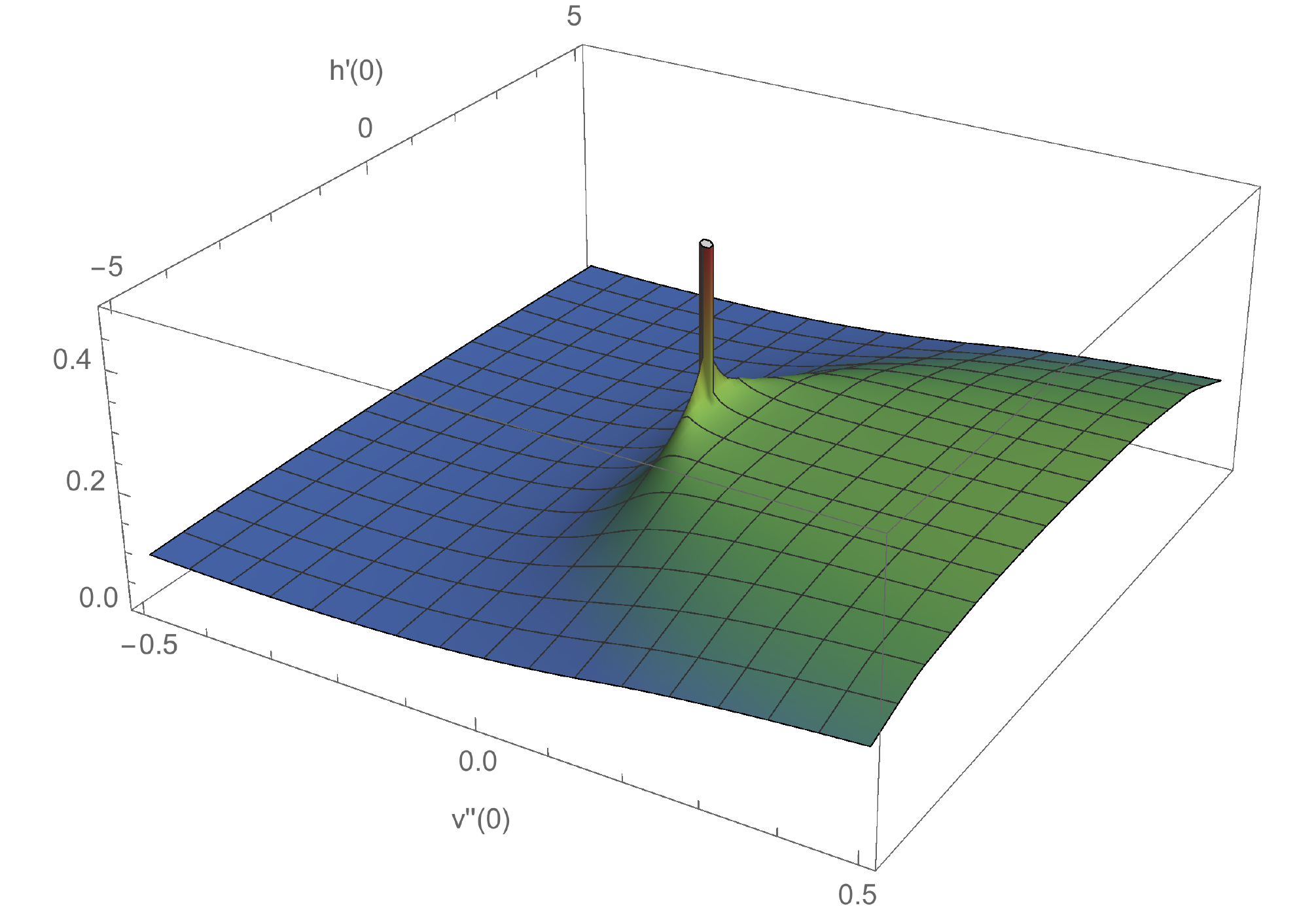}
\includegraphics[width=0.3\textwidth]{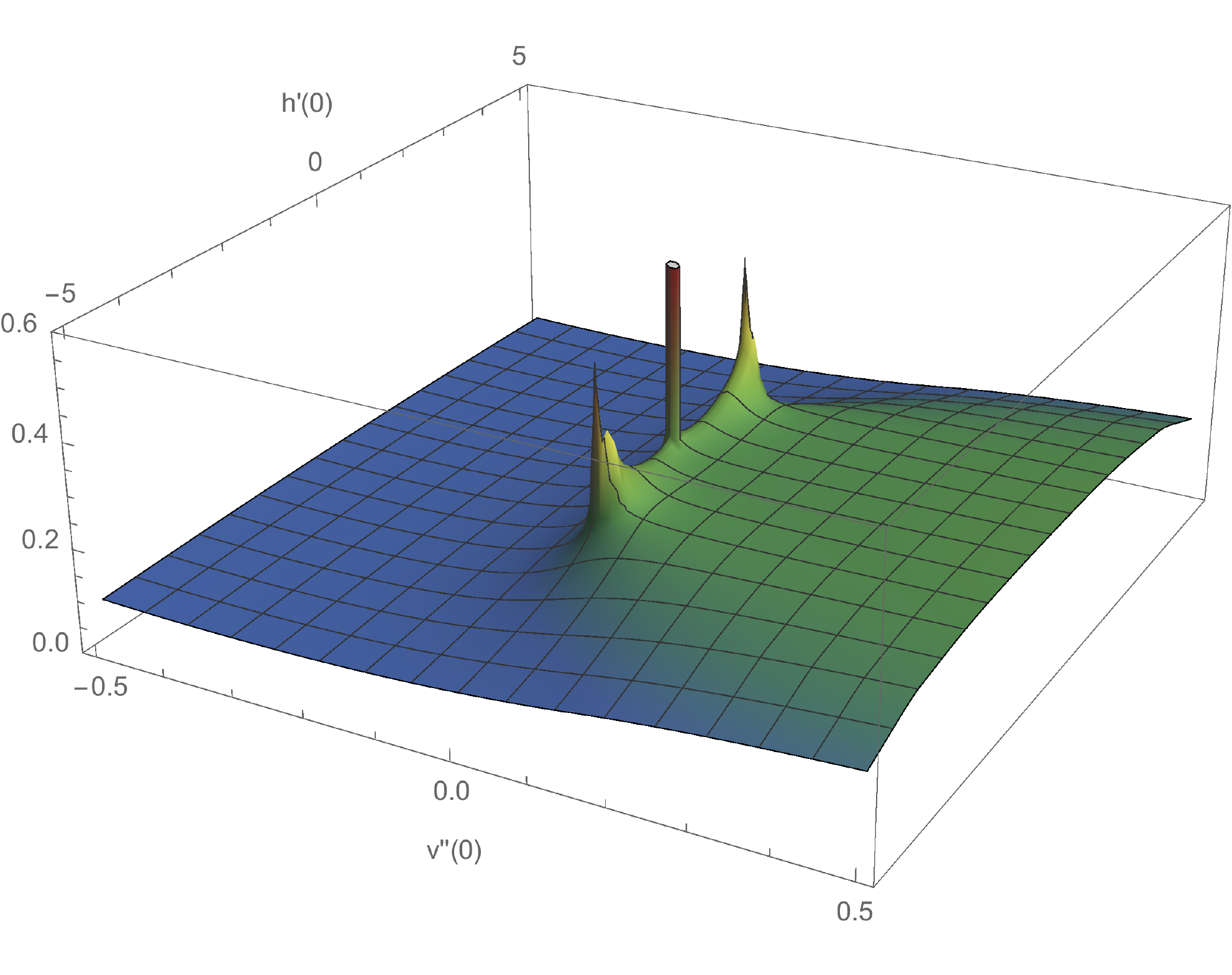}\\
\includegraphics[width=0.3\textwidth]{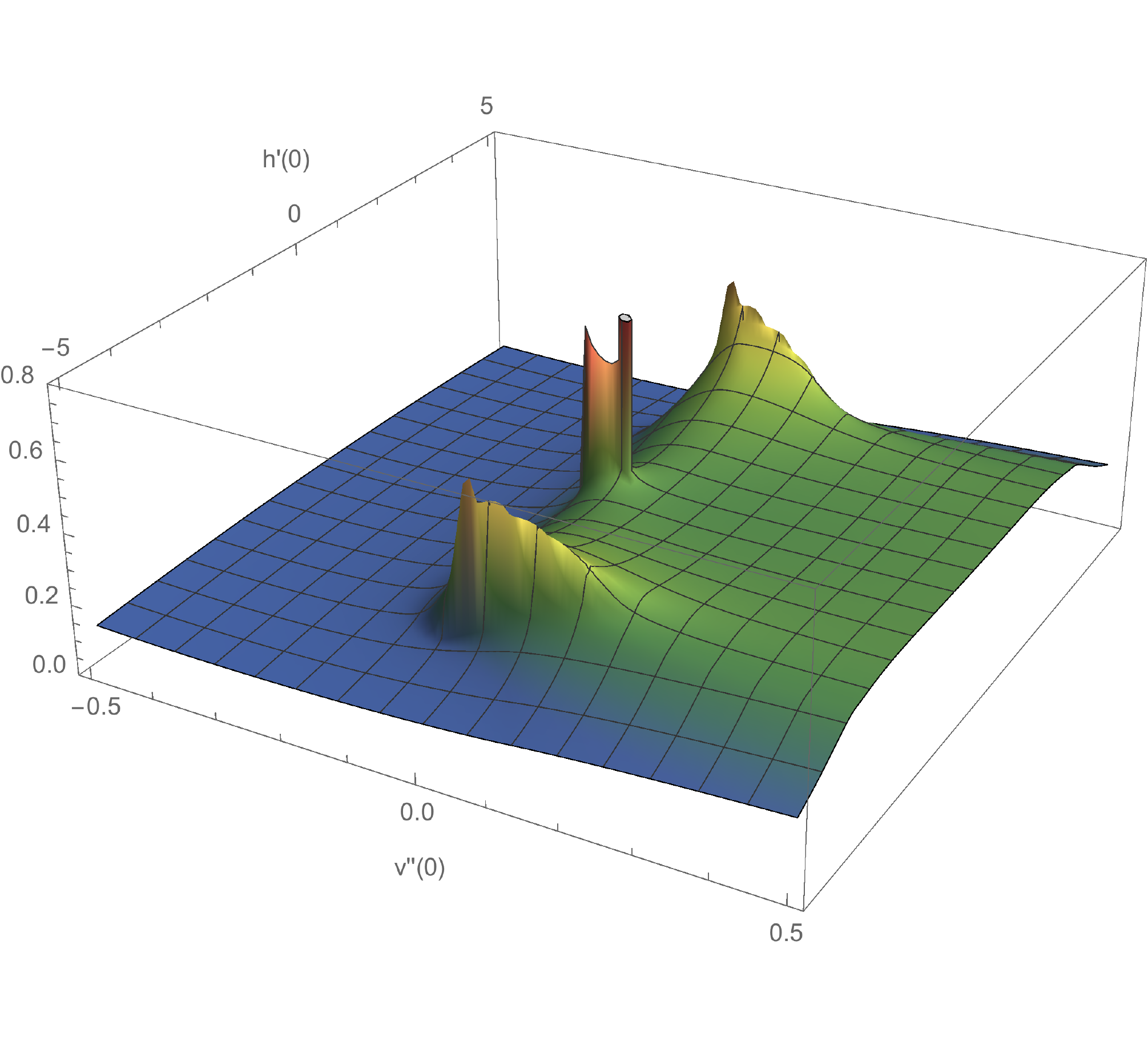}
\includegraphics[width=0.3\textwidth]{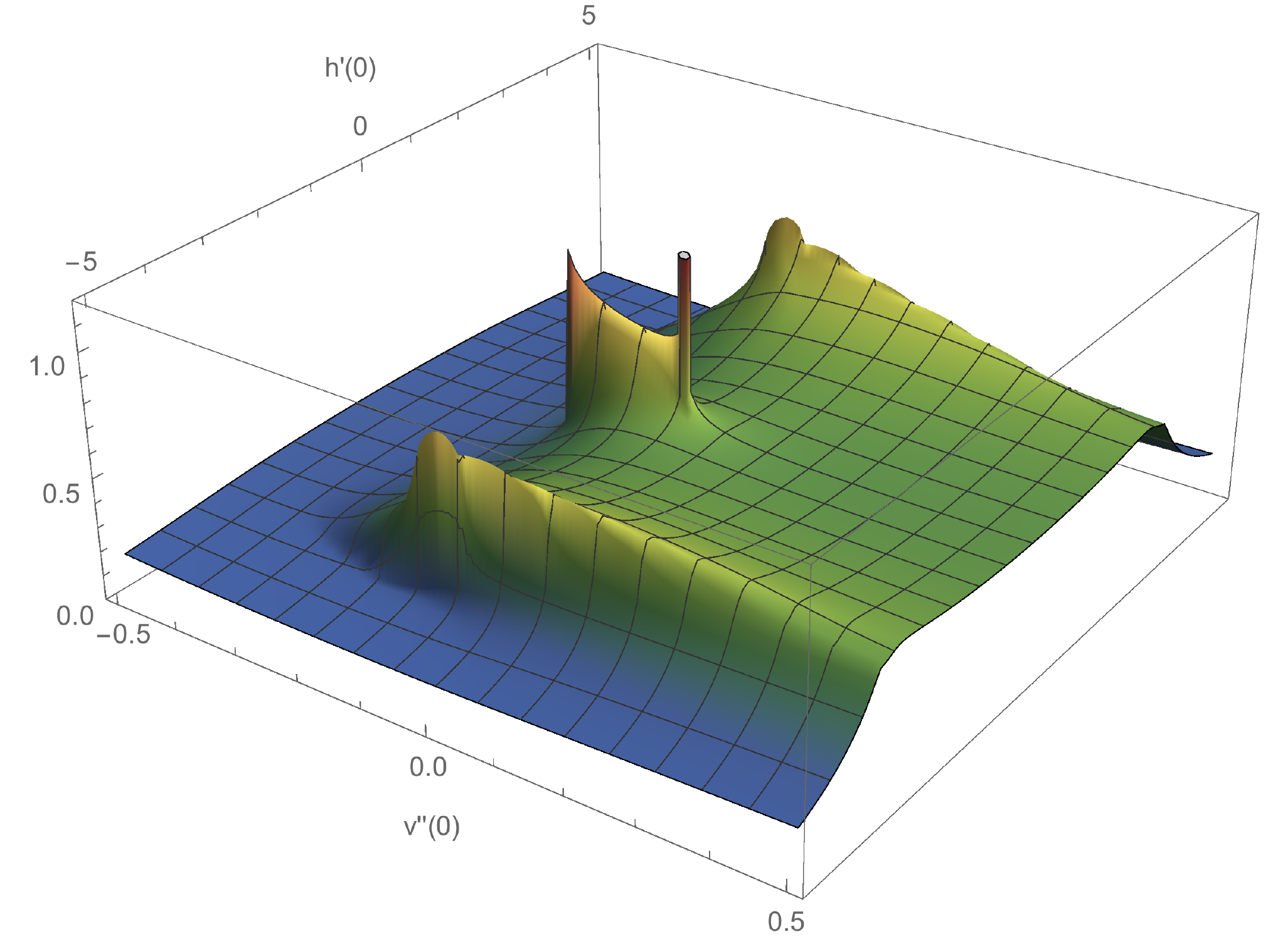}
\includegraphics[width=0.3\textwidth]{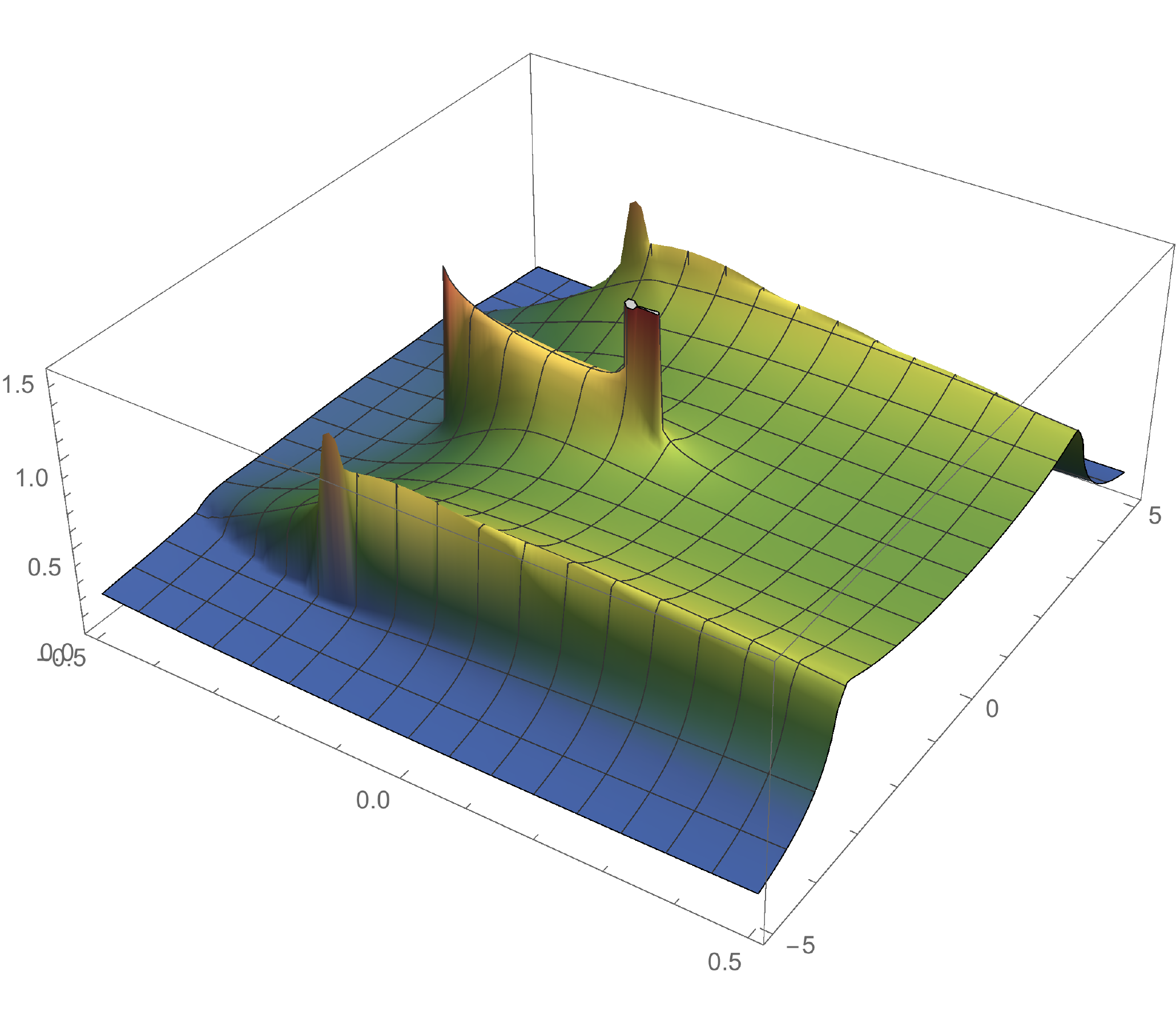}\\
\includegraphics[width=0.3\textwidth]{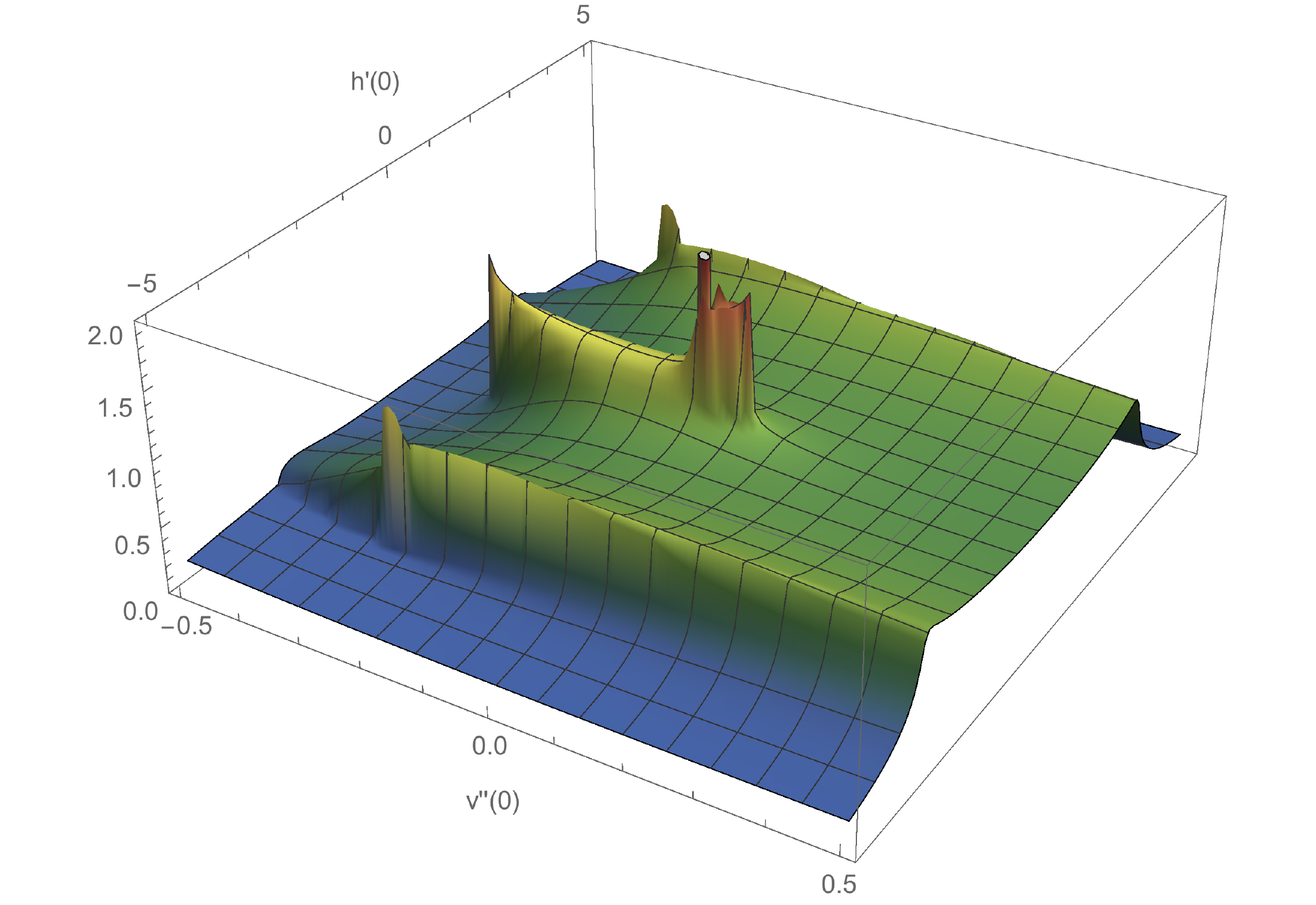}
\includegraphics[width=0.3\textwidth]{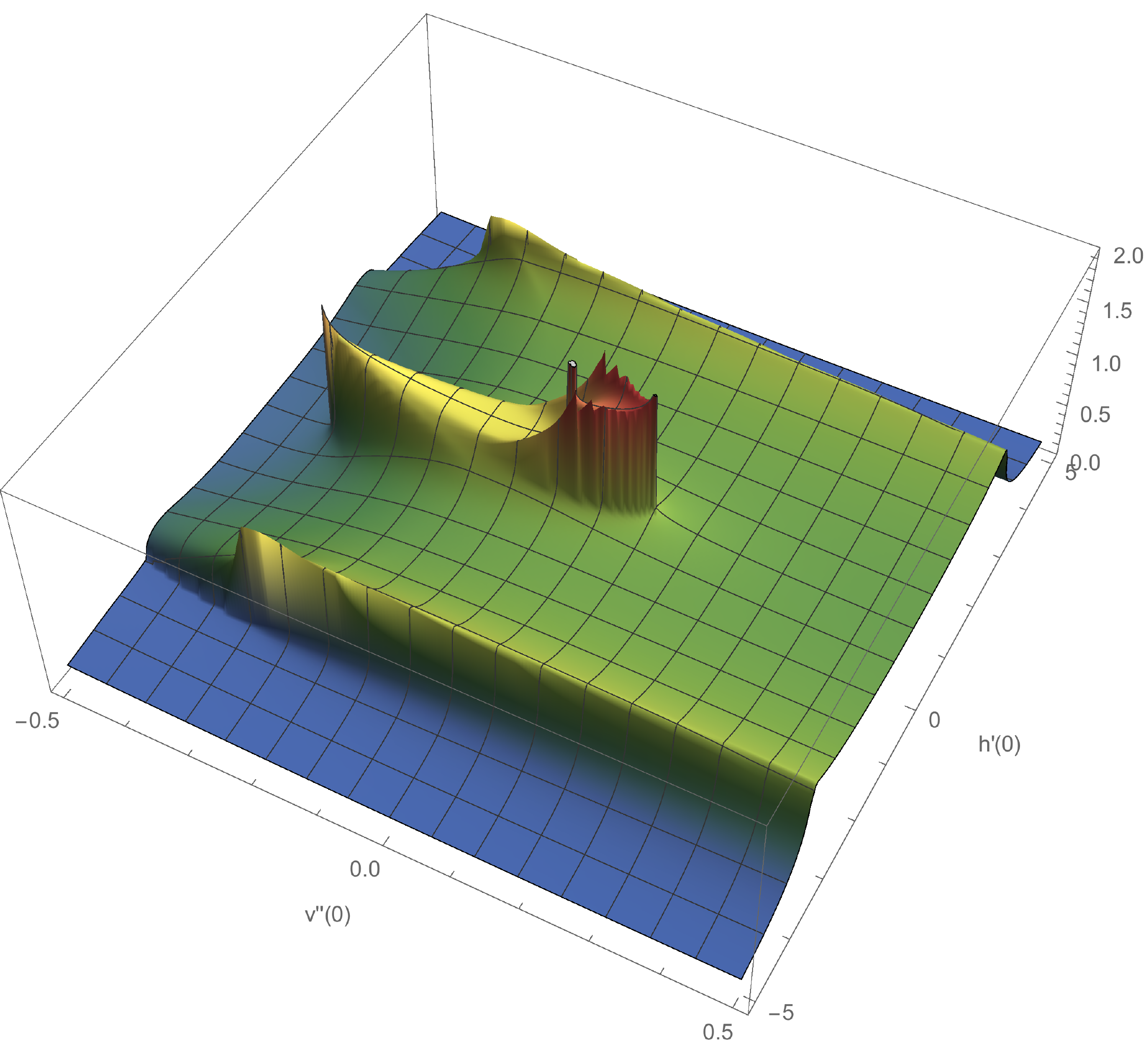}
\includegraphics[width=0.3\textwidth]{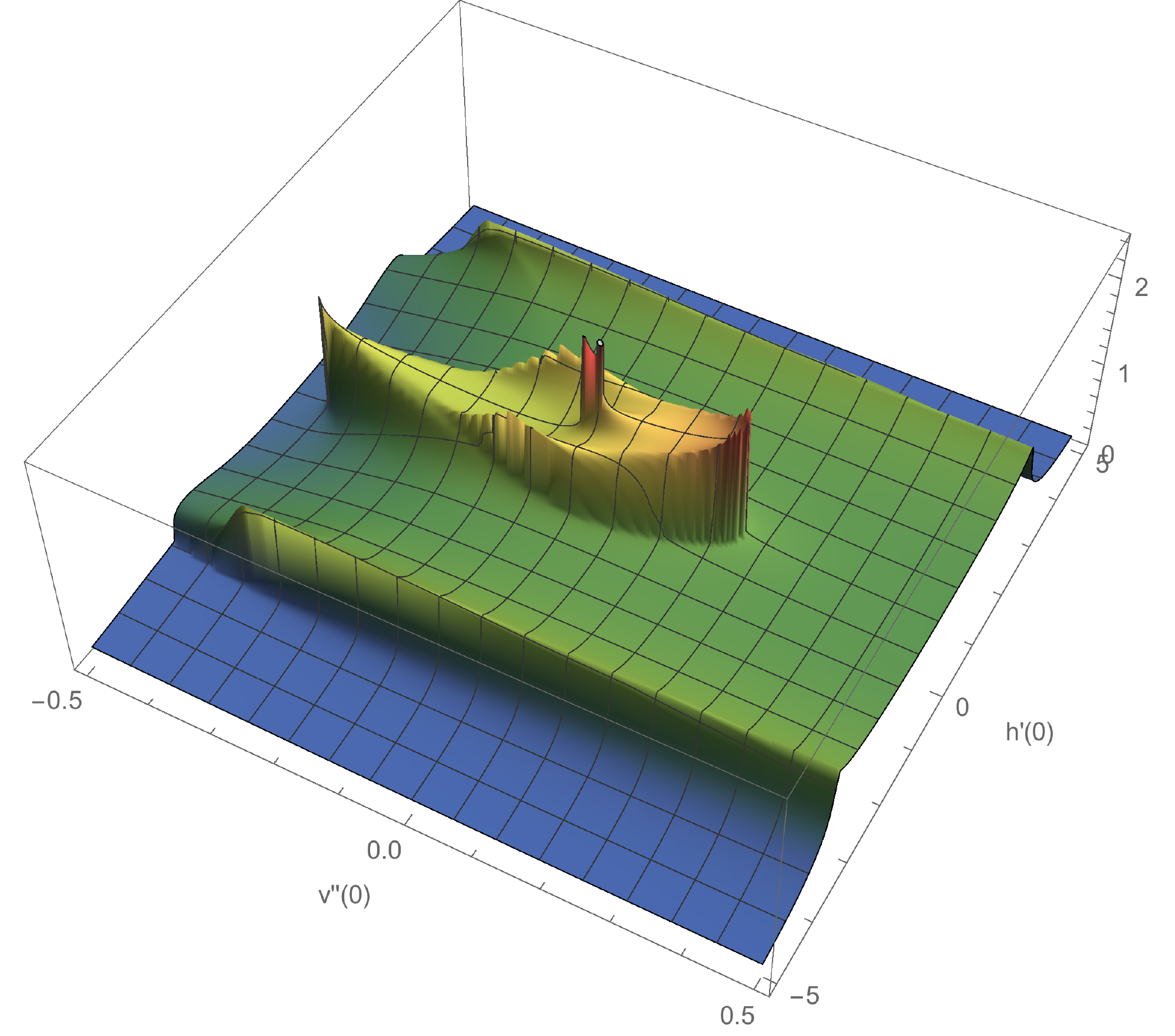}
 \caption{Spike plots for $X_f=1$ on varying the dimension: $d=5,4,3.9,3.5,3,\frac{8}{3},\frac{8}{3}-\frac{1}{10},\frac{5}{2},\frac{12}{5}$, 
 from left to right and from top to bottom.}
\label{fig:spike1}
\end{center}
\end{figure}

In Fig.~\ref{fig:spike1} we show the results of this analysis, for $X_f=1$ and for several dimensions:  
$d=5,4,3.9,3.5,3,\frac{8}{3},\frac{8}{3}-\frac{1}{10},\frac{5}{2},\frac{12}{5}$.
For $d=5$ and $d=4$, as it is expected, we see a single spike in $(\sigma,h_1)=(0,0)$ which corresponds to the Gau\ss ian solution.
More details on this are given, for $X_f<1$, in Sect~\ref{sec:d=4}.
In $3<d<4$ we have crossed the threshold below which both the operators $\phi^4$ and $\phi \bar{\psi}\psi$ become relevant, 
as is shown in Eq.~(\ref{eq:Z2_relevant_dimensions}).
In this interval, it is evident from the figure that we find three new spikes.
One is characterized by $h_1=0$ and $\sigma<0$ and corresponds to the Ising critical solution.
It is clearly visible in the fourth and fifth panels of Fig.~\ref{fig:spike1},
but not in the third, since it is very close to the Gau\ss ian FP.
The other two are physically equivalent, since they lie at opposite values of $h_1$,
and correspond to the chiral Ising universality class.
They have $\sigma<0$, which suggests that also these scaling solutions are in a broken regime for $X_f=1$, at least in the LPA approximation.
Moving to $\frac{8}{3}<d<3$ we cross the marginality-threshold for the operator $\phi^6$, but no other operators involving fermions have to be added to the set of the relevant ones.
This corresponds to the appearance of the tricritical theory in the pure scalar sector, as we see from the new spike which develops with $\sigma>0$ and $h_1=0$.
Once $d<\frac{8}{3}$ also the new operators $\phi^8$ and $\phi^3 \bar{\psi}\psi$ become relevant and new critical solutions may appear. 
Indeed, in the left and the central plot of the third line of Fig.~\ref{fig:spike1} we see two new spikes, 
which again occur at opposite values of $h_1$ and are therefore equivalent, this time with $\sigma>0$.  
Finally in the lower-right plot, where we present the case $d=\frac{12}{5}$, which is lower than $\frac{5}{2}$ enough to clearly see the effects of the new relevant scalar operator $\phi^8$, one can appreciate the third new spike at $\sigma<0$ and $h_1=0$.
The latter FP corresponds to the quadricritical scalar model as described for example in~\cite{Morris:1994jc,Codello:2012sc}.
The former solutions, already assuming that they globally exist, define what one could call the chiral quadricritical Ising universality class,
since they originate from the Gau\ss ian FP together with the purely scalar quadricritical model.

\begin{figure}[!t]
\begin{center}
\includegraphics[width=0.5\textwidth]{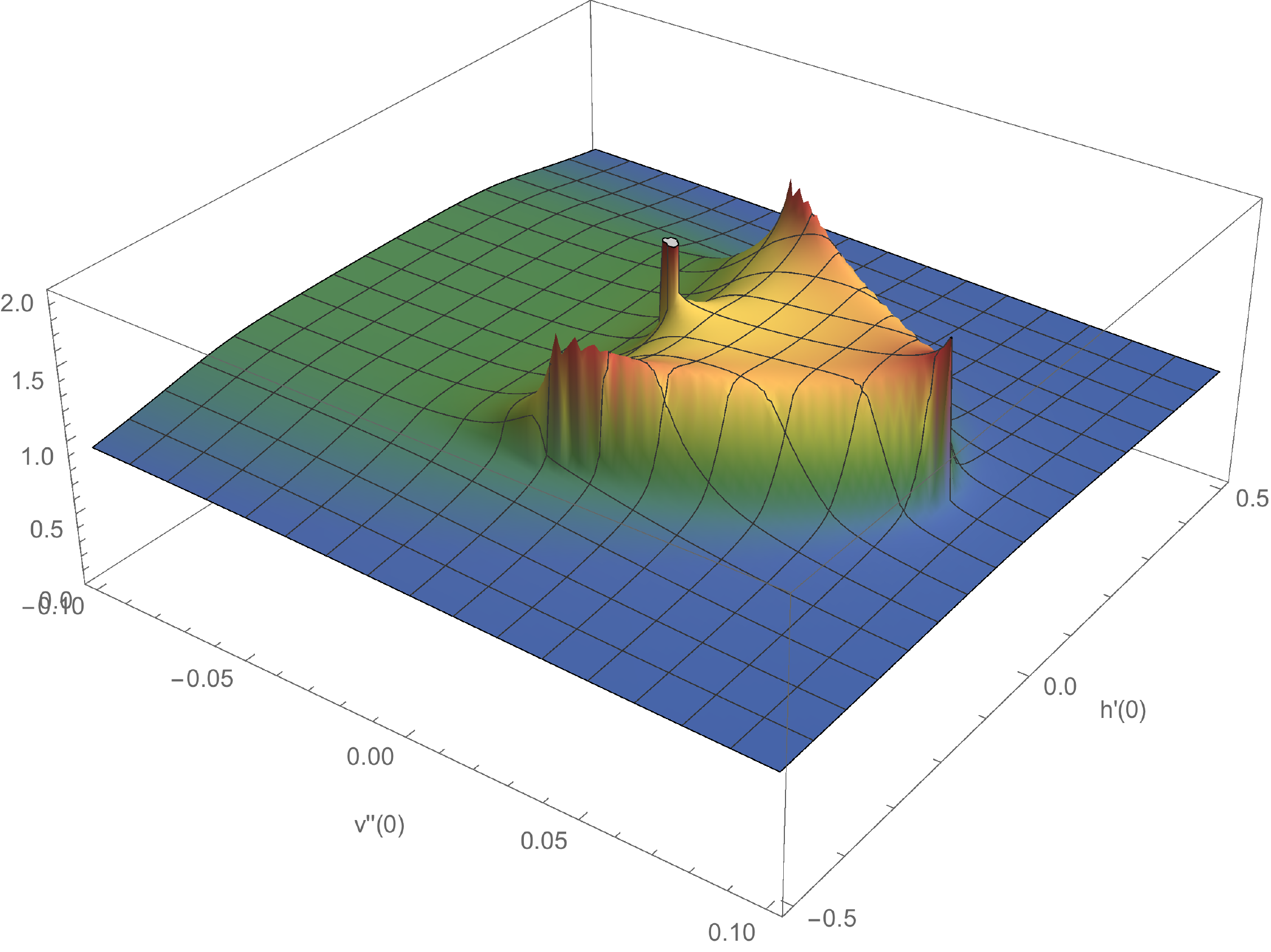}
 \caption{Spike plot for $d=\frac{8}{3}-\frac{1}{10}$ and $X_f=1$, zoomed area around the origin.}
\label{fig:spike2}
\end{center}
\end{figure}
\begin{figure}[!t]
\begin{center}
\includegraphics[width=0.5\textwidth]{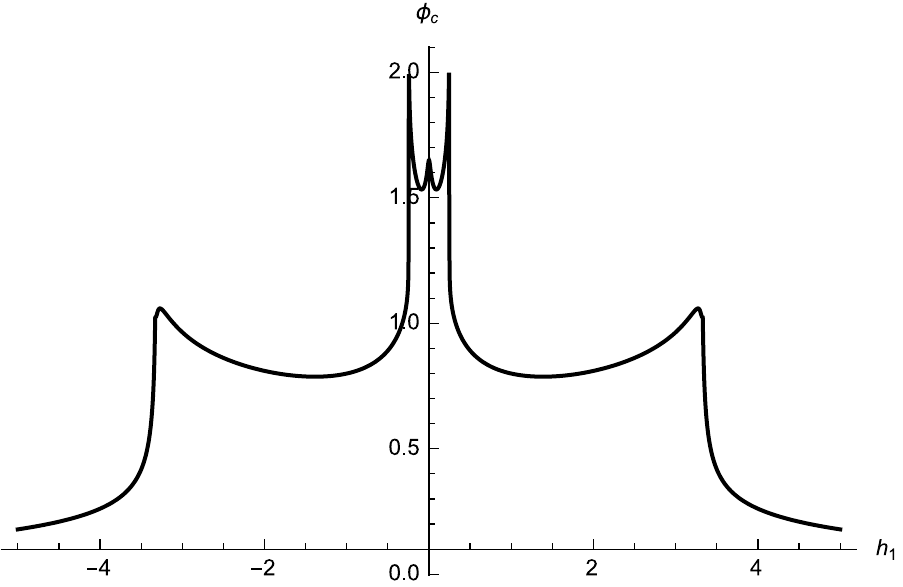}
 \caption{Spike plot for $d=\frac{8}{3}-\frac{1}{10}$ and $X_f=1$, zoomed area around the origin.}
\label{fig:spike3}
\end{center}
\end{figure}
We don't show more plots with lower values of $d$, since the pattern is pretty clear.
Pushing further this analysis towards dimensions close to $d=2$, though conceptually straightforward, would probably anyway require
more than the LPA.
To provide the reader with some more details, in Fig.~\ref{fig:spike2} we 
zoom in the panel of Fig.~\ref{fig:spike1} that refers to $d=\frac{8}{3}-\frac{1}{10}$ .
The three non trivial spikes which appeared at higher values of $d>3$ are now out of this graph.
From this figure one can see with more accuracy the presence of the three new nontrivial solutions.
The two of them which lie at $h_1\ne0$, can also be visualized by a plot at constant value of $\sigma$,
approximately corresponding to the position of the peaks, see Fig.~\ref{fig:spike3}.
Here the range of $h_1$ is wider than in Fig.~\ref{fig:spike2}, so that one can see also a trace
of the FPs generated at $d<4$, which are nevertheless located at a different value of $\sigma$.

The analysis we discussed in this Section can be repeated for other values of $X_f$, thus getting a
qualitative understanding of the position of the FPs as a function of both $d$ and $X_f$.
However, because of the uncertainties in the location of these peaks, it is hard to get a good qualitative
knowledge of this function. Nevertheless, the latter is needed to prove that the arguments presented
in this Section are rigorous, that each of the peaks corresponds to one FP, and to compute the corresponding
critical exponents. For this reason, in the next Section we are going to adopt a different numerical method that
will allow us to precisely answer these questions, focusing on $d=3$ for definiteness, but allowing for
a generic $X_f$.

\section{$d=3$ LPA at finite $X_f$. Numerical solution of the FP equations}
\label{sec:d=3_numerical_FP}
In this Section we construct, for some specific cases, the numerical solutions for $v$ and $h$ of the FP
differential equations, obtained by setting Eqs.~\eqref{LPAeqv} and \eqref{LPAeqh} equal to zero,
 in a domain for the dimensionless field $\phi$ that covers the asymptotic region.
 This is what might be called a global scaling solution.
 For convenience, we have actually considered the equivalent system for the quantities $v(\phi)$ and $y(\phi)=h^2(\phi)$.
We focus here on $d=3$ for which, from the analysis at $X_f=1$ performed in the previous Section, 
we expect a FP with non-trivial scalar potential and Yukawa function.
In the following we are going to take several values of $X_f$ into account.
After having found the corresponding nontrivial FP potentials, we determine the associated critical exponents and eigenperturbations.
The knowledge of the global scaling solutions will be important for a study of the quality of polynomial expansions,
presented in Sect.~\ref{sec:polynomial} . The latter approach is very useful especially in the case of the LPA${}^\prime$, 
which gives us access to a self-consistent computation of the anomalous dimensions without enlarging the truncation to a full
next-to-leading order of the derivative expansion. 
Clearly this programmatic analysis can be repeated for other values of $d$.

\begin{figure}[!t]
\begin{center}
\includegraphics[width=0.45\textwidth]{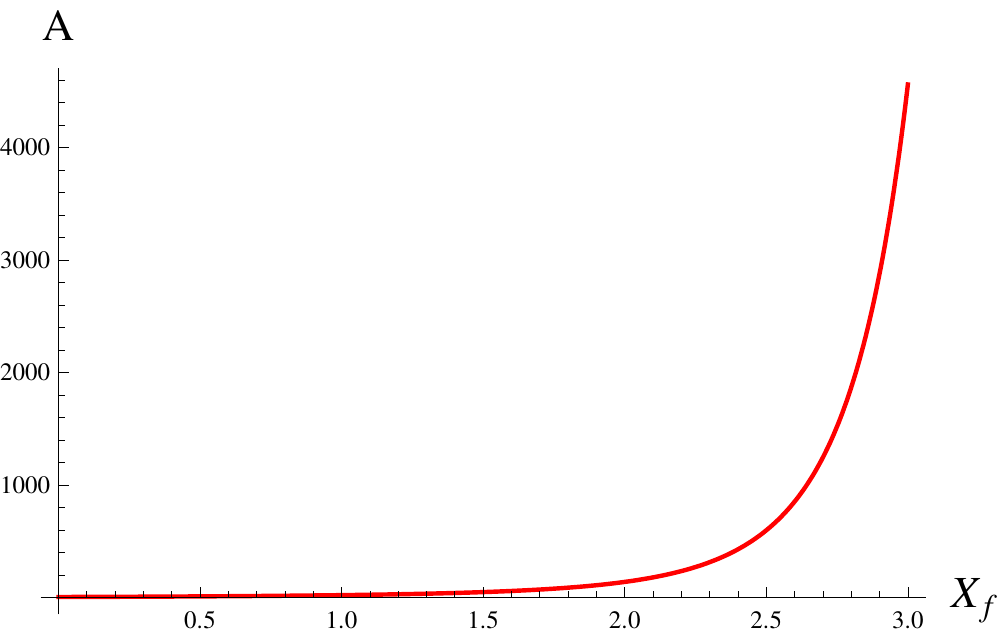}
\includegraphics[width=0.45\textwidth]{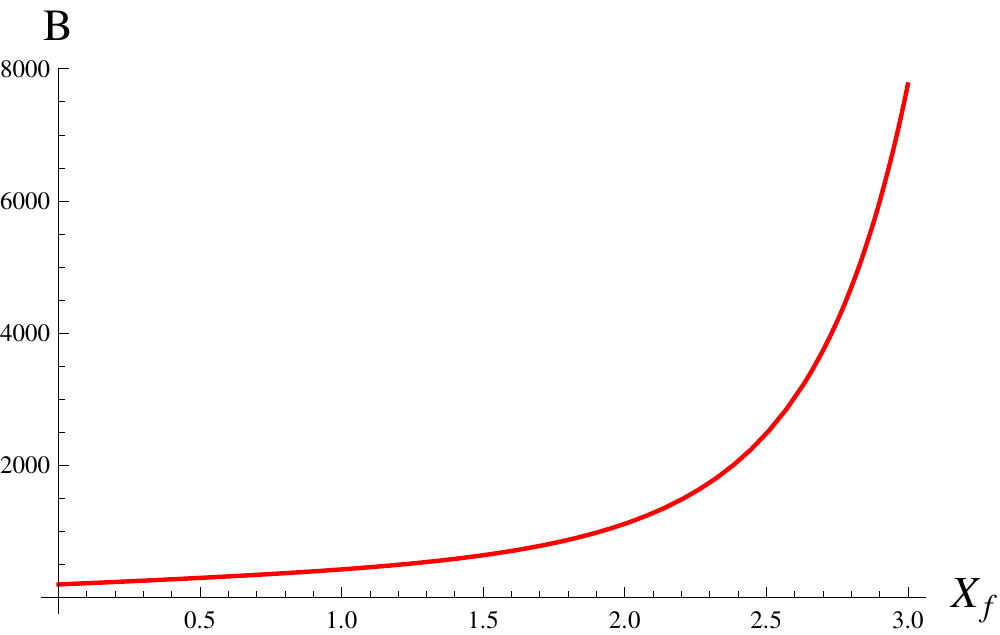}
 \caption{The values of the asymptotic parameters ($A$,$B$) defined by~\Eqref{vh_asympt} at the scaling solutions,
 varying $X_f$ in the range $10^{-3}<X_f<3$.}
\label{fig:locusAB}
\end{center}
\end{figure}

\begin{figure}[!t]
\begin{center}
\vspace{ -1.5cm}
\includegraphics[width=0.45\textwidth]{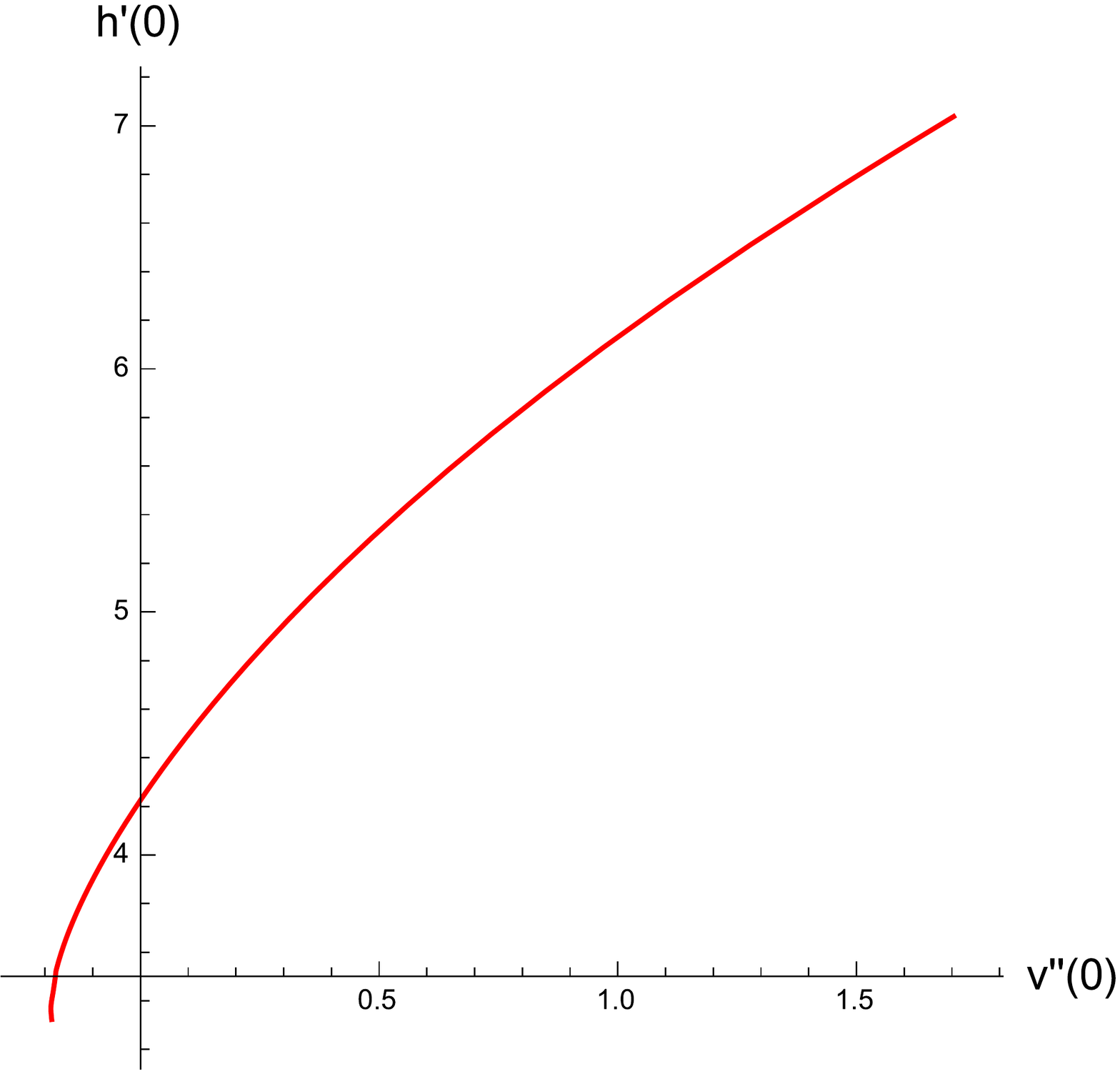}
\vspace{ -1.5cm}
 \caption{The values of ($v''(0)$,$h'(0)$) from the numerical global scaling solutions, varying $X_f$ in the range $10^{-3}<X_f<3$. 
One can notice the transition from the broken to the symmetric regime, which occurs at $X_f \simeq 1.64$ for the present LPA.}
\label{fig:locus}
\end{center}
\end{figure}

\begin{figure}[!t]
\begin{center}
\vspace{ -2.5cm}
\includegraphics[width=0.45\textwidth]{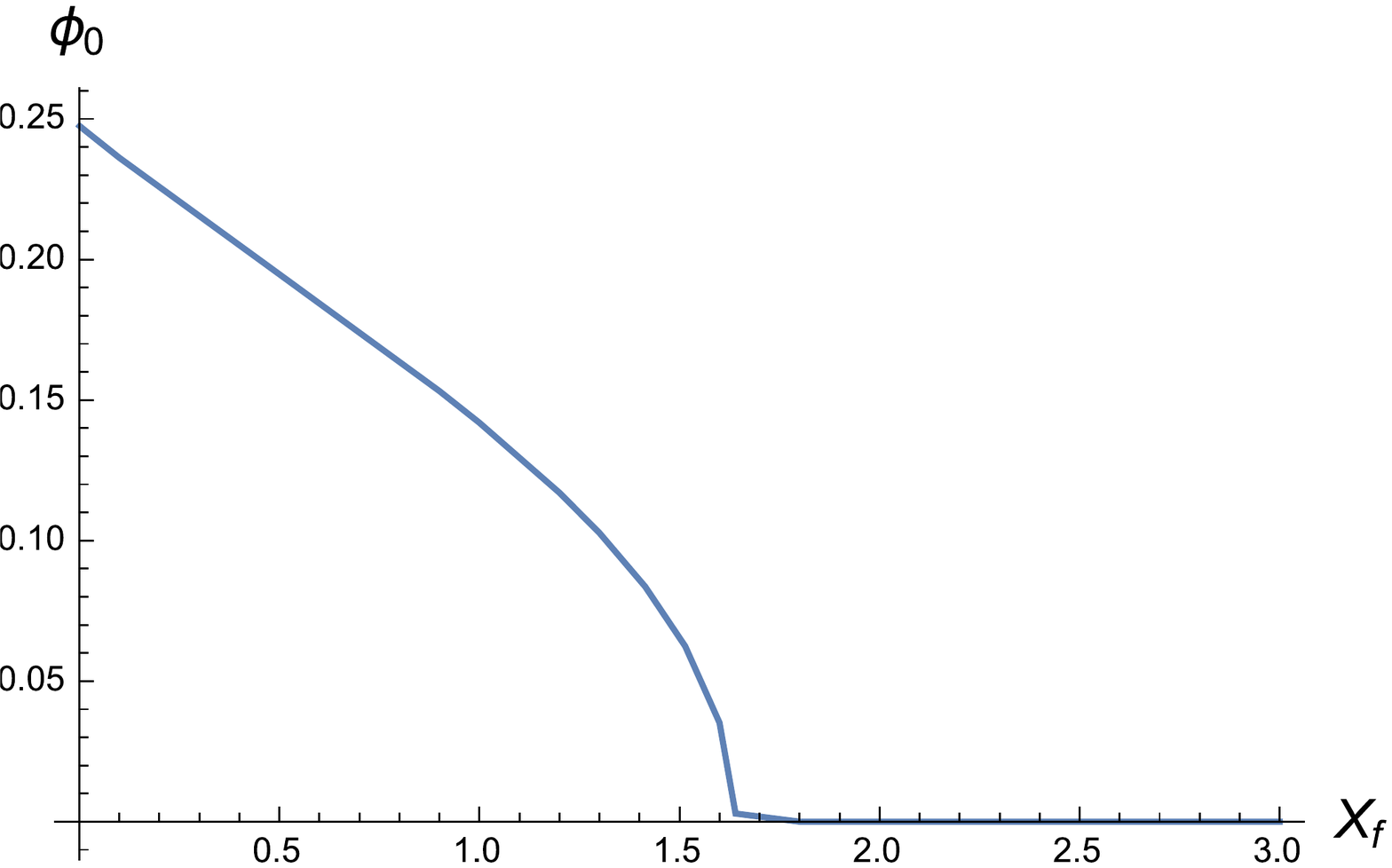}
\includegraphics[width=0.45\textwidth]{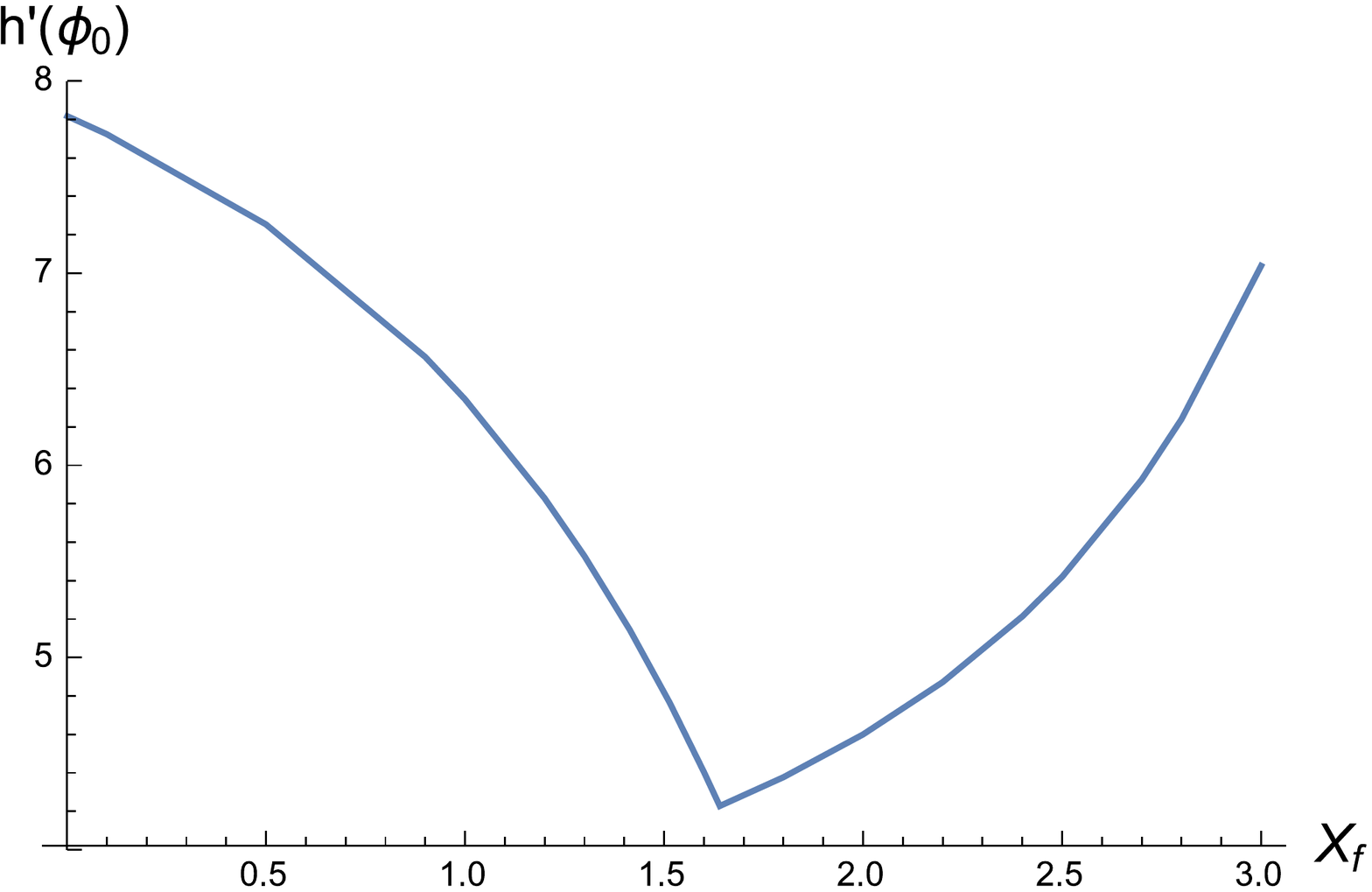}
\vspace{ -2.5cm}
 \caption{The vacuum expectation value $\phi_0(X_f)$ from the numerical 
global scaling solutions is shown in the left panel, while in the right panel we plot the corresponding value of $h'(\phi_0)(X_f)$, both in the LPA.}
\label{fig:vacuum}
\end{center}
\end{figure}

We choose to construct a global numerical solution by starting from the knowledge of the asymptotic behavior allowed by the FP equations.
Once the asymptotic expansions are determined with sufficient accuracy we proceed, with a shooting method, to the numerical integration
from the asymptotic region towards the origin.
The properties of the solutions which reach the origin depend on the free parameters in the asymptotic expansions. 
By requiring the solutions to transform correctly under $\mathbb{Z}_2$, one can uniquely fix the latter 
parameters to their FP values~\cite{Morris:1994ie}
The leading term of the asymptotic expansion for both $v$ and $h$ is determined, in the LPA with vanishing anomalous dimensions, 
by the classical scaling.
Here we report the first correction to it. Denoting $\alpha=2/(d-2)$,
the asymptotic behavior of the solution of the FP equations in the LPA reads
\be
\label{vh_asympt}
v_{\rm asympt}(\phi)\simeq 
A\, \phi ^{2 \alpha +2}+ \phi ^{-2 \alpha} \frac{C_d \,(B-2 A X_f(\alpha +1) (2 \alpha
   +1) )}{2 A B (\alpha +1) (2 \alpha +1) (d+2) } + \cdots
\ee
\be
h^2_{\rm asympt}(\phi)\simeq 
B\, \phi ^{2 \alpha} +\phi ^{-2-2 \alpha} \frac{ C_d \,\alpha  (4 \alpha  (2 \alpha +1)
   A+B)}{2A^2(\alpha +1) (2 \alpha  +1)^2 (d+2) }+ \cdots \nonumber
\ee
and depends on two real parameters $A$ and $B$.
In our analysis we have computed and used asymptotic expansions with eight terms for each potential.
Starting the numerical evolution from some large value for $\phi=\phi_{\rm max}$, we have then investigated
$v'(0)$ and $h(0)$ as functions of $A$ and $B$.
Computing numerically the gradient of these two functions, we were able to employ a kind of Newton-Raphson 
method to determine their zeros, i.e. the values of $A$ and $B$ corresponding to $\mathbb{Z}_2$-symmetric
scaling solutions.
%
\begin{figure}[!t]
\begin{center}
\vspace{ -3cm}
 \includegraphics[width=0.45\textwidth]{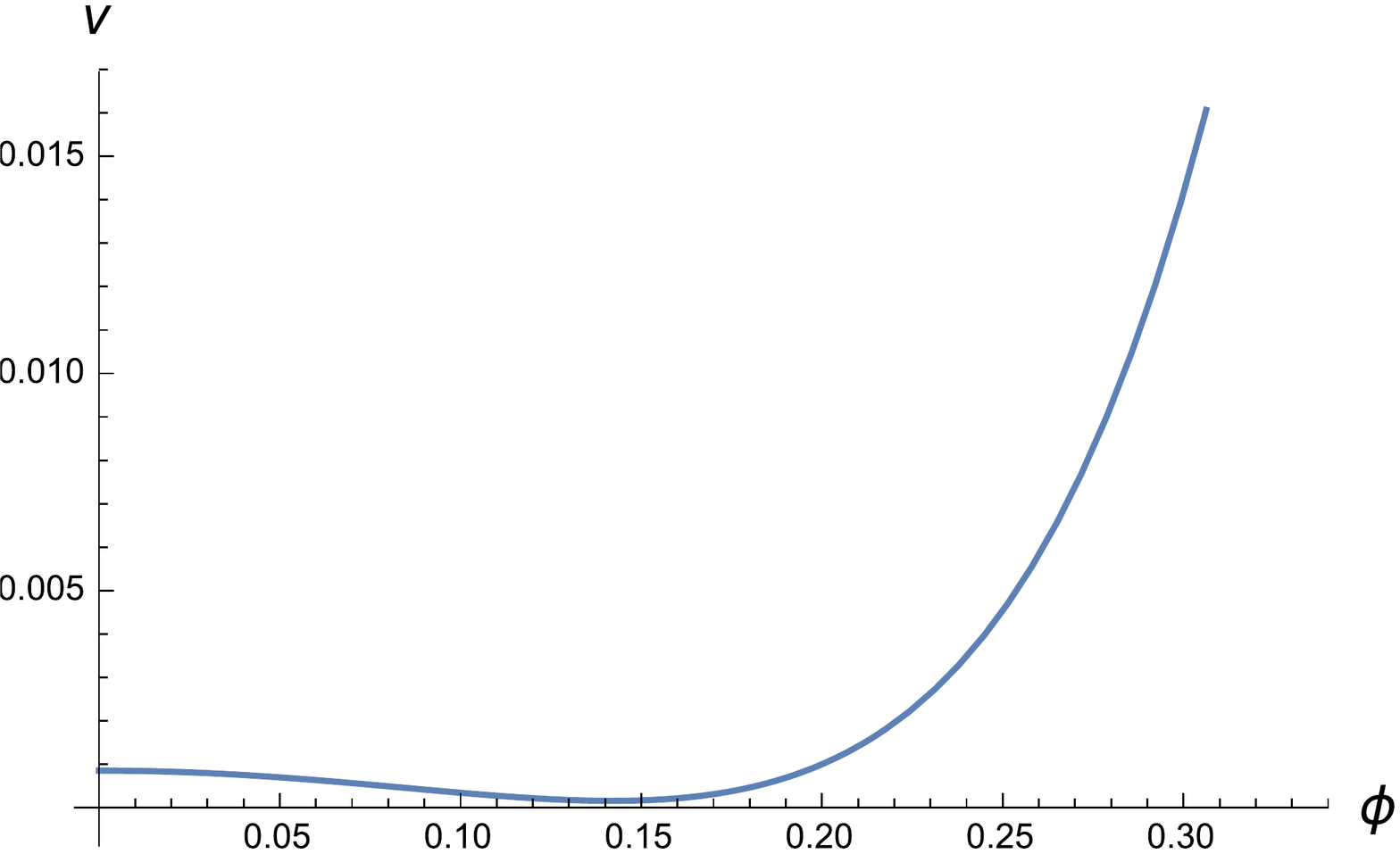} \quad
\includegraphics[width=0.45\textwidth]{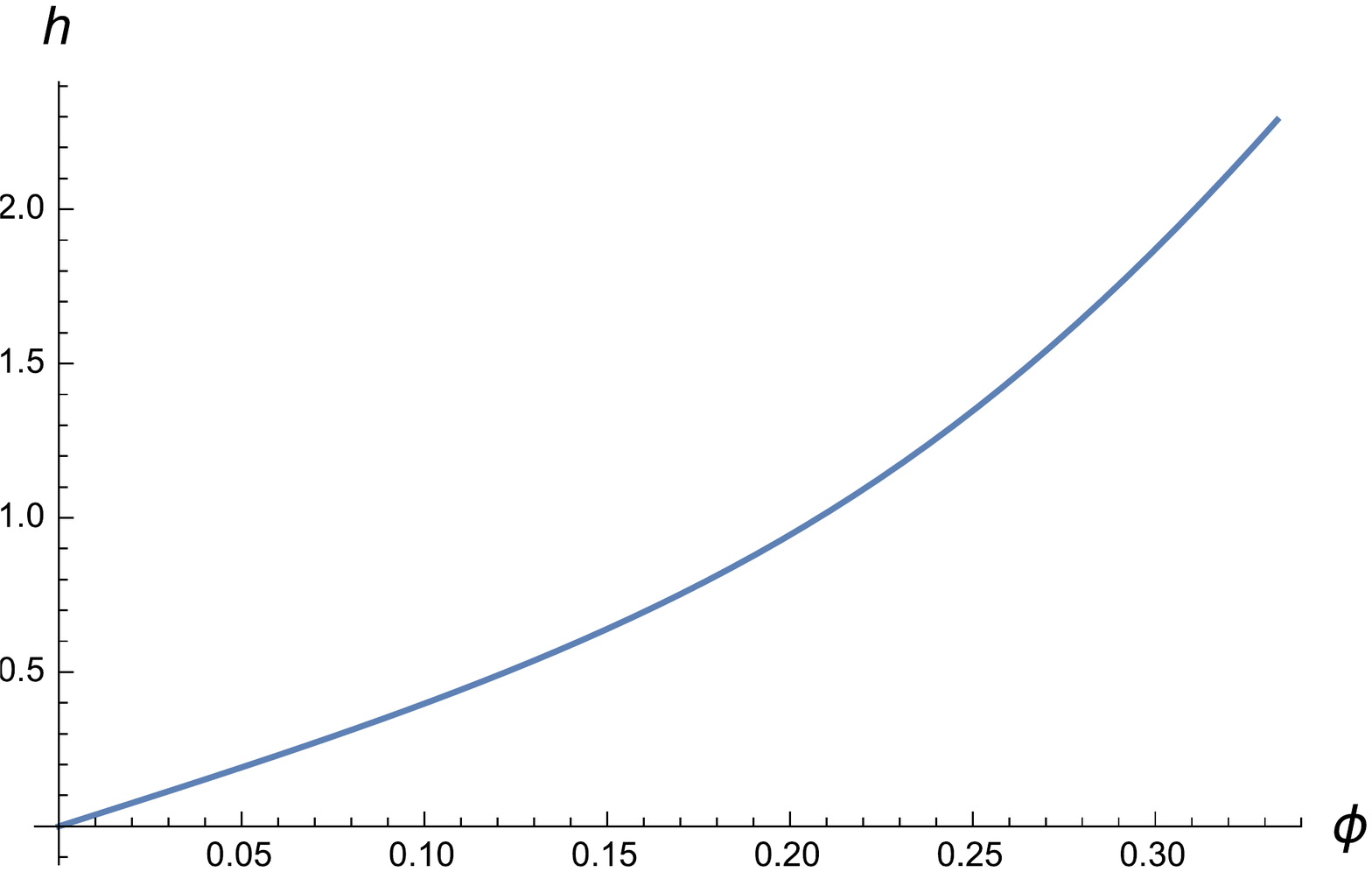}\\
\vspace{ -5cm}
\includegraphics[width=0.45\textwidth]{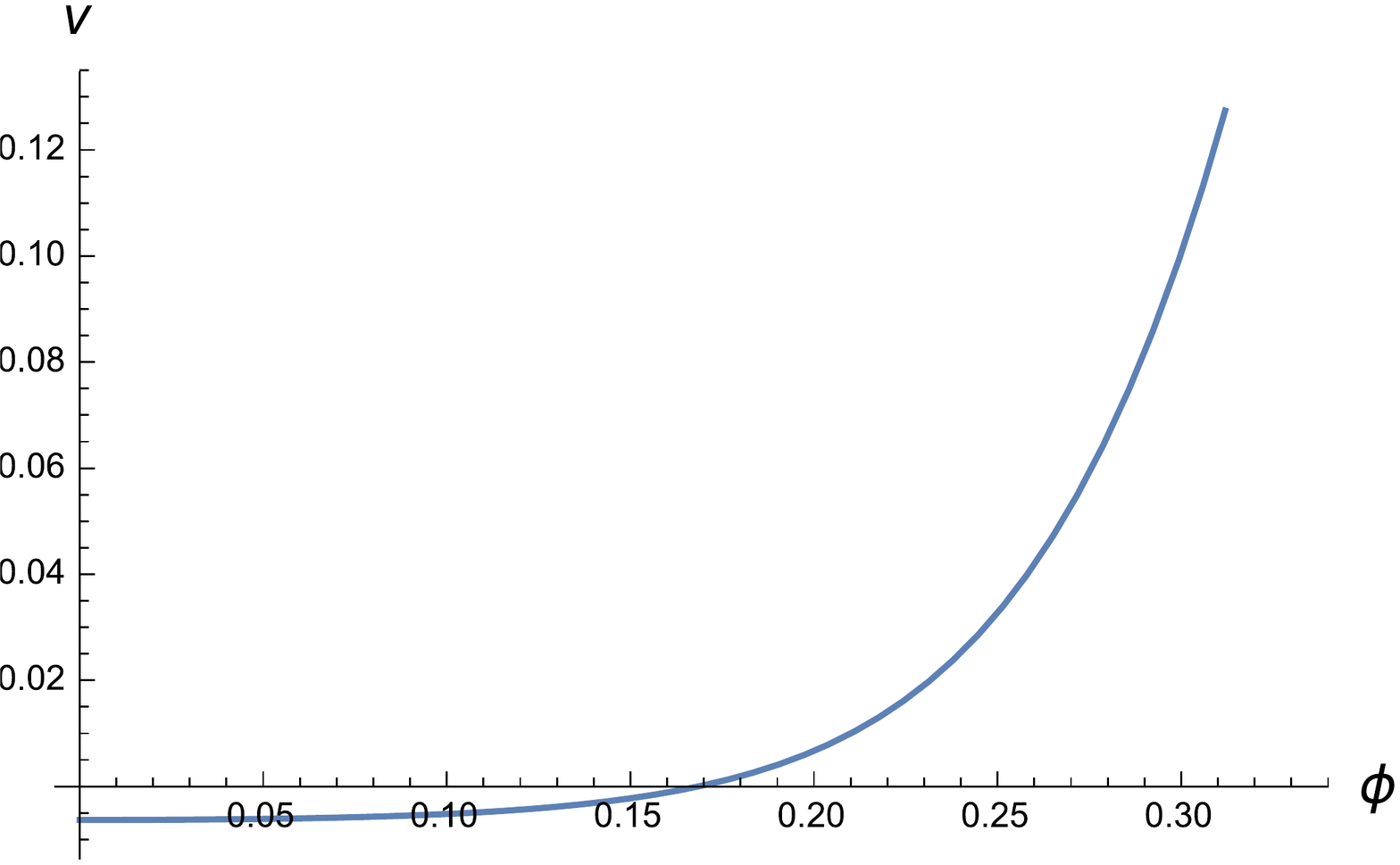} \quad
\includegraphics[width=0.45\textwidth]{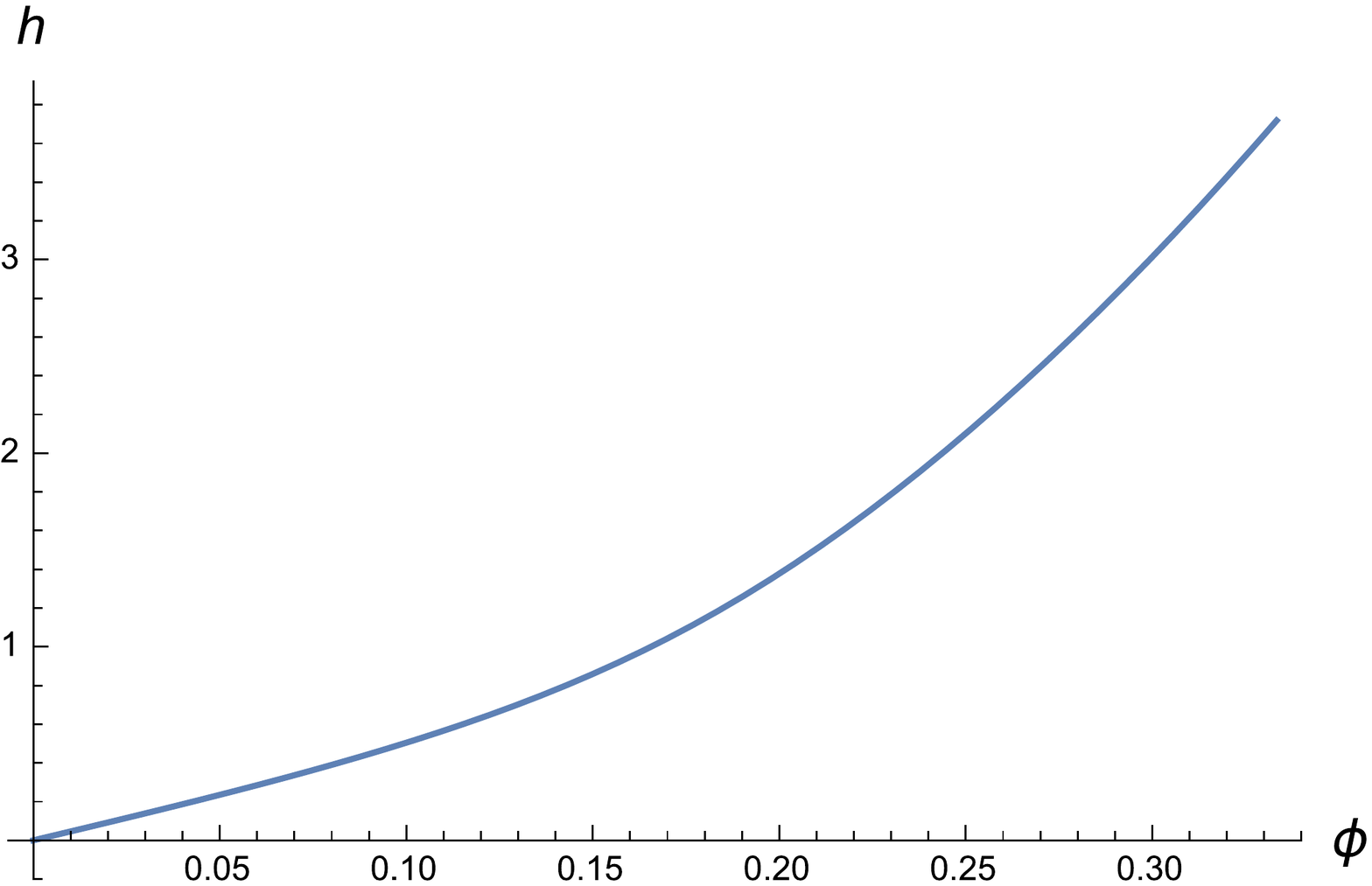}
\vspace{ -2.5cm}
 \caption{The potentials $v$ and $h$ at the global scaling solution, computed numerically within the LPA.
The case $X_f=1$, which is in the broken regime, appears in the first two panels (top),
while $X_f=2$, in the symmetric regime, is shown in the last two panels (bottom).}
\label{fig:scalingXF1-2}
\end{center}
\end{figure}
%
In Fig.~\ref{fig:scalingXF1-2} we present two examples of global solutions for the cases $X_f=1$ and $X_f=2$. 
The former is in the broken regime, since the $\mathbb{Z}_2$ symmetric scalar potential has a non trivial minimum, 
while the latter is in the symmetric regime.
Any solution ($v$, $h$) is characterized by two parameters, such as for example
$A$ and $B$, or $v''(0)$ and $h'(0)$, which indeed fix completely the Cauchy problem once they are complemented by the symmetry conditions.
In Fig.~\ref{fig:locusAB} we show the FP values of the integration constants $A$ and $B$ as defined by~\Eqref{vh_asympt}.
The locus of the FP solutions in the plane ($v''(0),\, h'(0)$) as a function of $X_f \in [10^{-3},3]$ is instead presented in Fig.~\ref{fig:locus}. 
Notice that as $X_f$ approaches zero, in the lower left end of the curve, $h'(0)$ attains a finite value, which is 
situated around 3.3.
It is evident that the two regimes, broken and symmetric, are realized in two complementary intervals of $X_f$. 
The transition between the two occurs at $X_f \simeq 1.64$ for the LPA.
In the next Section we will see that this value is slightly modified in the LPA${}^\prime$, and becomes $X_f \simeq 1.62$.
The vacuum expectation value $\phi_0$  and the value of $h'(\phi_0)$ as functions of $X_f$ are presented in Fig.~\ref{fig:vacuum}.
%

The critical exponents of these scaling solutions and the corresponding eigenperturbations are an important piece of information.
This is obtained by studying the evolution of the small perturbations around the FPs.
Therefore the linearized flow equations are the main tool to study such a problem.
They are constructed, taking advantage of the separation of variables in $\phi$ and $k$, by substituting into the flow equations
\be
v_k(\phi)=v^*(\phi)+\epsilon \delta v(\phi) \,e^{\lambda t} \quad , \quad y_{k}(\phi)=y^*(\phi)+\epsilon \delta y(\phi)\,e^{\lambda t}
\ee
and then keeping the first term in $\epsilon$, for $\epsilon\ll1$. 
Such a procedure leads to the following eigenvalue problem
\be
0= (\lambda -d)\delta v+\frac{1}{2} (d-2)
   \phi \,\delta v'+C_d \left(\frac{X_f}{(1+y)^2}\delta y-\frac{1}{\left(1+v''\right)^2}\delta v''\right)
\ee
and
\bea
0=\!\!&{}&\!\!(\lambda -2)\delta y +\left(\frac{d}{2}-1\right)\phi\, \delta y'+
C_d \Biggl[
\delta v''
   \frac{\left(2 y
   \left(y+1\right){}^2 y''-\left(y'\right){}^2 \left(y \left(v''+5\right)+3
   y^2+1\right)\right)}{y \left(1+y\right){}^2
   \left(1+v''\right)^3}
\nonumber\\
&{}& -\delta y \, (y')^2 \left(\frac{2}{\left(1+y\right){}^3 \left(v''+1\right)}+\frac{\left(3 y^2+2
   y+1\right) }{2 y^2 \left(1+y\right){}^2 \left(1+v''\right)^2}\right)
\nonumber\\
&{}&+
\delta y'  \, y' \left(\frac{2}{\left(1+y\right){}^2
   \left(1+v''\right)}+\frac{\left(3 y+1\right)}{y \left(1+y\right)
   \left(1+v''\right)^2}\right)-
   \frac{\delta y''}{\left(1+v''\right)^2}\Biggr]
\eea
where for simplicity we have renamed $v^*$ and $y^*$ as $v$ and $y$.
This system is of the form
\be
\left(\hat{O}  -\lambda \right) \delta f=0 \,.
\ee
if $\delta f$ is the vector of perturbations, $\delta f^T=(\delta v, \delta y)$, and $\hat{O}$ is the corresponding differential operator.
We have considered two different ways to solve this eigenvalue problem.

The first approach is a direct generalization of the one we have already discussed for scaling solutions, 
in this case applied to the full set of equations: FP plus linearized flow.
The asymptotic behavior of the eigenperturbations is computed by solving the asymptotic form of the linearized equations for large field, 
which is obtained using the known asymptotic expansion for $v$ and $y$ at the FP, given in Eq.~\eqref{vh_asympt}. In $d=3$ one finds
\be
\delta v_{\rm asympt}=\phi ^{6-2 \lambda }+\phi ^{-2 \lambda -4}\frac{ \left(450 A^2 \beta  X_f+B^2 \left(-2 \lambda ^2+11
   \lambda -15\right)\right)}{13500 \pi ^2 A^2 B^2}+{\cal O}\left(\phi^{-8-2\lambda}\right)
\ee
\bea
\delta y_{\rm asympt}=\!\!\!\!&{}&\beta  \phi ^{4-2 \lambda }-\phi ^{-2 \lambda -6} \left(\frac{\left(2 \lambda ^2\!-\!11 \lambda +15\right) (20
   A+B)}{16875 \pi ^2 A^3}+\frac{\beta  \left(240 A \lambda +B \left(2 \lambda ^2+5
   \lambda \!-\!6\right)\right)}{13500 \pi ^2 A^2 B}\right)\nonumber\\
   &{}&+{\cal O}\left(\phi^{-10-2\lambda}\right)\nonumber
\eea
In practice we used an asymptotic expansion with up to three terms per perturbation.
We note that in a linear homogeneous problem the overall normalization of the eigenvector $\delta f$ plays no role.
Therefore the asymptotic form of $\delta f$ depends only on a relative real parameter $\beta$, which we choose to be
a constant multiplying the leading term of $\delta y$. 
One more free parameter is needed for tuning the behavior of the solutions at the origin,
such that they fulfill the symmetry requirements $\delta v'(0)=0$ and $\delta y(0)=0$.
This can be interpreted as the eigenvalue $\lambda$ itself.
As a consequence, one expects a discrete spectrum of allowed values for $\lambda$ and $\beta$.
Unfortunately, due to numerical uncertainties, with this method we have been able only to restrict the eigenvalues to an interval described by a continuous function $\lambda(\beta)$.
Indeed one has to remember that the global numerical solutions have been constructed on some bounded neighborhood of the origin,
even if the latter overlaps with the region were the large field asymptotic behavior becomes dominant.
Moreover, the linearized equations depend on derivatives of the numerical global FP solutions, for which the accuracy is reduced.

\begin{figure}[!t]
\begin{center}
\vspace{ -3cm}
\includegraphics[width=0.45\textwidth]{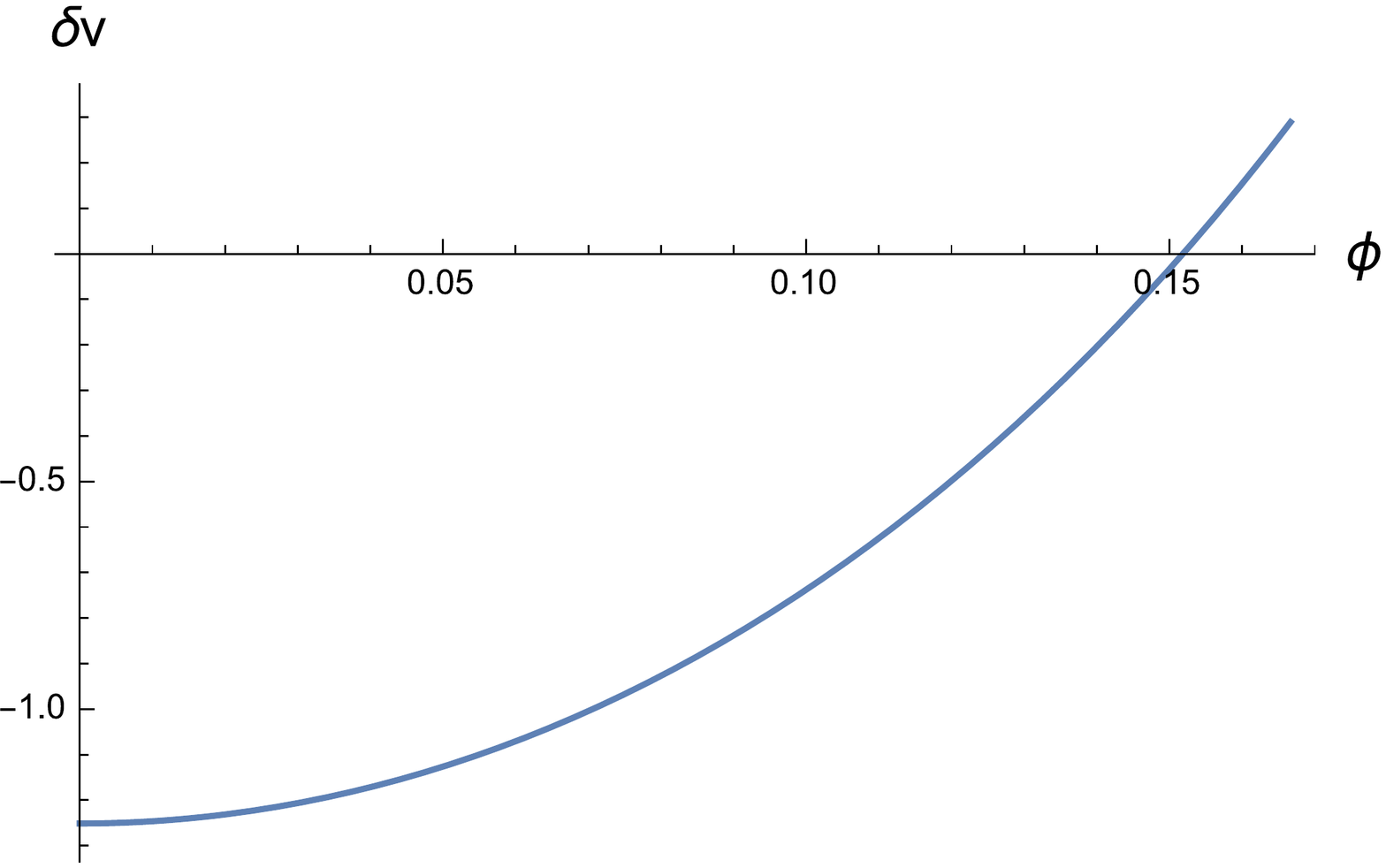}
\includegraphics[width=0.45\textwidth]{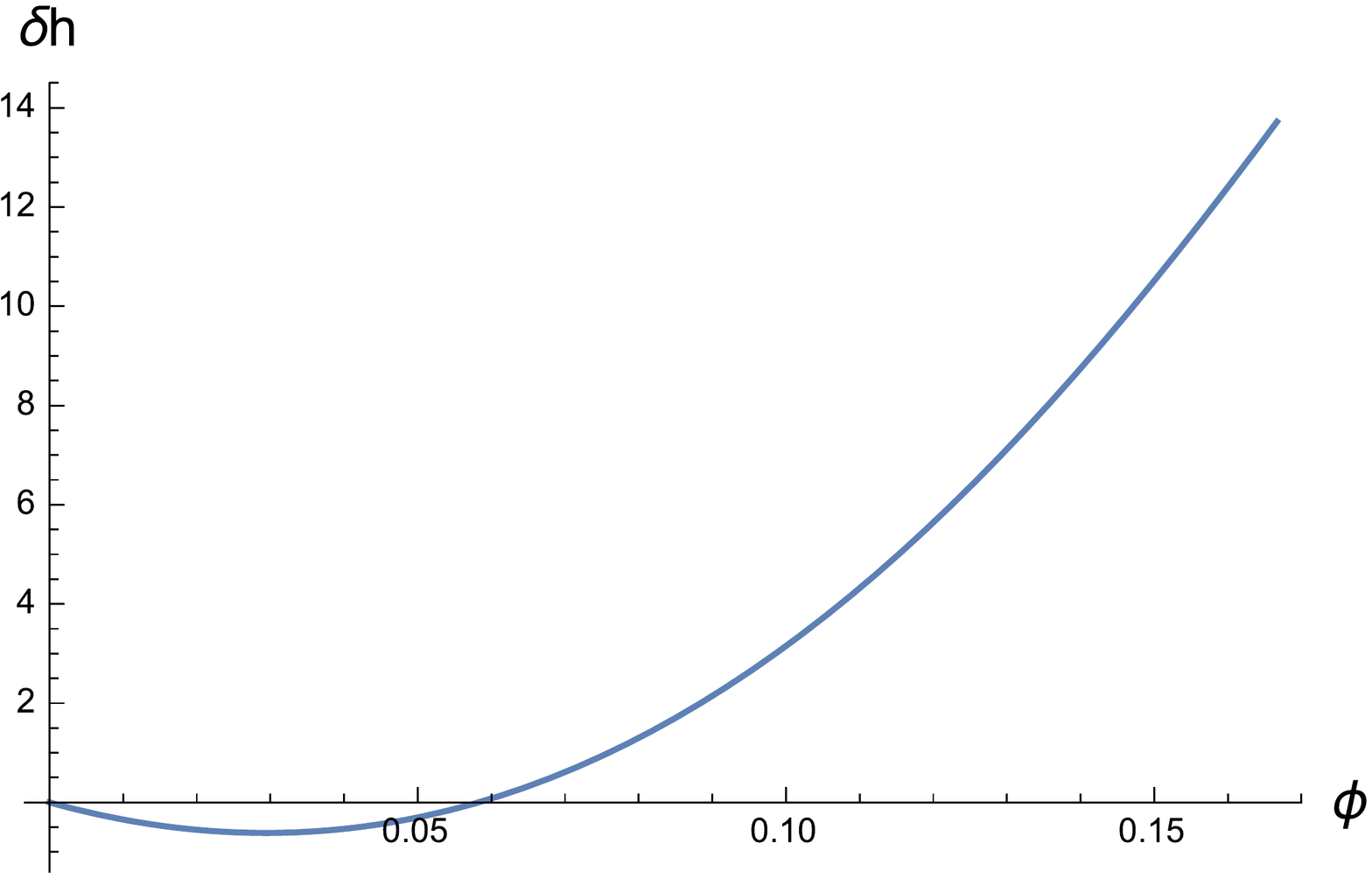}
\vspace{ -2.5cm}
\caption{Case $d=3$ and $X_f=1$: the components $\delta v$ and $\delta h$ of the relevant eigenperturbation, from the global numerical analysis of the LPA.}
\label{fig:eigenv1}
\end{center}
\end{figure}

The second approach we considered consists in inserting the known numerical FP solutions in the linearized equations,
computing a numerical expression for all the  $\phi$-dependent coefficients of this eigenvalue problem,
and then solving them by means of a pseudo-spectral method based on Chebyshev polynomials. 
Also in this case some uncertainties remain, for the same reasons mentioned above. 
As an example, for $X_f=1$ the leading critical exponent we find is $\theta_1=-\lambda_1=1.2279$, which refers to the only relevant direction
(we do not consider $\theta_0=3$, since it is related to an additive constant in the potential and it is unphysical in flat space).
All the other eigenvalues $\lambda_i$ are positive and associated to irrelevant operators, for instance 
$\theta_2 =-\lambda_2=-0.6236$ and $\theta_3=-\lambda_3=-1.5842$.
The relevant direction corresponds to the eigenperturbation $\delta f_1=(\delta v,\delta h)$ shown in Fig~\ref{fig:eigenv1}.
Notice the fact that the relevant eigenpertubation has $\delta h(\phi)\neq0$ unlike in the large-$X_f$ analysis,
where the only relevant perturbation compatible with symmetry requirements is $\delta v(\phi)=\delta c_v \phi^2$,
which corresponds to $\theta_1=1$. Even if $X_f=1$ is quite away from this limit,
it is know that in this case the FP theory is a ${\cal N}=1$ Wess-Zumino model~\cite{Sonoda:2011qd,Grover:2013rc},
and that the supersymmetry-preserving relevant perturbation is a change in the mass of the scalar field~\cite{Sonoda:2011qd,Synatschke:2010ub},
which therefore leaves the Yukawa sector unchanged.
Hence $\delta h\neq 0$ is probably a consequence of the explicit breaking of supersymmetry  introduced by
our regularization scheme.

We do not push further here the spectral analysis of the critical exponents and associated perturbations as a function of $X_f$, 
leaving it for a future study based on algorithms giving better control on the numerical errors.
In the present work, these global numerical computations at $X_f=1$ will serve as a reference for the development of
a different, local, approximation method, based on polynomial truncations of the functions $v(\phi)$ and $h(\phi)$.
The latter will be discussed in the next Section, and will be also used for a more reliable discussion of
the dependence of the critical exponents on the number of fermion degrees of freedom.
%

\section{Polynomial analysis in $d=3$}
\label{sec:polynomial}

In this Section we are going to discuss the use of polynomial
parameterizations and consequent truncations of the 
functions $v(\phi)$ and $h(\phi)$.
Though for definiteness we will address the specific case
of the unique $d=3$ nontrivial critical Yukawa theory, similar techniques
can be applied to the other scaling solutions in $2<d<3$, 
presumably with the same degree of success.
Sect.~\ref{subsec:polynomial_LPA} will present results obtained
within the LPA, which can be directly compared to the full functional analysis
developed in the previous Section. This will make us confident about the effectiveness
and soundness of polynomial truncations, as well as of the necessity to go beyond
a simple linear Yukawa coupling for an accurate description of critical properties
of the theory.
On these grounds, Sect.~\ref{subsec:polynomial_LPAprime} will push forward
the analysis to a self-consistent inclusion of the wave function renormalization
of the fields, which is essential for quantitative estimates of the critical exponents,
which will be compared with some literature for several values of $X_f$.
Polynomial truncations will be also used in Sect.~\ref{sec:d=4} for some comments
on the four-dimensional model.

Let us start by presenting the truncation schemes we are going to analyze.
Since we restrict ourselves to $d=3$, we will demand 
$v(\phi)$ and $h(\phi)$ to be even and odd respectively.
We will use the common notation $\rho=\phi^2/2$, and we will
adopt only one name for one and the same quantity, 
regarless of whether it is considered as a function of $\phi$ or as a function of $\rho$.
In the symmetric regime, the physically meaningful parameterization of the 
scalar potential is a Taylor expansion around vanishing field
\be
\label{polv_0}
v(\rho)=\sum_{n=0}^{N_v}\frac{\lambda_n}{n!} \rho^{n}\ .
\ee
Regarding the Yukawa potential, we are interested in two possible
Taylor expansions, one for $h(\phi)$, already adopted in \cite{Pawlowski:2014zaa},
and one for $y(\rho)=\left[ h(\phi)\right]^2$. In the symmetric regime
they read
\bea
\label{polh_01}
h(\phi)&=&\phi \sum_{n=0}^{N_h-1}\frac{h_n}{n!}\rho^{n}\label{eq:poly_h_SYM}\\
\label{polh2_02}
y(\rho)&=&\sum_{n=1}^{N_h}\frac{y_n}{n!} \rho^{n}\label{eq:poly_h2_SYM}\ .
\eea
In the regime of spontaneous symmetry breaking (SSB) the potential $v(\rho)$ develops a nontrivial minimum
$\kappa=\phi_0^2/2$, which becomes the preferred reference point for a different 
Taylor expansion
\be
\label{polv_ntv}
v(\rho)=\lambda_0+\sum_{n\ge 2}^{N_v}\frac{\lambda_n}{n!} (\rho-\kappa)^{n}\ .
\ee
Though, in general, $\kappa$ is no special point for the function $h(\phi)$,
it still enters in the definition of the vertex functions, 
from which one extracts the physical multi-meson Yukawa couplings.
As a consequence, in this regime it is necessary to change also the parameterizations
of $h(\phi)$ and $y(\rho)$, as follows
\bea
\label{polh_ntv1}
h(\phi)&=&\phi \sum_{n=0}^{N_h-1}\frac{h_n}{n!}
\left(\rho-\kappa\right)^{n}\label{eq:poly_h_SSB}\\
\label{polh2_ntv2}
y(\rho)&=&\sum_{n=1}^{N_h}\frac{y_n}{n!}\left[
\left(\rho-\kappa\right)^{n}-\left(-\kappa\right)^{n} \right]\label{eq:poly_h2_SSB}\ .
\eea

The pair $(N_v, N_h)$, or more generally an ordering of the polynomial couplings by priority of inclusion in the truncations, can be chosen by relying on naive dimensional counting, as in an effective field theory setup, 
or on the knowledge of the dynamics at a deeper level, e.g.
a global numerical solution for the FP functionals and the critical exponents.
In the latter strategy one would sort the critical exponents in order of relevance and would try to accurately describe the corresponding perturbations. Alternatively, and maybe less efficiently, one could scan over the results produced by
different pairs $(N_v, N_h)$ and select them on the base of
a comparison to the global numerical solution.
In the former strategy instead, since the dimension of a scalar self-interaction $\phi^{2n}$
is $n$, and the one of a multi-meson Yukawa coupling  
$\bar{\psi}\phi^{2n+1} \psi$ is $5/2+n$, we would expect that
the pairs $(N_v=D,N_h=D-2)$, for the truncation of $h(\phi)$ given in
Eqs. (\ref{eq:poly_h_SYM},\ref{eq:poly_h_SSB}),
 correspond to including operators up to dimension $D$.
However, since by truncating at level $N_h=D-2$ we loose information about an operator of
dimension $D+1/2$, if we want to be slightly more accurate we could include the latter and
consider the pairs $(N_v=D,N_h=D-1)$.
In our analysis we did perform to some extent a random scan over
different pairs $(N_v,N_h)$, and we found that the two strategies nicely agree,
so that $(N_v=D,N_h=D-1)$
is a very good systematic choice for polynomial truncations.
For similar reasons, as well as for the sake of comparison, we made the same
choice also for the truncation of  $y(\rho)$ 
given in Eqs. (\ref{eq:poly_h2_SYM},\ref{eq:poly_h2_SSB}).

It is necessary to stress that, in both the parameterizations given above, even at lowest order in the truncation for the Yukawa coupling,
 the beta-functions for $h_0$ or $y_1$ are different from the classic
result~\cite{Jungnickel:1995fp} illustrated in the 
reviews~\cite{ReviewRG} and used for the present $d=3$ critical
theory for instance in~\cite{Rosa:2000ju,Hofling:2002hj,Braun:2010tt,Janssen:2014gea,Borchardt:2015rxa}.
This happens because $\partial_t h(\phi)$, which comes from the projection of the r.h.s. of the flow equation on the term $i\bar{\psi}\psi$, 
is a nonlinear function of $\phi$, independently of the parameterization of $h(\phi)$, be it linear in $\phi$ or not.
Hence, in order to define the running of a linear Yukawa coupling,
a further projection is needed.
The prescription adopted by the above-mentioned studies is to identify the beta function of the linear Yukawa
coupling with the first $\phi$-derivative of $\partial_t h(\phi)$
at the minimum of the potential.
For the truncations under consideration in this work instead, 
$\partial_t h_0$ comes from the zeroth
order $\phi$-derivative of $\partial_t h(\phi)/\phi$, while 
$\partial_t y_1$ is defined as the first order $\rho$-derivative
of $\partial_t y(\rho)=2h(\phi)\partial_t h(\phi)$,
always evaluated at the minimum of the potential.
Simplicity is our main motivation for choosing a parameterization of the running
Yukawa sector which does not include the traditional Yukawa
beta-function, as we are now going to explain.

The traditional projection has the structure of a Taylor expansion
of $\partial_t h(\phi)$ about $\phi=\phi_0$ ($\phi_0$ being the minimum of $v(\phi)$).
The choice of such an expansion for the parameterization of $h(\phi)$
would entail an explicit breaking of $\mathbb{Z}_2$ symmetry, which
requires this function to be odd. Ideally, one would need to match two 
Taylor expansions, one about $\phi=\phi_0$ and another one about $\phi=-\phi_0$,
by imposing suitable conditions at the origin.
These are just provided by  $\mathbb{Z}_2$ symmetry.
The result of this construction however is not a simple Taylor expansion any more
\be
h(\phi)=\frac{1}{2}\sum_{n=1}^{N_h}\frac{g_n}{n!}\left[\left(\phi-\phi_0\right)^{n}
+(-1)^{n+1}\left(\phi+\phi_0\right)^{n}\right]\label{eq:poly_h_third}
\ee
and the projection rule on the generic coupling $g_n$ is
more involved than simply taking the $n$-th $\phi$-derivative
and evaluating it at $\phi=\phi_0$.
Yet, it is true that the latter projection works for the $N_h$-th coupling,
such that this truncation does include the traditional beta-function 
of the linear Yukawa coupling as the $N_h=1$ case.
In this work we preferred to consider and compare only
the two truncation schemes presented in Eqs. (\ref{eq:poly_h_SYM},\ref{eq:poly_h_SSB})
and Eqs. (\ref{eq:poly_h2_SYM},\ref{eq:poly_h2_SSB}),
leaving the one in \Eqref{eq:poly_h_third} aside.
In the next Sections we are going to show that both polynomial truncations
converge to the same results for large enough $N_v$ and $N_h$,
an observation that clearly should apply to all possible parameterizations. 
Furthermore, in both polynomial truncations simply by setting $N_h=1$ 
one gets estimates that are 
significantly different from the full truncation-independent results.
That the latter statement also applies to the truncation in \Eqref{eq:poly_h_third},
can be assessed by comparison to the literature, which the reader
can find in Sect.~\ref{subsec:polynomial_LPAprime}.

\subsection{LPA}
\label{subsec:polynomial_LPA}

In Sect.~\ref{sec:d=3_numerical_FP} we looked for the $d=3$ nontrivial critical theories at varying $X_f$ within the LPA,
by means of numerical solvers for the ODEs defining the FP potentials.
Here we repeat this analysis with the different method of polynomial truncations and
we compare the results with the ones we previously found.
The FPs emerge from the solution of a system of coupled nonlinear algebraic equations
for the couplings. The critical exponents are defined by (minus) the eigenvalues of 
the stability matrix at the FP, i.e. the matrix of derivatives of the beta-functions
with respect to the couplings~\cite{ReviewRG}. The anomalous dimensions are 
computed in a non-self-consistent way, by neglecting them in the FP
equations descending from Eqs. (\ref{eq:floweq_v},\ref{eq:floweq_h}), 
and then by evaluating the flow equations for the wave function renormalizations
Eqs. (\ref{eq:floweq_etaphi},\ref{eq:floweq_etapsi}) at this FP position.

Let us start from the standard way of describing the Yukawa models, that is by
approximating the Yukawa potential $h(\phi)$ with a single linear coupling.
On the grounds of the results of the full functional analysis presented in
Sect.~\ref{sec:d=3_numerical_FP}, one could expect that this approximation performs well, 
since far enough from the large-field region the FP function $h(\phi)$ 
does not strongly deviate from a straight line, see Fig.~\ref{fig:scalingXF1-2}.
For a linear Yukawa function, the expansions around the origin of $h(\phi)$ and $y(\rho)$ give results which are identical order by order in $N_v$,
both in the shape of the FP functions (in the sense that $y_1=2 h_0^2$ at the FP) and in the critical exponents. 
As a consequence we can present them in a single table for the
former parameterization, the latter providing the same results. 
This is Tab.~\ref{tab:LPA_Xf1_SYM_linearh}, where we set e.g. $X_f=1$.
The first two critical exponents form a complex
conjugate pair, which is clearly unsatisfactory. This is produced by the
expansion around a trivial minimum of $v(\phi)$, that for $X_f=1$ is not justified.
Once we turn to the SSB parameterization of $h(\phi)$, which is given on the
left panel of Tab.~\ref{tab:LPA_Xf1_SSB_h}, they become real. 
However, things become cumbersome for
the single-coupling SSB parameterization of $y(\rho)$, since we were not able to
find any FP at all (which might nevertheless exist). 
Let us recall that, even in the case of a single Yukawa coupling, the beta functions
descending from the two different polynomial truncations of $h(\phi)$ and $y(\rho)$
are different, hence one cannot simply translate the FP position from one parameterization
to the other. 
As soon as we add $y_2$ the FP can be easily found.
This then stimulates to consider the general effect of allowing for higher polynomial
Yukawa couplings. 

The immediate observation is that their inclusion significantly alters
the position of the FP and the critical exponents. Some degree of convergence 
is observed in several systematic strategies for the increase of $N_v$ and/or $N_h$,
but this can be convergence to the wrong results, i.e. to FP functions that do not
agree with the numerical global solution. The linear Yukawa truncations provide
one example of this fact. This is visible by comparing the two panels 
of Tab.~\ref{tab:LPA_Xf1_SSB_h},
where on the r.h.s. we show the results provided by the $(N_v=D,N_h=D-1)$
systematic choice that we have already discussed above. The latter turns out
to converge to the correct value of the FP couplings, as we are now going to argue.
In Tab.~\ref{tab:LPA_Xf1_SSB_h2} we show the results obtained by the 
systematic $(D,D-1)$-extension of polynomial truncations for $y(\rho)$.
Comparing the two panels one can see how the critical exponents can be 
computed by large polynomial truncations independently of whether these are
around the origin or a nontrivial vacuum. Furthermore, comparing the right panels
of Tab.~\ref{tab:LPA_Xf1_SSB_h2} and Tab.~\ref{tab:LPA_Xf1_SSB_h} it can be
observed how both the FP potentials and the critical exponents converge to 
values that are independent of the chosen parameterization.
That these values are the ones corresponding to the full global solution
provided in Sect.~\ref{sec:d=3_numerical_FP}, is shown in the right
panel of Tab.~\ref{tab:LPA_Xf1_SSB_h2}.
Notice however that there is a 0.6\% difference between the relevant exponent
computed with the polynomial truncations and the one obtained by the global numerical analysis.
Even if we feel that we have the former method under a better control, we cannot 
give our preference to any of these estimates.

\begin{table}
\resizebox{12cm}{2cm}{
\begin{tabular}{|l|l|l|l|l|l|l|l|l|l|l|l|l|}
 \hline
 $(N_v,N_h)$ 		&\,\, (2,1) 	&\,\, (3,1) 	&\,\, (4,1)	&\,\, (5,1) 	&\,\, (6,1)	&\,\, (7,1)	&\,\, (8,1)	&\,\, (9,1)	&\,\, (10,1) \\
 \hline
 $\, \,\lambda _1$ 	&$-$0.04901	&$-$0.1225	&$-$0.1602 	&$-$0.1743 	&$-$0.1765	&$-$0.1740	&$-$0.1720	&$-$0.1716	&$-$0.1721\\ 
 $\,\,\lambda _2$ 	&5.887		&6.841		&7.128 		&7.204		&7.214		&7.203		&7.193		&7.191		&7.193\\
 $\,\,\lambda _3$ 	&\ \ \ ---		&84.22		&121.9		&134.7		&136.7		&134.5		&132.7		&132.4		&132.8\\ 
\hline
 $\,\,h_0$			&2.620		&2.464		&2.382 		&2.351 		&2.347		&2.352		&2.356		&2.357		&2.356\\ 
\hline
 $\,\,\theta_1$  		&1.701		&1.546		&1.438 		&1.378	 	&1.358		&1.362		&1.372		&1.376		&1.375\\ 
 $\,\,\theta_2 $ 		&$-$1.050	&$-$1.156	&$-$1.246$+i$ 0.2686 &$-$1.068$+i$ 0.3386 &$-$0.9602$+i$ 0.3238	&$-$0.9119$+i$ 0.2933 &$-$0.9150$+i$ 0.2844 &$-$0.9386$+i$ 0.2941 &-0.9526$+i$ 0.3044\\ 
 $\,\,\theta_3 $ 		&\ \ \ ---	&$-$1.864	&$-$1.246$-i$ 0.2686	 &$-$1.068$-i$ 0.3386 &$-$0.9602$-i$ 0.3238 &$-$0.9119$-i$ 0.2933 &$-$0.9150$-i$ 0.2844 &$-$0.9386$-i$ 0.2941 &$-$0.9526$-i$ 0.3044\\ 
 \hline
 $\,\,\eta_\psi$ 		&0.2395		&0.2510		&0.2572		&0.2595		&0.2599 	&0.2595		&0.2591		&0.2591		&0.2592\\ 
 $\,\,\eta_\phi$ 		&0.2620		&0.2306		&0.2150		&0.2092 	&0.2083		&0.2093		&0.2101		&0.2103		&0.2101\\ 
 \hline
\end{tabular}
}

\caption{Case $d=3$ and $X_f=1$, polynomial expansion of $h(\phi)$ around a trivial vacuum of the potential,
 with a fixed linear Yukawa function (standard Yukawa interaction), in the LPA.}
\label{tab:LPA_Xf1_SYM_linearh}
\end{table}

\begin{table}
\resizebox{6cm}{2cm}{
\begin{tabular}{|l|l|l|l|l|l|l|l|l|l|l|l|l|}
 \hline
 $(N_v,N_h)$ 		&\,\, (5,1)	&\,\, (6,1) 	&\,\, (7,1) 	&\,\, (8,1)	&\,\, (9,1)	\\ \hline
 $\, \,\kappa$ 		& 0.01114	& 0.01115	& 0.01114	& 0.01114 	& 0.01114 	\\ 
 $\,\,\lambda _2$ 	& 25.08 		& 24.88 		& 24.80		& 24.84 		& 24.85 		\\
 $\,\,\lambda _3$ 	& 813.8		& 800.3		& 793.33	& 796.5		& 797.5		\\ \hline
 $\,\,h_0$			& 5.716 		& 5.690		& 5.674		& 5.681 		& 5.683		\\ \hline
 $\,\,\theta_1$  		& 1.338 		& 1.333		& 1.336 		& 1.336 		& 1.335		\\ 
 $\,\,\theta_2 $ 		& $-$0.2461	& $-$0.2466	& $-$0.2490 	& $-$0.2484	& $-$0.2483 	\\ 
 $\,\,\theta_3 $ 		& $-$2.232 	& $-$2.060		& $-$2.033 	& $-$2.067		& $-$2.075\\ 
\hline
$\,\,\eta_\psi$ 		&0.2629		& 0.2288	&0.2288		&0.2288		&0.2288 	\\ 
 $\,\,\eta_\phi$ 		&0.5259		& 0.5166	&0.5155		&0.5160		&0.5162	 	\\
 \hline
\end{tabular}
}
\resizebox{6cm}{2cm}{
\begin{tabular}{|l|l|l|l|l|l|l|l|l|l|l|l|l|}
 \hline
 $(N_v,N_h)$ 		&\,\, (5,4)	&\,\, (6,5) 	&\,\, (7,6) 	&\,\, (8,7)	&\,\, (9,8)	\\ \hline
 $\, \,\kappa$ 		& 0.01002	& 0.01009	& 0.01008	& 0.01007 	& 0.01007 	\\ 
 $\,\,\lambda _2$ 	& 15.34 		& 15.32 		& 15.30		& 15.28 		& 15.28 		\\
 $\,\,\lambda _3$ 	& 508.3		& 506.8		& 503.6		& 502.1		& 502.1		\\ \hline
 $\,\,h_0$			& 4.220 		& 4.211		& 4.207 		& 4.206 		& 4.207		\\ 
 $\,\,h_1$  			& 48.23 		& 47.73		& 47.46 		& 47.43 		& 47.48 		\\ \hline
 $\,\,\theta_1$  		& 1.231 		& 1.234		& 1.236 		& 1.236 		& 1.235		\\ 
 $\,\,\theta_2 $ 		& $-$0.6144	& $-$0.6078	& $-$0.6080 	& $-$0.6106	& $-$0.6117 	\\ 
 $\,\,\theta_3 $ 		& $-$1.649 	& $-$1.551		& $-$1.520 	& $-$1.521		&$-$1.531		\\ 
\hline
$\,\,\eta_\psi$ 		&0.3435		& 0.3409	&0.3402		&0.3404		&0.3407 	\\ 
 $\,\,\eta_\phi$ 		&0.4916		& 0.4910	&0.4899		&0.4895		&0.4895	 	\\
 \hline
\end{tabular}
}
\caption{Case $d=3$ and $X_f=1$, polynomial expansion of $h(\phi)$ around a non trivial vacuum
 for both the potential and the Yukawa function, in the LPA, with or without the inclusion 
of multiple-meson-exchange interactions (right and left panel respectively).}
\label{tab:LPA_Xf1_SSB_h}
\end{table}

\begin{table}
\resizebox{6cm}{2cm}{
\begin{tabular}{|l|l|l|l|l|l|l|l|l|l|l|l|l|}
 \hline
 $(N_v,N_h)$ 		&\,\, (4,3) 	& \,\, (5,4) &\,\, (6,5) &\,\, (8,7) &\,\, (9,8)\\
 \hline
 $\, \,\lambda _1$ 	& $-$0.1209 	& $-$0.1315 & $-$0.1339 & $-$0.1315 & $-$0.1309 \\ 
 $\,\,\lambda _2$ 	& 10.60 		& 11.05 	& 11.16 	& 11.09 	& 11.06 	\\
 $\,\,\lambda _3$ 	& 293.2		& 339.6	& 351.0	& 342.7	& 340.1	\\ \hline
 $\,\,y_1$			& 26.84 		& 28.38	& 28.76 	& 28.53 	& 28.44	 \\ 
 $\,\,y_2$  			& 986.6 		& 1161	& 1206 	& 1178 	& 1167 	\\ \hline
 $\,\,\theta_1$  		& 1.324 		& 1.253	& 1.226	& 1.230 	& 1.236 \\ 
 $\,\,\theta_2 $ 		&$-$0.8293	& $-$0.7186& $-$0.6410 & $-$0.5892 & $-$0.5989 \\ 
 $\,\,\theta_3 $ 		& $-$2.690 	& $-$2.215	& $-$1.838 & $-$1.460 & $-$1.446 \\ 
 \hline
$\,\,\eta_\psi$ 		& 0.5209	&0.5615	& 0.5716 & 0.5642 &0.5618  \\ 
 $\,\,\eta_\phi$ 		& 0.4486	&0.4645	& 0.4683 & 0.4663 &0.4654  \\ 
 \hline
\end{tabular}
}
\qquad
\resizebox{6cm}{2cm}{
\begin{tabular}{|l|l|l|l|l|l|l|l|l|l|l|l|l|}
 \hline
 $(N_v,N_h)$ 		&\,\, (5,4)	& \,\, (6,5)	&\,\, (7,6) 	&\,\, (8,7)	&\,\, (9,8)	& ($\infty,\infty$)\\
 \hline
 $\, \,\kappa$ 		& 0.01000	& 0.01013	& 0.01006 	& 0.01006 	& 0.01007 	&0.01007 \\ 
 $\,\,\lambda _2$ 	& 15.58 		& 15.17 		& 15.30 		& 15.28 		& 15.28		& 15.28 \\
 $\,\,\lambda _3$ 	& 521.8		& 498.9		& 503.0		& 502.0		& 502.3		& 502.8\\ 
\hline
 $\,\,y_1$			& 44.59 		& 43.00		& 43.51 		& 43.44 		& 43.43		& 43.45\\ 
 $\,\,y_2$  			& 1925 		& 1818		& 1842 		& 1837 		& 1837		& 1839\\ 
\hline
 $\,\,\theta_1$  		& 1.260		& 1.221 		& 1.236 	  	& 1.236 		& 1.235		& 1.228\\  
 $\,\,\theta_2 $ 		& $-$0.6849 	&$-$0.7738 	&$-$0.5964  	& $-$0.6111 	& $-$0.6127  	&$-$0.624 \\ 
 $\,\,\theta_3 $ 		& $-$1.693 	& $-$1.077	& $-$1.511  	& $-$1.522	& $-$1.537 	&$-$1.584 \\ 
 \hline
 $\,\,\eta_\psi$ 		&0.3458 	&0.3384 	&0.3410  	&0.3406  	&0.3406  	&\ \ \ \ --- \\ 
 $\,\,\eta_\phi$ 		&0.4955 	&0.4887 	&0.4897  	&0.4894  	&0.4895		&\ \ \ \ ---\\ 
 \hline
\end{tabular}
}

\caption{Case $d=3$ and $X_f=1$, polynomial expansion of $y(\rho)$ in the LPA.
Left panel: expansion around the origin, for which the global numerical solution provides 
$\lambda _1=-0.1313$, $y_1=28.47$, and unstable higher couplings. 
Right panel: expansion around a nontrivial vacuum and, in the last column, the corresponding couplings extracted from the global numerical solution.
}
\label{tab:LPA_Xf1_SSB_h2}
\end{table}

\begin{figure}[!t]
\begin{center}
\vspace{ -2.5cm}
\includegraphics[width=0.45\textwidth]{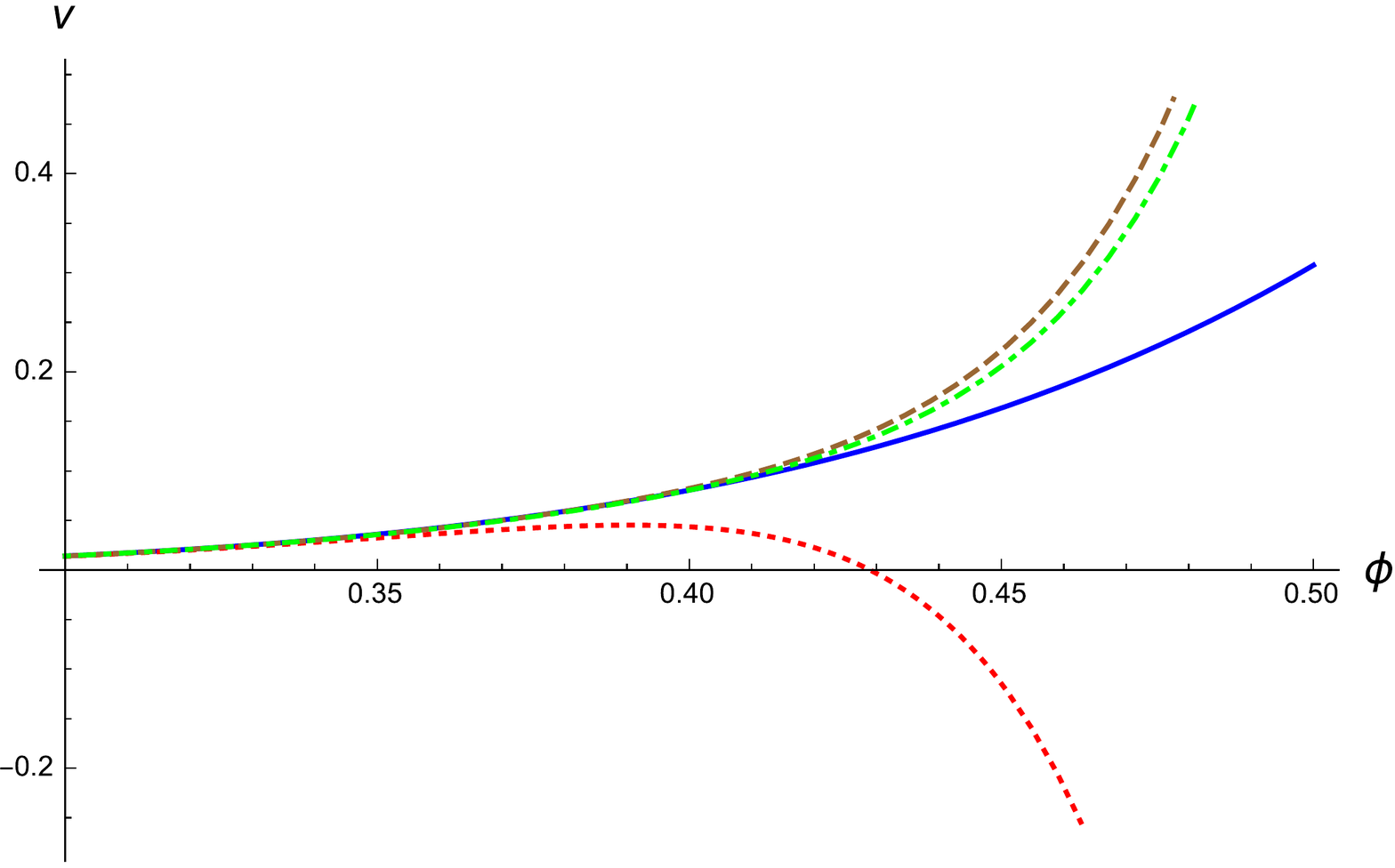}
\includegraphics[width=0.45\textwidth]{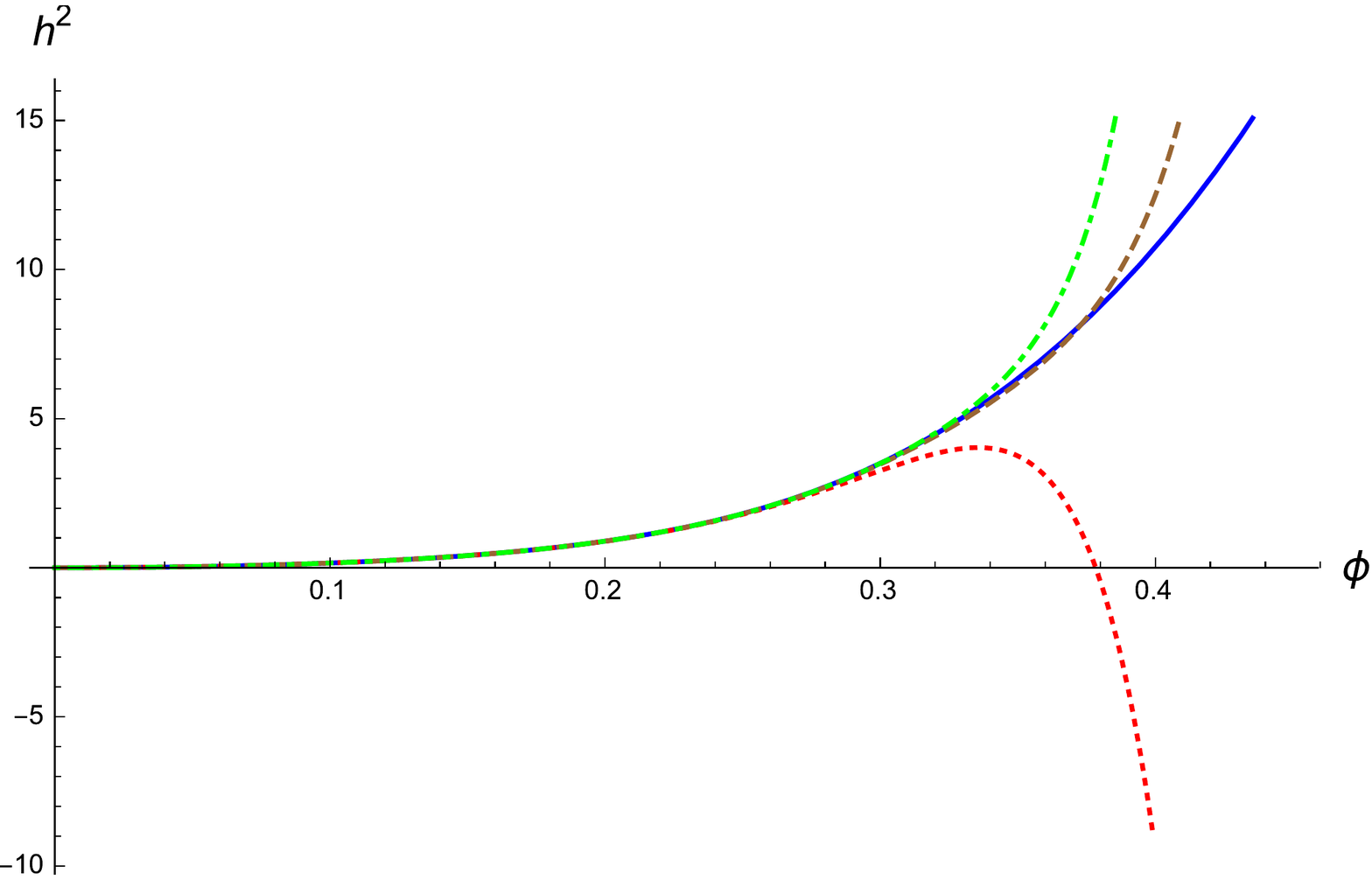}
\vspace{ -2.5cm}
\caption{Comparison of the $X_f=1$ global numerical solution in the LPA (blue, continuous) with the corresponding $(N_v=9,N_h=8)$ polynomial solutions, 
 around the origin as in Eqs.~(\ref{polv_0})-(\ref{polh2_02}) (red, dotted), around a non trivial vacuum as in Eqs.~(\ref{polv_ntv})-(\ref{polh2_ntv2}) (brown, dashed) and in Eqs.~(\ref{polv_ntv})-(\ref{polh_ntv1})
  (green, dot-dashed), for the potential $v(\phi)$ (left panel) and the Yukawa function $y(\phi)=h^2(\phi)$ (right panel).}
\label{fig:comparepol}
\end{center}
\end{figure}
\begin{figure}[!t]
\begin{center}
\vspace{ -2.5cm}
\includegraphics[width=0.45\textwidth]{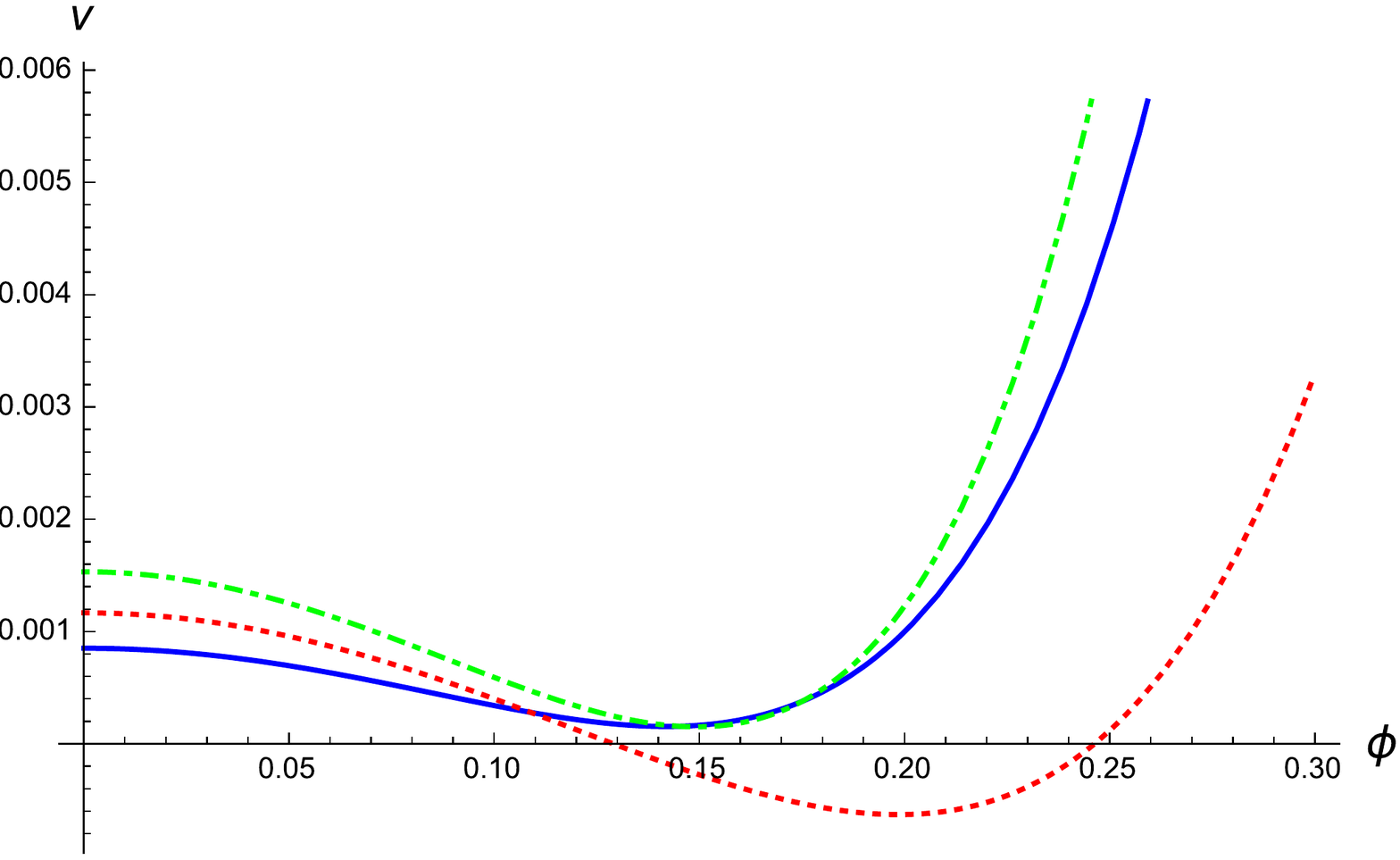}
\includegraphics[width=0.45\textwidth]{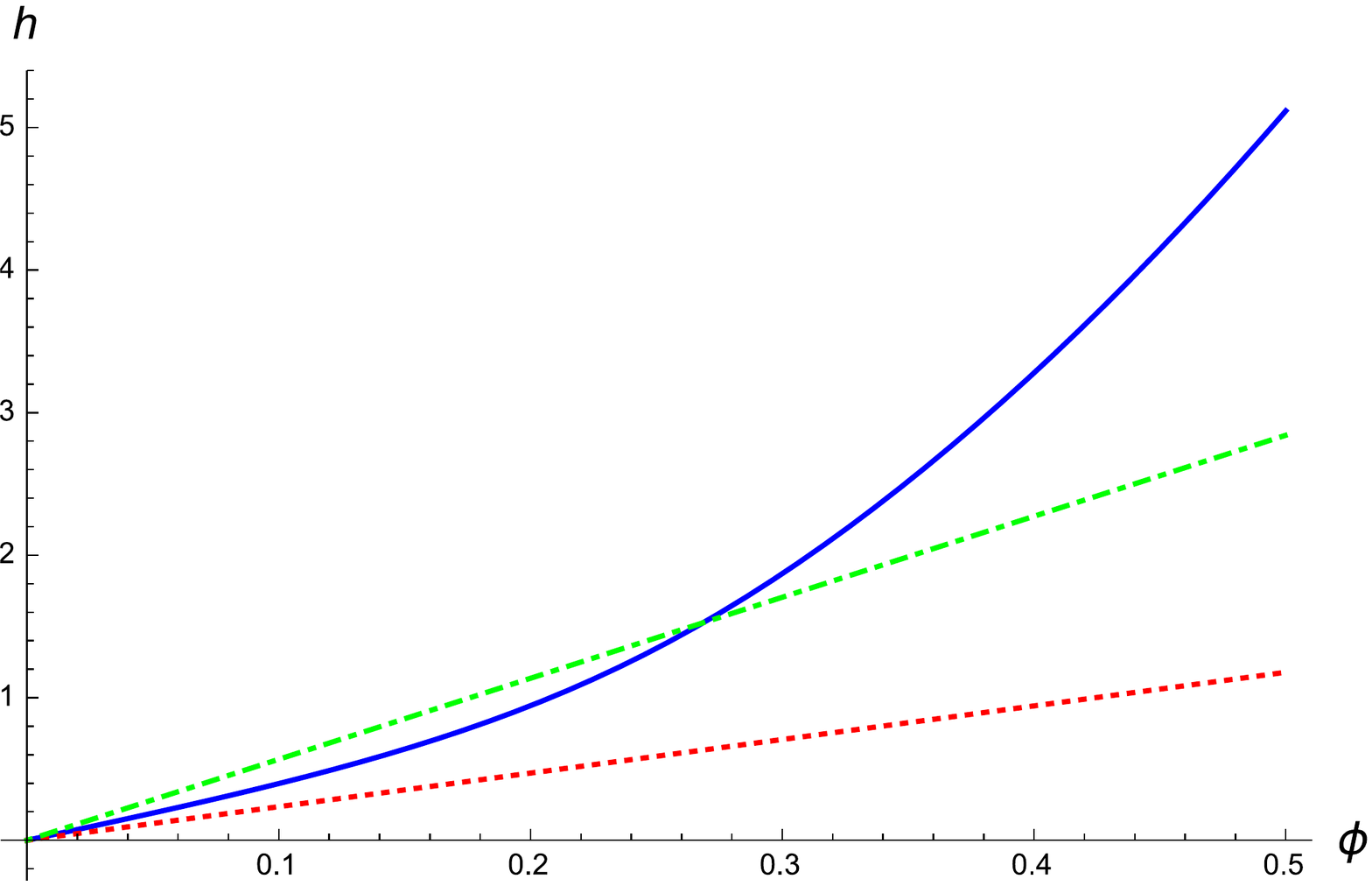}
\vspace{ -2.5cm}
 \caption{Comparison of the $X_f=1$ global numerical solution in the LPA (blue, continuous) with the corresponding $(N_v=9,N_h=1)$ polynomial solutions, 
 around the origin as in Eqs.~(\ref{polv_0})-(\ref{polh2_02})  (red, dotted) and around a non trivial vacuum as  in Eqs.~(\ref{polv_ntv})-(\ref{polh_ntv1})  (green, dot-dashed), for the potential $v(\phi)$ (left panel) and the Yukawa function $h(\phi)$ (right panel).}
\label{fig:compareyuk1}
\end{center}
\end{figure}

In Fig.~\ref{fig:comparepol} we plot different kinds of polynomial solutions, all in
a $(N_v=9,N_h=8)$ truncation,
against the numerical global FP functions, still for $X_f=1$.
For the potential $v$ we show only the domain $\phi\ge 0.3$, the agreement among all the curves 
being perfect for smaller values.
The expansion around the origin has a smaller domain of validity as expected.
Regarding the two set of expansions around a non trivial vacuum, 
the scalar potentials for the two cases are almost indistinguishable,
while for the Yukawa function we obtain a slightly better result employing the one of 
Eq.~(\ref{polh2_ntv2}), as it is shown in the right panel of the figure.
The same kind of plots can be obtained for the polynomial truncations based on a single Yukawa coupling, 
corresponding to a linear Yukawa function.
These are shown in Fig.~\ref{fig:compareyuk1}, were we consider both polynomial expansions,
 around the origin and the non trivial minimum, for $N_v=9$.
The left panel is especially interesting since it shows how, if one forces a linear Yukawa function, 
even with the SSB expansion, the shape of the potential is poorly reproduced.

\begin{table}
\resizebox{7cm}{2cm}{
\begin{tabular}{|l|l|l|l|l|l|l|l|l|l|l|l|l|}
 \hline
 $X_f$ 				&$\,\, 0.3$	& \,\, 0.6	&\,\, 0.9	&\,\, 1.2	&\,\, 1.5 	&\,\, 1.64 \\
 \hline
 $\, \,\kappa$ &$2.311\ 10^{-2}$ &$1.704\ 10^{-2}$ &$1.173\ 10^{-2}$	&$6.845\ 10^{-3}$ &$2.219\ 10^{-3}$ &$1.126\ 10^{-4}$ \\ 
 $\,\,\lambda _2$ 	&9.872 	&12.21	&14.52	&16.75	&18.77	&19.61	\\
 $\,\,\lambda _3$ 	&183.6 	&294.3	&443.4	&632.5	&856.0	&967.6	\\ \hline
 $\,\,h_0$			&4.154 	&4.178	&4.200	&4.218	&4.227	&4.230	\\ 
 $\,\,h_1$  			&35.08	&40.29	&45.66	&51.12	&56.52	&59.04	\\ \hline
 $\,\,\theta_1$  		&1.435 	&1.344	&1.261	&1.185 	&1.117	&1.087	\\ 
 $\,\,\theta_2 $ 		&$-$0.6683 &$-$0.6481 &$-$0.6216 &$-$0.5896 &$-$0.5466 &$-$0.5212\\ 
 $\,\,\theta_3 $ 		&$-$1.022 &$-$1.250 &$-$1.464 &$-$1.656 &$-$1.887 &$-$2.096	\\ \hline
 $\,\,\eta_\psi$ 		&0.2780	& 0.3000 &0.3292 &0.3667 &0.4164 &0.4482 \\ 
 $\,\,\eta_\phi$ 		&0.2366	& 0.3111 &0.4342 &0.6249 &0.8850 &1.014 \\  \hline
\end{tabular}
}
\resizebox{7cm}{2cm}{
\begin{tabular}{|l|l|l|l|l|l|l|l|l|l|l|l|l|}
 \hline
 $X_f$ 			&\,\, 1.64	& \,\, 2	& \,\, 2.5 &\,\, 3	&\,\, 3.5	\\
 \hline
 $\, \,\lambda_1$ &$-2.267\ 10^{-3}$ &0.1403 &0.5480 &1.705 &6.165\\ 
 $\,\,\lambda _2$ 	&19.50 	&29.48	&65.58	&232.9	&1698\\
 $\,\,\lambda _3$ 	&960.5 	&1955	&7265	&$5.313\ 10^4$	&$1.090\ 10^6$	\\ 
\hline
 $\,\,h_0$			&4.223 	&4.600	&5.422	&7.041	&10.88	\\ 
 $\,\,h_1$  			&58.84	&79.82	&142.6	&353.6	&1505	\\ 
\hline
 $\,\,\theta_1$  		&1.071 	&0.9976	&0.9336	&0.9538	&1.041	\\ 
 $\,\,\theta_2 $ 		&$-$0.5212 &$-$0.4661 &$-$0.3727	&$-$0.2725 &$-$0.1783 \\ 
 $\,\,\theta_3 $ 		&$-$2.063 &$-2.725\pm 0.2953$ &$-2.763\pm 0.8557$ &$-2.507\pm 1.242$	&$-1.956\pm 1.695$ \\ 
\hline
 $\,\,\eta_\psi$ 		&0.4521	&0.3372	&0.1066 &-0.1522 &-0.3048	\\ 
 $\,\,\eta_\phi$ 		&1.012	&1.545	&2.971	&6.660	&19.64	\\  
\hline
\end{tabular}
}

\caption{Case $d=3$ and varying $X_f$, polynomial expansion of $h(\phi)$ around the non-trivial (left panel) or trivial (right panel) minimum for both the potential and the Yukawa function, with $N_h=8$ and $N_v=9$ in the LPA.}
\label{tab:LPA_Xfs_h}
\end{table}
\begin{table}
\resizebox{7cm}{2cm}{
\begin{tabular}{|l|l|l|l|l|l|l|l|l|l|l|l|l|}
 \hline
 $X_f$ 				&$\,\, 0.3$	& \,\, 0.6	&\,\, 0.9	&\,\, 1.2	&\,\, 1.5 &\,\, 1.64 \\
 \hline
 $\, \,\kappa$ &$2.310\ 10^{-2}$ &$1.705\ 10^{-2}$ &$1.174\ 10^{-2}$	&$6.846\ 10^{-3}$ &$2.187\ 10^{-3}$ &$3.115\ 10^{-5}$ \\ 
 $\,\,\lambda _2$ 	&9.889 	&12.21	&14.52	&16.75	&18.75	&19.56	\\
 $\,\,\lambda _3$ 	&184.1 	&294.1	&443.4	&632.6	&853.3	&961.5	\\ \hline
 $\,\,y_1$			&48.27 	&46.33	&44.26	&41.48	&37.82	&35.78	\\ 
 $\,\,y_2$  			&1413	&1600	&1783	&1927	&1997	&1997	\\ \hline
 $\,\,\theta_1$  		&1.436 	&1.344	&1.261	&1.184 	&1.112	&1.077	\\ 
 $\,\,\theta_2 $ 		&$-$0.6818 &$-$0.6643 &$-$0.6245 &$-$0.5897 &$-$0.5459 &$-$0.7877\\ 
 $\,\,\theta_3 $ 		&$-$1.021 &$-$1.242 &$-$1.467 &$-$1.665 &$-$1.864	&$-$0.5190\\ \hline
 $\,\,\eta_\psi$ 		&0.2789	& 0.2998 &0.3290 &0.3667 &0.4171 &0.4498 \\ 
 $\,\,\eta_\phi$ 		&0.2367	& 0.3111 &0.4342 &0.6249 &0.8850 &1.014 \\  \hline
\end{tabular}
}
\resizebox{7cm}{2cm}{
\begin{tabular}{|l|l|l|l|l|l|l|l|l|l|l|l|l|}
 \hline
 $X_f$ 			&\,\, 1.64	& \,\, 2	& \,\, 2.5 &\,\, 3	&\,\, 3.5 \\
 \hline
 $\, \,\lambda_1$ &$-6.085\ 10^{-4}$ &0.1424 &0.5501 &1.706 &6.164 \\ 
 $\,\,\lambda _2$ 	&19.53 	&29.52	&65.65	&232.8	&1698	\\
 $\,\,\lambda _3$ 	&959.5 	&1954	&7258	&$5.301\ 10^4$ &$1.089\ 10^6$\\ 
\hline
 $\,\,y_1$			&35.72 	&42.37	&58.84	&99.13	&236.9	\\ 
 $\,\,y_2$  			&1993	&2944	&6192	&$1.990\ 10^4$ &$1.310\ 10^5$\\ 
\hline
 $\,\,\theta_1$  		&1.076 	&1.003	&0.9374	&0.9551	&1.041	\\ 
 $\,\,\theta_2 $ 		&$-$0.5196 &$-$0.4652 &$-$0.3727	&$-$0.2726 &$-$0.1783 \\ 
 $\,\,\theta_3 $ 		&$-$2.006 &$-$2.582 &$-2.794\pm 0.8023$ &$-2.520\pm 1.231$ &$-1.958\pm 1.694$\\ 
\hline
 $\,\,\eta_\psi$ 		&0.4509	&0.3360	&0.1061 &$-$0.1520 &$-$0.3048 \\ 
 $\,\,\eta_\phi$ 		&1.014	&1.548	&2.974	&6.659	  &19.64	\\  
\hline
\end{tabular}
}

\caption{Case $d=3$ and varying $X_f$, polynomial expansion of $y(\rho)$ around the non-trivial (left panel) or trivial (right panel) minimum for both the potential and the Yukawa function, with $N_h=8$ and $N_v=9$ in the LPA.}
\label{tab:LPA_Xfs_h2}
\end{table}
\begin{figure}[!t]
\begin{center}
\includegraphics[width=0.45\textwidth]{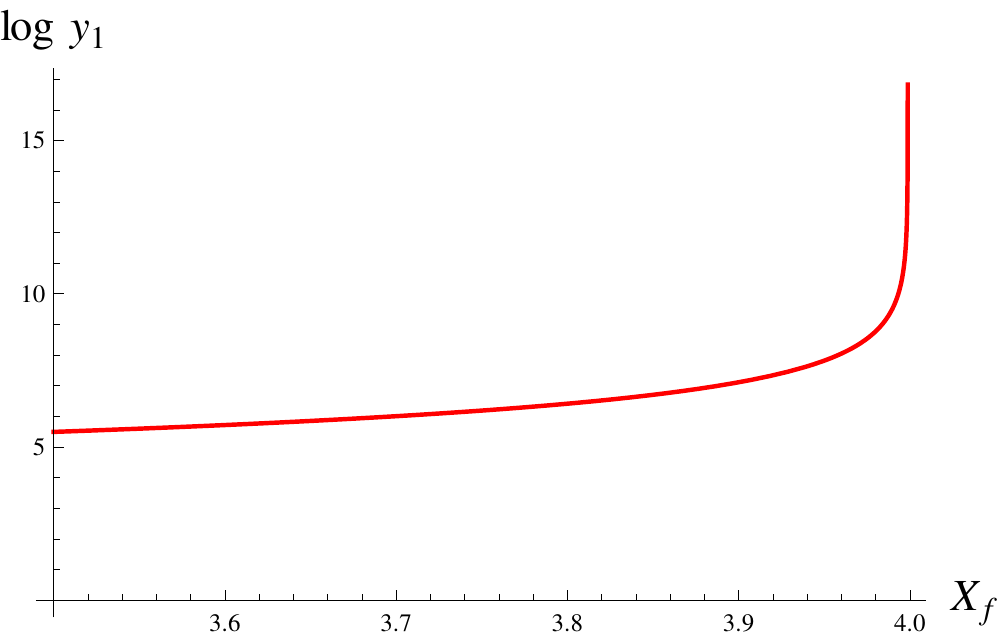}
\caption{Behavior of the coupling $y_1$ in a $N_h=5$, $N_v=6$ polynomial truncation
of $y(\rho)$ around a trivial vacuum, within the LPA. The curve is a fit of data from $X_f=3.5$ to $X_f=4-10^{-7}$.}
\label{fig:LPA_y1_pole}
\end{center}
\end{figure}
Having observed that in the LPA the $(D,D-1)$-systematic polynomial expansions
converge to the global solution for $X_f=1$, we assume that this is 
always the case, and make use of them for addressing how the FP and 
the critical exponents depend on $X_f$ within the LPA.
In Sect.~\ref{sec:largeXf} we have argued that when $X_f$ is not small,
there is no reason to trust the LPA for the $d=3$ critical theory, since
$\eta_\phi$ should approch unity as $X_f$ increases.
This is what the global numerical analysis also indicates.
Indeed in Sect.~\ref{sec:d=3_numerical_FP} we found that the constants 
$A$ and $B$ wildly grow from $X_f=3$ on, in practice making the construnction
of FP potentials harder and harder.
This problem is easily addressed by means of the polynomial expansions.
The results obtained with a $(9,8)$-truncation, both for $h(\phi)$ and $y(\rho)$,
are shown in Tab.~\ref{tab:LPA_Xfs_h} and Tab.~\ref{tab:LPA_Xfs_h2}.

As expected, the anomalous dimensions show a very different $X_f$-dependence.
Starting with $\eta_\psi>\eta_\phi$ for very small $X_f$, the former decreases
and the latter increases as $X_f$ is increased.
Still for $X_f$ around one, the two are small enough for qualitatively trusting
the LPA, though for estimates of the critical exponents the LPA${}^\prime$
provides different and more accurate results.
The polynomial truncations agree with the global analysis and locate around $X_f=1.64$
the transition from the SSB to the SYM regime for the FP potential.
Around this value $\eta_\phi$ reaches unity thus signalling the inconsistent
use of the LPA.
Yet if we insist on using this approximation for larger values of $X_f$,
the breakdown of the approach is signalled by different phenomena.
First of all the critical exponents become complex, from about $X_f=2$ on.
Then the anomalous dimensions $\eta_\phi$ and $\eta_\psi$, which are determined in a somehow un-legitimate
way, become much bigger than unity and negative respectively.
At the same time the couplings at the FP increase very rapidly,
similarly to what was observed in Fig.~\ref{fig:locusAB}.
Actually in LPA it is easier than in the global numerical analysis to understand
how quickly they grow.
The result of a $(6,5)$-polynomial truncation of $y(\rho)$ around a trivial minimum
is shown in Fig.~\ref{fig:LPA_y1_pole}.
It is quite accurate to fit the behavior of the coupling $y_1$ close to $X_f=4$
with a simple pole $y_1\approx121.2/(3.999-X_f)$.
Also the remaining couplings have a rate of growth that
is compatible to a divergence at a finite value of $X_f$,
but these values would lie beyond the pole of $y_1$.

\begin{figure}[!t]
\begin{center}
\vspace{ -2.5cm}
\includegraphics[width=0.45\textwidth]{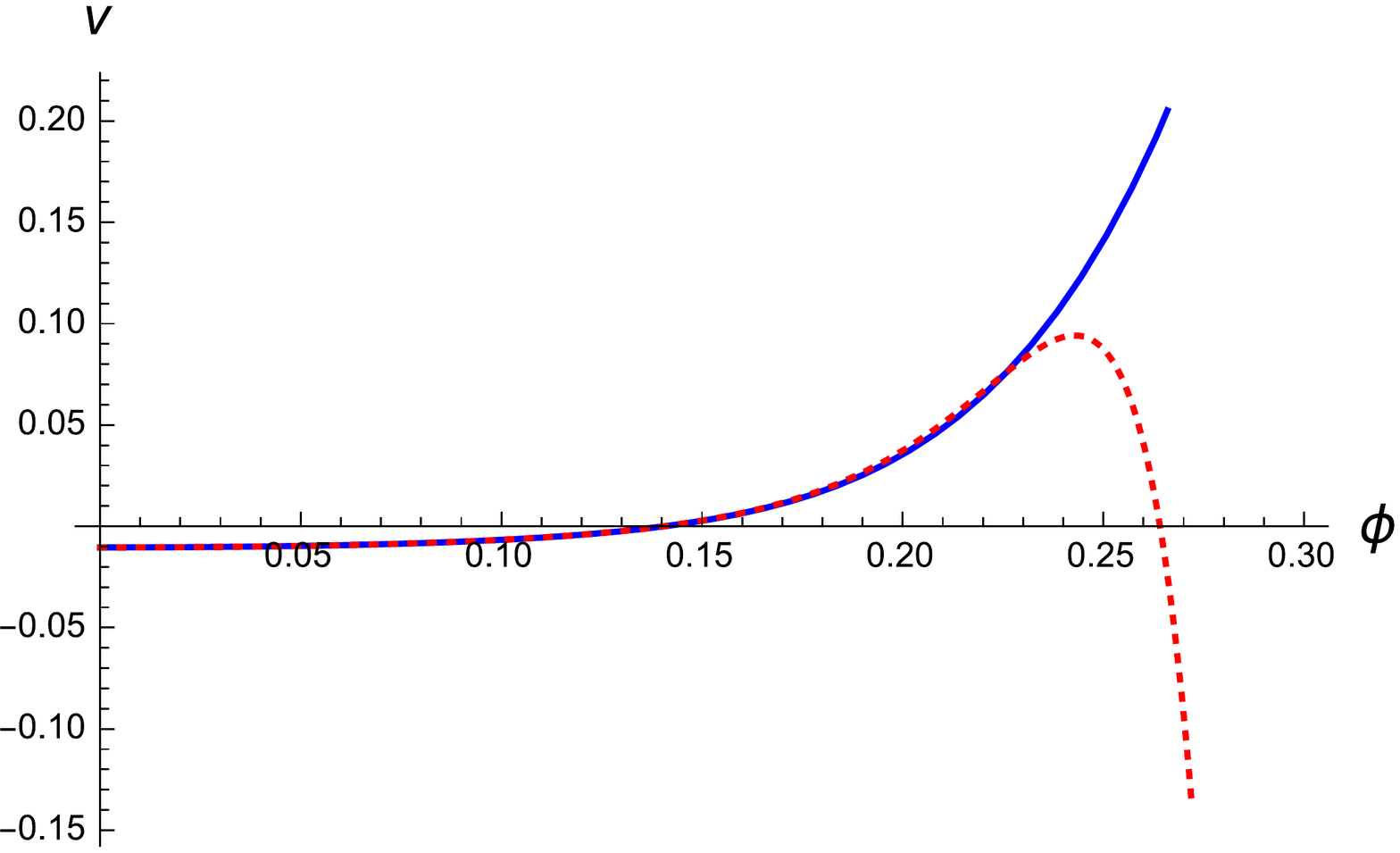}
\includegraphics[width=0.45\textwidth]{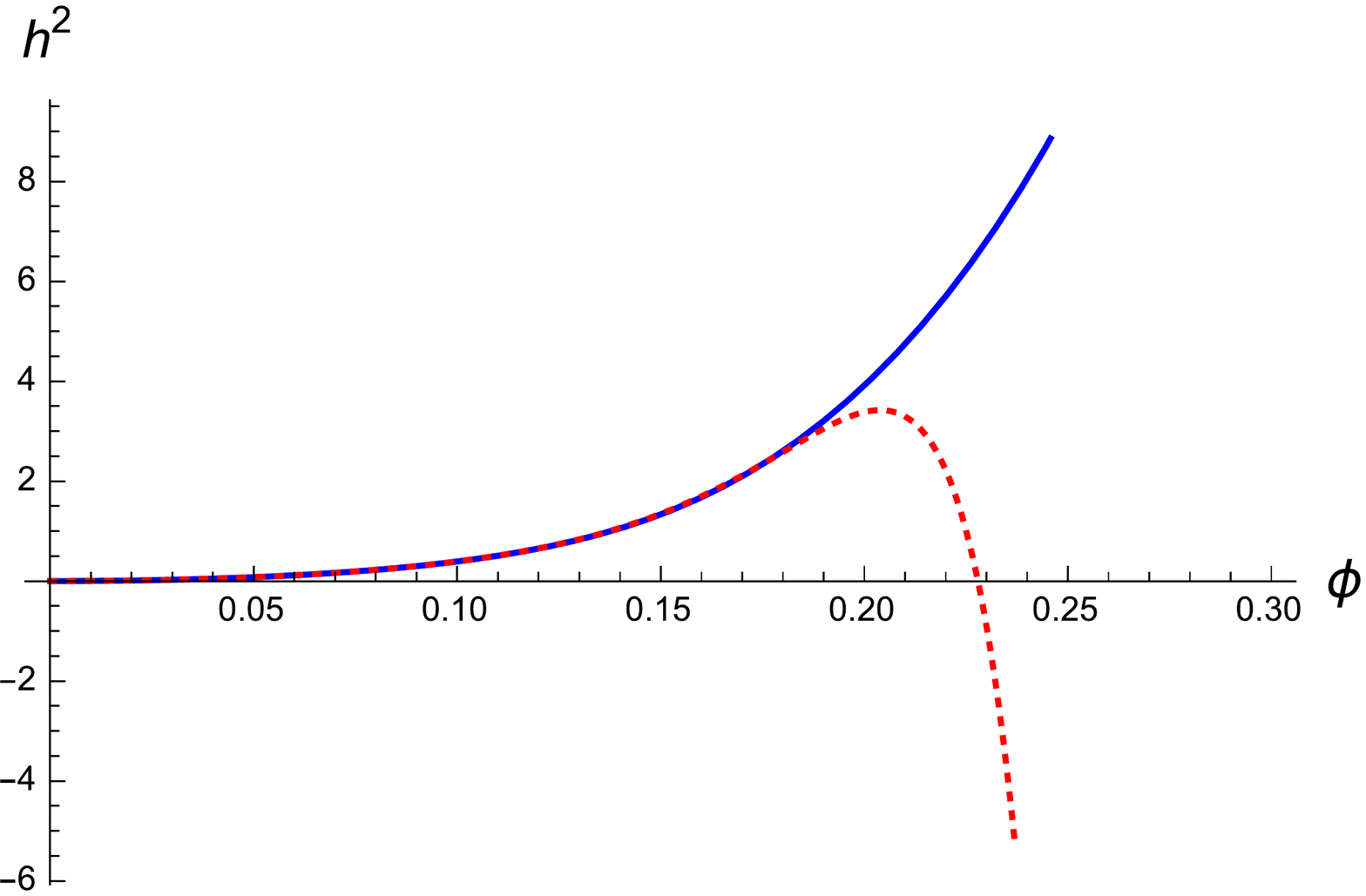}
\vspace{ -2.5cm}
\caption{Comparison of the numerical solution in the LPA (blue, continuous) with the corresponding $(N_v=9,N_h=8)$-polynomial solutions, for $X_f=2$,
 around the origin as in Eqs.~(\ref{polv_0})-(\ref{polh2_02}) (red, dotted), around a non trivial vacuum as in Eqs.~(\ref{polv_ntv})-(\ref{polh2_ntv2}) (brown, dashed) and in Eqs.~(\ref{polv_ntv})-(\ref{polh_ntv1})
  (green, dot-dashed), for the potential $v$ (left panel) and the Yukawa function $y(\phi)=h^2(\phi)$ (right panel).}
\label{fig:comparepol_d3_Xf25}
\end{center}
\end{figure}

Also the comparison between the polynomial truncations and the global numerical results
illustrates the appearance of severe problems as $X_f$ increases. 
Moving to larger values of $X_f$ and entering 
the symmetric regime one sees, again comparing against the numerical solution of the ODEs, 
that the polynomial approximation has a smaller radius of convergence and therefore leads to a less trustworthy estimate of the LPA results.
As an example we present the case $X_f=2.5$ in Fig.~\ref{fig:comparepol_d3_Xf25}.
Here the two curves show a good overlap for $\phi<0.18$, both for $v(\phi)$ and $y(\phi)$,
while at $X_f=1$ the same grade of agreement was found for $\phi<0.28$. 
Again the strongest restriction is imposed by the Yukawa function.
Instead of interpreting these problems as a sign of the generic weakness of the polynomial truncations for
large-$X_f$, we take the point of view that they are the way in which these truncations manifest the failure of the LPA
for $X_f$ roughly bigger than $1.6$. We think that the results of the next Section support this interpretation.
%

\subsection{LPA${}^\prime$}
\label{subsec:polynomial_LPAprime}

In the LPA${}^\prime$ the anomalous dimensions are consistently determined by solving the FP equations together with
the flow equations for the wave function renormalizations. In the previous Sections we have shown
that this is necessary for a correct qualitative description of the dynamics of the model, roughly above $X_f\approx1.6$.
The expectation is that thanks to the wave functions renormalizations the system should gradually 
move towards the large-$X_f$ limit, as it was already checked for truncations with a linear Yukawa 
function~\cite{Rosa:2000ju,Hofling:2002hj,Braun:2010tt,Sonoda:2011qd}.
In this Section we want also to understand how big are the effects of the wave function renormalizations
on the critical exponents, already for small $X_f$.

\begin{table}
\resizebox{6cm}{2cm}{
\begin{tabular}{|l|l|l|l|l|l|l|l|l|l|l|l|l|}
 \hline
 $(N_v,N_h)$ 	&\,\, (5,1)	& \,\, (6,1) 	&\,\, (7,1)	&\,\, (8,1)	&\,\, (9,1)\\  \hline
 $\, \,\kappa$ &$6.250\ 10^{-3}$ &0.01261 &0.01262 	&0.01262 	&0.01262	\\ 
 $\,\,\lambda_2$ 	&6.299 		&6.995 		&7.000		&7.001 		&7.000	\\
 $\,\,\lambda_3$ 	&52.38 		&64.06		&64.28		&64.33 		&64.29	\\ 
\hline
 $\,\,h_0$		&2.139 		&2.533		&2.534 		&2.534 		&2.534	\\ 
\hline
 $\,\,\theta_1$  	&1.595 		&1.548		&1.548 		&1.548 		&1.548 	\\ 
 $\,\,\theta_2 $ 	&$-$0.7528	&$-$0.6828	&$-$0.6832  	&$-$0.6828 	&$-$0.6828 \\ 
 $\,\,\theta_3 $ 	&$-$1.241		&$-$1.289 		&$-$1.299		&$-$1.297		&$-$1.294	\\ 
\hline
 $\,\,\eta_\psi$ 	&0.1168		& 0.1273	&0.1272		&0.1272		&0.1272 \\ 
 $\,\,\eta_\phi$ 	&0.2807		& 0.2237	&0.2238		&0.2238		&0.2238	 \\  
\hline
\end{tabular}
}
\resizebox{6cm}{2cm}{
\begin{tabular}{|l|l|l|l|l|l|l|l|l|l|l|l|l|}
 \hline
 $(N_v,N_h)$ 		&\,\, (5,4)	&\,\, (6,5) 	&\,\, (7,6) 	&\,\, (8,7)	&\,\, (9,8)	\\ \hline
 $\, \,\kappa$ 		& 0.01080	& 0.01078	& 0.01077	& 0.01078 	& 0.01078 	\\ 
 $\,\,\lambda _2$ 	& 6.009		& 5.998 		& 5.997		& 5.998 		& 5.999 		\\
 $\,\,\lambda _3$ 	& 61.01		& 60.50		& 60.47		& 60.54		& 60.56		\\ \hline
 $\,\,h_0$			& 2.474 		& 2.473		& 2.474 		& 2.474 		& 2.474		\\ 
 $\,\,h_1$  			& 7.548 		& 7.530		& 7.542 		& 7.545 		& 7.544 		\\ \hline
 $\,\,\theta_1$  		& 1.444 		& 1.443		& 1.443 		& 1.443 		& 1.443		\\ 
 $\,\,\theta_2 $ 		& $-$0.7721	& $-$0.7739	& $-$0.7745 	& $-$0.7743	& $-$0.7741 	\\ 
 $\,\,\theta_3 $ 		& $-$1.078 	& $-$1.077	& $-$1.084 	& $-$1.086	& $-$1.085		\\ 
\hline
$\,\,\eta_\psi$ 		&0.1535		& 0.1535	&0.1536		&0.1536		&0.1536 	\\ 
 $\,\,\eta_\phi$ 		&0.2214		& 0.2211	&0.2211		&0.2212		&0.2212	 	\\
 \hline
\end{tabular}
}
\caption{Case $d=3$ and $X_f=1$, polynomial expansion of $h(\phi)$ around a non trivial vacuum
 for both the potential and the Yukawa function, in the LPA${}^\prime$, with or without the inclusion 
of multiple-meson-exchange interactions (right and left panel respectively).}
\label{tab:LPAprime_Xf1_SSB_h}
\end{table}

\begin{table}
\resizebox{6cm}{2cm}{
\begin{tabular}{|l|l|l|l|l|l|l|l|l|l|l|l|l|}
 \hline
 $(N_v,N_h)$ 	& $\,\, (5,1)$ & \,\, (6,1) 	&\,\, (7,1)	&\,\, (8,1)	&\,\, (9,1)\\  \hline
 $\, \,\kappa$ 		&$9.208\ 10^{-3}$ &$9.210\ 10^{-3}$	&$9.212\ 10^{-3}$ &$9.213\ 10^{-3}$ 	&$9.212\ 10^{-3}$	\\ 
 $\,\,\lambda _2$ 	&8.300 	&8.307 		&8.315		&8.316 		&8.314	\\
 $\,\,\lambda _3$ 	&72.23 	&72.45		&72.77		&72.82 		&72.71	\\ \hline
 $\,\,y_1$			&18.64 	&18.65		&18.67 		&18.67 		&18.67	\\ 
\hline
 $\,\,\theta_1$  		&1.732 	&1.731		&1.732 		&1.732 		&1.732 	\\ 
 $\,\,\theta_2 $ 		&$-$0.5319 &$-$0.5324	&$-$0.5325  	&$-$0.5318 	&$-$0.5321 \\ 
 $\,\,\theta_3 $ 		&$-$1.626 	&$-$1.657 		&$-$1.676		&$-$1.672		&$-$1.664 	\\ \hline
 $\,\,\eta_\psi$ 		&0.1886	& 0.1887	&0.1887		&0.1887		&0.1887 \\ 
 $\,\,\eta_\phi$ 		&0.2680	& 0.2681	&0.2683		&0.2684		&0.2683	 \\  \hline
\end{tabular}
}
\resizebox{6cm}{2cm}{
\begin{tabular}{|l|l|l|l|l|l|l|l|l|l|l|l|l|}
 \hline
 $(N_v,N_h)$ 	& $\,\, (5,4)$ & \,\, (6,5) 	&\,\, (7,6)	&\,\, (8,7)	&\,\, (9,8)\\  \hline
 $\, \,\kappa$ 		&0.01079& 0.01077	& 0.01078	&0.01078 	&0.01078 	\\ 
 $\,\,\lambda _2$ 	&6.005 	& 5.997 		& 5.997		&5.999 		&5.999	\\
 $\,\,\lambda _3$ 	&60.83 	& 60.43		& 60.50		&60.59 		&60.56	\\ \hline
 $\,\,y_1$			&13.05 	& 13.04		& 13.04 		&13.04 		&13.04	\\ 
 $\,\,y_2$  			&152.0 	& 151.4		& 151.7		&151.8 		&151.7	\\ \hline
 $\,\,\theta_1$  		&1.444 	& 1.443		& 1.443 		&1.443 		&1.443 	\\ 
 $\,\,\theta_2 $ 		&$-$0.7710&$-$0.7738	&$-$0.7745  	&$-$0.7743 	&$-$0.7741 \\ 
 $\,\,\theta_3 $ 		&$-$1.072 	&$-$1.077 		&$-$1.086		&$-$1.086		&$-$1.084 	\\ \hline
 $\,\,\eta_\psi$ 		&0.1536	& 0.1536	&0.1536		&0.1536		&0.1536 \\ 
 $\,\,\eta_\phi$ 		&0.2214	& 0.2211	&0.2211		&0.2212		&0.2212	 \\  \hline
\end{tabular}
}

\caption{Case $d=3$ and $X_f=1$, polynomial expansion of $y(\rho)$ around a non trivial vacuum
 for both the potential and the Yukawa function, in the LPA${}^\prime$, with or without the inclusion 
of multiple-meson-exchange interactions (right and left panel respectively).}
\label{tab:LPAprime_Xf1_SSB_h2}
\end{table}

\begin{table}
\resizebox{7cm}{2cm}{
\begin{tabular}{|l|l|l|l|l|l|l|l|l|l|l|l|l|}
 \hline
 $X_f$ 				&$\,\, 0.3$	& \,\, 0.6	&\,\, 0.9	&\,\, 1.2	&\,\, 1.5 &\,\, 1.62 \\
 \hline
 $\, \,\kappa$ &$2.377\ 10^{-2}$ &$1.793\ 10^{-2}$ &$1.253\ 10^{-2}$	&$7.316\ 10^{-3}$ &$2.171\ 10^{-3}$ &$1.164\ 10^{-4}$ \\ 
 $\,\,\lambda _2$ 	&5.719 	&6.028	&6.045	&5.849	&5.530	&5.385	\\
 $\,\,\lambda _3$ 	&55.00 	&61.19	&61.55	&57.38	&50.81	&47.92	\\ \hline
 $\,\,h_0$			&2.745 	&2.641	&2.518	&2.385	&2.252	&2.201	\\ 
 $\,\,h_1$  			&9.355	&8.798	&7.890	&6.831	&5.789	&5.400	\\ \hline
 $\,\,\theta_1$  		&1.537 	&1.490	&1.453	&1.427 	&1.411	&1.407	\\ 
 $\,\,\theta_2 $ 		&$-$0.8158 &$-$0.7883 &$-$0.7755 &$-$0.7751 &$-$0.7833 &$-$0.7879\\ 
 $\,\,\theta_3 $ 		&$-$0.9829 &$-$1.066 &$-$1.089 &$-$1.063 &$-$1.004 &$-$0.9742\\ \hline
 $\,\,\eta_\psi$ 		&0.1510	& 0.1529 &0.1537 &0.1531 &0.1514 &0.1505 \\ 
 $\,\,\eta_\phi$ 		&0.1366	& 0.1687 &0.2073 &0.2499 &0.2936 &0.3108 \\  \hline
\end{tabular}
}
\resizebox{7cm}{2cm}{
\begin{tabular}{|l|l|l|l|l|l|l|l|l|l|l|l|l|}
 \hline
 $X_f$ 			&\,\, 1.62	& \,\, 2	& \,\, 3 	&\,\, 4	&\,\, 6	&\,\, 8 \\
 \hline
 $\, \,\lambda_1$ &$-7.622\ 10^{-4}$ &$4.135\ 10^{-2}$ &0.1443 &0.2316 &0.3602 &0.4448 \\ 
 $\,\,\lambda _2$ 	&5.375 	&5.472	&5.604	&5.562	&5.185	&4.701	\\
 $\,\,\lambda _3$ 	&47.83 	&43.65	&32.95	&23.64	&11.05	&4.560	\\ 
\hline
 $\,\,h_0$			&2.198 	&2.157	&2.037	&1.915	&1.703	&1.538	\\ 
 $\,\,h_1$  			&5.388	&4.863	&3.635	&2.694	&1.537	&0.9481	\\ 
\hline
 $\,\,\theta_1$  		&1.277 	&1.229	&1.134	&1.077 	&1.024	&1.004	\\ 
 $\,\,\theta_2 $ 		&$-$0.7776 &$-$0.7742 &$-$0.7794	&$-$0.7962 &$-$0.8345 &$-$0.8649 \\ 
 $\,\,\theta_3 $ 		&$-$0.8944 &$-$0.9581 &$-$1.101 &$-$1.196	&$-$1.287	&$-$1.311	\\ 
\hline
 $\,\,\eta_\psi$ 		&0.1508	&0.1314	&$9.347\ 10^{-2}$ &$6.939\ 10^{-2}$ &$4.341\ 10^{-2}$ &$3.073\ 10^{-2}$	\\ 
 $\,\,\eta_\phi$ 		&0.3106	&0.3721	&0.5057	&0.6024	&0.7223	&0.7894	 \\  
\hline
\end{tabular}
}

\caption{Case $d=3$ and various $X_f$,  polynomial expansion of $h(\phi)$ around the non-trivial (left panel) or trivial (right panel) minimum for both the potential and the Yukawa function, with $N_h=8$ and $N_v=9$ in the LPA${}^\prime$.}
\label{tab:LPAprime_Xfs_h}
\end{table}
\begin{table}
\resizebox{7cm}{2cm}{
\begin{tabular}{|l|l|l|l|l|l|l|l|l|l|l|l|l|}
 \hline
 $X_f$ 				&$\,\, 0.3$	& \,\, 0.6	&\,\, 0.9	&\,\, 1.2	&\,\, 1.5 &\,\, 1.62 \\
 \hline
 $\, \,\kappa$ &$2.377\ 10^{-2}$ &$1.793\ 10^{-2}$ &$1.253\ 10^{-2}$	&$7.315\ 10^{-3}$ &$2.169\ 10^{-3}$ &$1.125\ 10^{-4}$ \\ 
 $\,\,\lambda _2$ 	&5.719 	&6.028	&6.045	&5.849	&5.530	&5.384	\\
 $\,\,\lambda _3$ 	&55.00 	&61.19	&61.55	&57.37	&50.79	&47.90	\\ \hline
 $\,\,y_1$			&17.51 	&15.62	&13.67	&11.85	&10.26	&9.690	\\ 
 $\,\,y_2$  			&214.7	&192.0	&162.1	&131.55	&104.5	&95.07	\\ \hline
 $\,\,\theta_1$  		&1.537 	&1.490	&1.453	&1.427 	&1.411	&1.407	\\ 
 $\,\,\theta_2 $ 		&$-$0.8152 &$-$0.7882 &$-$0.7755 &$-$0.7751 &$-$0.7831 &$-$0.7877\\ 
 $\,\,\theta_3 $ 		&$-$0.9833 &$-$1.066 &$-$1.088 &$-$1.062 &$-$1.003 &$-$0.9727\\ \hline
 $\,\,\eta_\psi$ 		&0.1510	& 0.1529 &0.1537 &0.1531 &0.1514 &0.1505 \\ 
 $\,\,\eta_\phi$ 		&0.1366	& 0.1687 &0.2073 &0.2499 &0.2936 &0.3108 \\  \hline
\end{tabular}
}
\resizebox{7cm}{2cm}{
\begin{tabular}{|l|l|l|l|l|l|l|l|l|l|l|l|l|}
 \hline
 $X_f$ 			&\,\, 1.62	& \,\, 2	& \,\, 3 	&\,\, 4	&\,\, 6	&\,\, 8 \\
 \hline
 $\, \,\lambda_1$ &$-7.366\ 10^{-4}$ &$4.137\ 10^{-2}$ &0.1443 &0.2316 &0.3602 &0.4448 \\ 
 $\,\,\lambda _2$ 	&5.374 	&5.471	&5.604	&5.562	&5.185	&4.701	\\
 $\,\,\lambda _3$ 	&47.81 	&43.63	&32.95	&23.64	&11.05	&4.560	\\ 
\hline
 $\,\,y_1$			&9.667 	&9.304	&8.296	&7.338	&5.804	&4.733	\\ 
 $\,\,y_2$  			&94.77	&83.91	&59.23	&41.28	&20.95	&11.67	\\ 
\hline
 $\,\,\theta_1$  		&1.277 	&1.229	&1.134	&1.077 	&1.024	&1.004	\\ 
 $\,\,\theta_2 $ 		&$-$0.7775 &$-$0.7742 &$-$0.7794	&$-$0.7962 &$-$0.8345 &$-$0.8649 \\ 
 $\,\,\theta_3 $ 		&$-$0.8935 &$-$0.9578 &$-$1.101 &$-$1.196	&$-$1.287	&$-$1.311	\\ 
\hline
 $\,\,\eta_\psi$ 		&0.1508	&0.1314	&$9.347\ 10^{-2}$ &$6.939\ 10^{-2}$ &$4.341\ 10^{-2}$ &$3.073\ 10^{-2}$	\\ 
 $\,\,\eta_\phi$ 		&0.3106	&0.3721	&0.5057	&0.6024	&0.7223	&0.7894	 \\  
\hline
\end{tabular}
}

\caption{Case $d=3$ and various $X_f$,  polynomial expansion of $y(\rho)$  around the non-trivial (left panel) or trivial (right panel) minimum for both the potential and the Yukawa function, with $N_h=8$ and $N_v=9$ in the LPA${}^\prime$.}
\label{tab:LPAprime_Xfs_h2}
\end{table}

As in the previous Section, let us start our discussion with the $X_f=1$ model.
Tab.~\ref{tab:LPAprime_Xf1_SSB_h} is the LPA${}^\prime$ version of 
Tab.~\ref{tab:LPA_Xf1_SSB_h},
which considers the truncation of $h(\phi)$ with or without higher Yukawa couplings.
If the effect of the inclusion of multi-meson exchange on the relevant exponent $\theta_1$
 was of the 8\% in the LPA, it gets reduced to the 7\%  in the LPA${}^\prime$.
However, in the truncation of $y(\rho)$ the effect is of the 20\%, see 
Tab.~\ref{tab:LPAprime_Xf1_SSB_h2}
Also, the convergence of the polynomial truncations seems quicker in the LPA${}^\prime$.
A comparison between the left panels of Tab.~\ref{tab:LPAprime_Xf1_SSB_h} and Tab.~\ref{tab:LPAprime_Xf1_SSB_h2}
illustrates how the predictions of the FRG can be made independent of the truncation scheme,
here in the form of a different definition of Yukawa couplings, only by including
full functions of field amplitudes, that is by allowing for higher polynomial couplings.

Once we turn to the dependence of the results on $X_f$, which is shown in 
Tab.~\ref{tab:LPAprime_Xfs_h} and Tab.~\ref{tab:LPAprime_Xfs_h2}, it becomes visible how the 
difference between the LPA and the LPA${}^\prime$ can be negligible 
only for unphysical very small values of $X_f$. For $\theta_1$, it is the 7\%
at $X_f=0.3$, and the 14\% already at $X_f=1$.
On the contrary, as we will see later in this Section by comparing our results to the literature,
the effect of the inclusion of higher Yukawa couplings decreases with incresing $X_f$.
The transition between the SSB and the symmetric regime for the FP potential in the LPA${}^\prime$
is around $X_f=1.62$, while it occurs at $X_f=2.31$ for truncations with a linear Yukawa
function~\cite{Borchardt:2015rxa}. From these tables it also seems reasonable to expect
 that in the $X_f\to 0$ limit the Yukawa couplings attain finite nonvanishing values, as
it was observed already in the LPA, see Fig.~\ref{fig:locus}.
Also, the trend in the change of $\theta_1$ and $\eta_\phi$ is compatible with an approach
to the corresponding Ising values, thus further supporting the discussion at the end of
Sect.~\ref{sec:model_and_flow}.
As far as the $X_f\to\infty$ limit is concerned instead, the smooth transition to the large-$X_f$ exponents 
is evident in the right panels of
Tab.~\ref{tab:LPAprime_Xfs_h} and Tab.~\ref{tab:LPAprime_Xfs_h2}.

Let's now come to the comparison of our results with the literature.
The classic methods for the investigation of the critical properties of the Gross-Neveu
and Yukawa models are the $\epsilon$- and the $1/N_f$-expansions~\cite{Rosenstein:1988dj,ZinnJustin:1991yn,Rosenstein:1993zf,Gracey:1990,Gracey:1993kb,Gracey:1993kc,Vasiliev:1997sk}.
The former can be of great utility since both expansions around the upper
and the lower critical dimensions give comparable results, such that $d=3$ does not
seem a too wild extrapolation.
Yet, some treatment for these asymptotic series is needed. 
Resummation is unfortunately out
of reach since they are known only up to the second or third order~\cite{Gracey:1990,Rosenstein:1993zf},
apart for the anomalous dimensions for which the computations have been pushed up to the 
fourth order~\cite{Vasiliev:1997sk}.
Polynomial interpolations of the two different $\epsilon$-expansions
have been studied in~\cite{Janssen:2014gea} for the case $X_f=8$,
and we report their results borrowing their notations, such that $P_{i,j}$ denotes a
polynomial which is $i$-loop exact near the lower critical dimension, and $j$-loop exact near the upper.
We also report the crude extrapolations that are obtained by simply setting $\epsilon=1$
in the expansions of $\theta_1=\nu^{-1}$, $\eta_\phi$ and $\eta_\psi$~\footnote{We made use of the
formulas reported in~\cite{Rosenstein:1993zf}, with typos corrected according to the observations of~\cite{Janssen:2014gea}.}.
Also the $1/N_f$ expansion clearly needs some care, since we are interested in low number of fermions.
Actually we are going to refer to this method only for $X_f=8$ and $X_f=4$, corresponding to
$N_f=2$ and $N_f=1$ respectively. 
Again only the second or third order is known~\cite{Gracey:1993kb,Gracey:1993kc}.
For the correlation-length exponent $\theta_1=\nu^{-1}$ we adopt the Pad\'e approximant used in~\cite{Janssen:2014gea}, 
while for the anomalous dimensions we refer to the Pad\'e - Borel treatment reported in~\cite{Hofling:2002hj}.

The available FRG literature is rich and it offers a precious background on which we
can measure the effects of the enlargement of the truncation discussed in this work.
Essentially all the past studies considered the LPA${}^\prime$,
including a scalar potential and a simple linear Yukawa 
coupling~\cite{Rosa:2000ju,Hofling:2002hj,Braun:2010tt,Sonoda:2011qd,Janssen:2014gea,Borchardt:2015rxa},
that can be considered as the first order in the truncation of \Eqref{eq:poly_h_third}. 
The only exception in this sense is provided by the supersymmetry-preserving scheme that
has been applied to the $X_f=1$ case, which retained a full superpotential~\cite{Synatschke:2010ub,Heilmann:2014iga,Zanusso:2015},
thus including multi-meson exchange in the Yukawa sector,
and sometimes was pushed to the next-to-next-to-leading order of the (supercovariant) derivative expansion.
Also the choice of regulators is diverse, comprehending the linear, the sharp and the exponential 
ones (which in the tables we abbreviate with lin, sha, exp).
In some studies the scalar potential was approximated by polynomial truncations
in the symmetric regime, for which we provide the corresponding $N_v$
($N_w$ in case of truncations of the superpotential for supersymmetric flows). 
In others, that we label by $N_v=\infty$ (or $N_w=\infty$), the differential equations for the FP and the perturbations
around it were solved by numerical methods, which are different from paper to paper.
Our results are labeled by $N_h>1$.

Other methods to which we can compare in special cases are Monte-Carlo simulations
and the conformal bootstrap. Both of them can give high-precision computations of the
critical exponents, but so far they have had a limited application to low-$X_f$ Yukawa models.
For $X_f=8$ two lattice calculations of the critical exponents are available.
One based on staggered fermions~\cite{Karkkainen:1993ef}, though ignoring a sign problem, 
provides results which are in good agreement
with continuum methods, as it appears from Tab.~\ref{tab:comparison_Xf8}.
An independent work applying the fermion bag approach~\cite{Chandrasekharan:2013aya},
that is free from the sign problem, is instead offering very different results:
$\nu=0.83(1)$, $\eta_\phi=0.62(1)$, $\eta_\psi=0.38(1)$.
In both cases it is not clear whether the symmetry of the lattice model
is the expected one in the continuum limit~\footnote{We are grateful to H. Gies for informing us 
about these discussions.}. 
Recently, another sign-problem-free simulation adopting the continuous time quantum Monte-Carlo method
for a model of spinless fermions on a honeycomb lattice, 
provides estimates of the critical exponents of the chiral Ising universality class
for $X_f=4$, i.e. a single Dirac field~\cite{Wang:2014cbw}.
These results are compared to those emerging from the continuum
methods in Tab.~\ref{tab:comparison_Xf4}.
Surprisingly they are much closer to our estimates for the case $X_f=2$, see 
Tab.~\ref{tab:comparison_Xf2}.

Regarding the latter case, notice that the results from~\cite{Hofling:2002hj} are affected
by the absence of some terms in the flow equations that,
being proportional to the vev of the scalar, become important for $X_f\leq2$~\footnote{See
the discussion in~\cite{Borchardt:2015rxa}.}.
Their effect significantly reduces the value of $\nu$.
Since upon inclusion of multi-meson exchange the transition from the symmetric to the SSB 
regime occurs at lower values of $X_f$, our computations are still in the symmetric regime.
This might qualitatively explain the drastic departure from the results of~\cite{Borchardt:2015rxa}.

\begin{table}
\resizebox{7cm}{2.5cm}{
\begin{tabular}{|l|l|l|l|l|l|l|l|l|l|l|l|l|}
 \hline
								&$\nu$			&$\theta_1$		&$\eta_\phi$			&$\eta_\psi$	\\
 \hline
 FRG $(N_v=9,N_h=8)$ lin (this work)				&1.004			&0.996			&0.789				&0.031		\\ 
 FRG $(N_v=3,N_h=1)$ exp~\cite{Hofling:2002hj}			&1.016			&0.984			&0.786				&0.028		\\ 
 FRG $(N_v=6,N_h=1)$ sha~\cite{Janssen:2014gea}			&1.022			&0.978			&0.767				&0.033		\\ 
 FRG $(N_v=11,N_h=1)$ lin~\cite{Braun:2010tt}			&1.018			&0.982			&0.760				&0.032		\\ 
 FRG $(N_v=\infty,N_h=1)$ lin~\cite{Hofling:2002hj}		&1.018			&0.982			&0.756				&0.032		\\ 
 FRG $(N_v=\infty,N_h=1)$ lin~\cite{Borchardt:2015rxa}		&1.018			&0.982			&0.760				&0.032		\\ 
 Monte-Carlo~\cite{Karkkainen:1993ef}				&1.00(4)		&1.00(4)		&0.754(8)			&\ \ --- 	\\
 $1/N_f$ 2nd/3rd order~\cite{Gracey:1993kc,Janssen:2014gea}	&1.040			&0.962			&0.776			&0.044		\\
 $(2+\epsilon)$ 3rd order~\cite{Gracey:1990}			&1.309			&0.764			&0.602				&0.081		\\
 $(4-\epsilon)$ 2nd order~\cite{Rosenstein:1993zf}		&0.948			&1.055			&0.695				&0.065		\\
 $P_{2,2}$ interpolated 
 $\epsilon$-expansion~\cite{Janssen:2014gea}			&1.005			&0.995			&0.753				&0.034		\\
 $P_{3,2}$ interpolated 
 $\epsilon$-expansion~\cite{Janssen:2014gea}			&1.054			&0.949			&0.716				&0.041		\\
\hline
\end{tabular}
}

\caption{Critical exponents for $X_f=8$. For a short description of the approximations involved in each method, see the main text.}
\label{tab:comparison_Xf8}
\end{table}

\begin{table}
\resizebox{7cm}{1.5cm}{
\begin{tabular}{|l|l|l|l|l|l|l|l|l|l|l|l|l|}
 \hline
								&$\nu$			&$\theta_1$		&$\eta_\phi$			&$\eta_\psi$	\\
 \hline
  FRG $(N_v=9,N_h=8)$ lin (this work)				&0.929			&1.077			&0.602				&0.069		\\ 
 FRG $(N_v=3,N_h=1)$ exp~\cite{Hofling:2002hj}			&0.962			&1.040			&0.554				&0.067		\\ 
 FRG $(N_v=\infty,N_h=1)$ lin~\cite{Hofling:2002hj,Borchardt:2015rxa}		&0.927			&1.079			&0.525				&0.071		\\ 
 Monte-Carlo~\cite{Wang:2014cbw}				&0.80(3)		&1.25(3)		&0.302(7)			&\ \ --- 	\\
 $1/N_f$ 2nd/3rd 
 order~\cite{Gracey:1993kb,Gracey:1993kc,Janssen:2014gea}	&0.955			&1.361			&0.635				&0.105		\\
 $(4-\epsilon)$ 2nd order~\cite{Rosenstein:1993zf}		&0.862			&1.160			&0.502				&0.110		\\
\hline
\end{tabular}
}

\caption{Critical exponents for $X_f=4$. For a short description of the approximations involved in each method, see the main text.}
\label{tab:comparison_Xf4}
\end{table}

\begin{table}
\resizebox{7cm}{1.65cm}{
\begin{tabular}{|l|l|l|l|l|l|l|l|l|l|l|l|l|}
 \hline
								&$\nu$			&$\theta_1$		&$\eta_\phi$			&$\eta_\psi$	\\
 \hline
  FRG $(N_v=9,N_h=8)$ lin (this work)				&0.814			&1.229			&0.372				&0.131		\\ 
 FRG $(N_v=3,N_h=1)$ exp~\cite{Hofling:2002hj}			&0.633			&1.580			&0.319				&0.113		\\ 
 FRG $(N_v=3,N_h=1)$ lin~\cite{Hofling:2002hj}			&0.623			&1.605			&0.308				&0.112		\\ 
 FRG $(N_v=\infty,N_h=1)$ exp~\cite{Hofling:2002hj}		&0.640			&1.563			&0.319				&0.114		\\ 
 FRG $(N_v=\infty,N_h=1)$ lin~\cite{Hofling:2002hj}		&0.621			&1.610			&0.308				&0.112		\\ 
 FRG $(N_v=\infty,N_h=1)$ lin~\cite{Borchardt:2015rxa}				&0.4836			&2.068			&0.3227				&0.1204		\\
  $(4-\epsilon)$ 2nd order~\cite{Rosenstein:1993zf}		&0.773			&1.293			&0.317				&0.154		\\
\hline
\end{tabular}
}

\caption{Critical exponents for $X_f=2$. For a short description of the approximations involved in each method, see the main text.}
\label{tab:comparison_Xf2}
\end{table}

\begin{table}
\resizebox{9.5cm}{2cm}{
\begin{tabular}{|l|l|l|l|l|l|l|l|l|l|l|l|l|}
 \hline
							&$\nu$		&$\theta_1$	&$\theta_2$		&$\eta_\phi$		&$\eta_\psi$	&3-2$\theta_1$\\
 \hline
  FRG $(N_v=9,N_h=8)$ lin (this work)			&0.693		&1.443		&$-$0.796			&0.154			&0.221			&0.114		   \\ 
 SUSY FRG $(N_w=\infty)$ 
opt $n=2$ NLO~\cite{Heilmann:2014iga}		&0.711		&$1.407$	&$-0.771$		&0.186			&0.186			&0.186		   \\
 SUSY FRG $(N_w=\infty)$ 
opt $n=2$ NNLO~\cite{Heilmann:2014iga}		&0.710		&$1.410$	&$-0.715$		&0.180			&0.180			&0.180		   \\
 SUSY FRG $(N_w=\infty)$ 
opt $n=1$~\cite{Zanusso:2015}		&0.708		&1.413		&$-$0.381			&0.174			&0.174			&0.174		   \\
 SUSY FRG $(N_w=\infty)$ 
opt $n=2$~\cite{Zanusso:2015}				&0.706		&1.417		&$-$0.377			&0.167			&0.167			&0.167		   \\
 FRG $(N_v=2,N_h=1)$ 
1-loop~\cite{Sonoda:2011qd}				&0.72		&1.39		&$-$0.71			&0.15			&0.15			&0.22		   \\
$(4-\epsilon)$ 1st 
order~\cite{Grover:2013rc}				&\ ---		&\ ---		&\ ---			&0.143			&0.143			&\ ---		   \\
 $(4-\epsilon)$ 
 2nd order~\cite{Rosenstein:1993zf}		&0.710		&1.408		&\ ---			&0.184			&0.184			&0.184		\\
 Conformal Bootstrap
\cite{Bashkirov:2013vya}				&\ ---		&\ ---		&\ ---			&0.13			&0.13			&\ ---		   \\
\hline
\end{tabular}
}

\caption{Critical exponents for $X_f=1$. About the FRG results, the schemes, the regulators, and the approximations are very different, see the main text. }
\label{tab:comparison_Xf1}
\end{table}

Also the comparison for $X_f=1$, which is presented in Tab.~\ref{tab:comparison_Xf1},
requires some comments. Let us recall that for this field-content the system at criticality is described by a
${\cal N}=1$ Wess-Zumino model~\cite{Sonoda:2011qd,Grover:2013rc}.
Hence, if the regularization does not break supersymmetry,
the critical anomalous dimensions of the scalar and of the spinor should be equal.
Furthermore, a superscaling relation $\nu^{-1}=(d-\eta)/2$, which was first observed in~\cite{Gies:2009az}
and later proved to hold at any order in the supercovariant derivative expansion in~\cite{Heilmann:2014iga},
is expected to hold.
This is what happens for example in the $\epsilon$-expansions or in the SUSY FRG.
Since the scheme adopted in the present work explicitely breaks supersymmetry,
we expect and we observe violations of these properties.
Also in~\cite{Sonoda:2011qd} supersymmetry is broken by regularization, and these violations are present,
but they could be partially reduced or canceled by tuning the regulator.
This tuning gives the results reported in Tab.~\ref{tab:comparison_Xf1}.
A similar analysis of the regulator dependence of universal quantities and of the consequent breaking of
supersymmetry could be performed in future studies for the present family of truncations.
Yet, even by explicitly breaking the FP supersymmetry, we get exponents which are not
very far from the ones produced by the above mentioned methods.
Let us add few details on the SUSY FRG results shown in Tab.~\ref{tab:comparison_Xf1}.
They are obtained by setting one of the regulators to zero, 
and choosing a shape similar to the linear regulator for the other, with an exponent $n$ that differentiates
between the conventional linear regulator (opt $n=2$) and a slight variant (opt $n=1$).
Also the truncation scheme is different from the one discussed in the present paper, since it is related to an
expansion in powers of the supercovariant derivative, that has been considered at the level of the LPA${}^\prime$~\cite{Synatschke:2010ub,Zanusso:2015},
at next-to-leading order (NLO) or at next-to-next-to-leading order (NNLO)~\cite{Heilmann:2014iga}.
For the case $X_f=1$ we can also compare with a pioneering study based on the conformal bootstrap~\cite{Bashkirov:2013vya}.
In Tab.~\ref{tab:comparison_Xf1} we included the one-loop computations of~\cite{Sonoda:2011qd,Grover:2013rc},
even if two-loop results are on the market~\cite{Rosenstein:1993zf},
on the base of the naive observation that for Yukawa systems with complex scalars and spinors, 
whose FP should effectively show ${\cal N}=2$ supersymmetry~\cite{Balents:2008vd},
the anomalous dimensions obtained from the first-order of the $(4-\epsilon)$
expansion, $\eta_\phi=\eta_\psi=1/3$, agree with the available exact results~\cite{Aharony:1997bx}.

\section{d=4}
\label{sec:d=4}

\begin{figure}[!t]
\begin{center}
\includegraphics[width=0.45\textwidth]{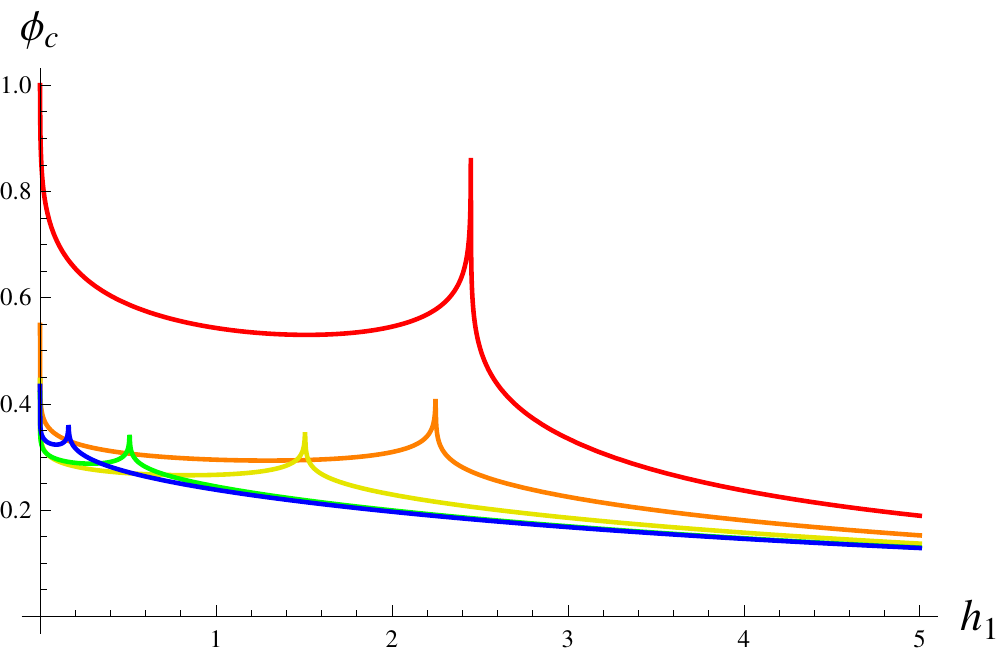}
 \caption{Spike plots for $X_f=0$, $v(\phi)=0$ and $d\in\{3.5, 3.7, 3.9, 3.99, 3.999\}$ from red (upper) to blue (lower) in the LPA${}^\prime$.}
\label{fig:d=4_merging}
\end{center}
\end{figure}

From the leading order of the $1/X_f$-expansion one expects that for large enough $X_f$
the chiral Ising FP merges with the Gau\ss ian FP in the $d\to4$ limit.
Also at $X_f=0$, for which we know from the discussion at the end of Sect.~\ref{sec:model_and_flow}
that only mirrored images of the purely scalar FPs can exist,
one can observe that the latter merge with the Gau\ss ian FP for $d\to4$,
compatibly with the presumed triviality of four-dimensional scalar theory.
This is illustrated in Fig.~\ref{fig:d=4_merging}, which is produced as Fig.~\ref{fig:spike3}
but integrating only the FP equation for $h(\phi)$ at $v(\phi)=0$ and $X_f=0$ in the LPA${}^\prime$.
Yet, it remains to be shown what happens for a small non-vanishing number of fermions.
Dimensional analysis indicates $d=4$ as the upper critical dimension
for any $X_f$. This can be checked by means of the FRG, either by numerical integration of the FP equation,
as it was shown for example in Sect.~\ref{sec:LPA_Xf1_genericd} for $X_f=1$, 
or by the polynomial truncations discussed in the last Sections.
Indeed, the latter have already been used in the past, precisely to address this question.

In fact, an exploratory study of what happens to the $d\to4$ limit in a $\mathbb{Z}_2$-symmetric Yukawa model with very small $X_f$
was performed in~\cite{Gies:2009hq}, in order to test a mechanism for the generation of nontrivial FPs in
fermion-boson models, that has subsequently found in chiral-Yukawa models some natural candidates~\cite{Gies:2009sv}.
That analysis pointed out that within a $(N_v=2,N_h=1)$ polynomial truncation, according to the scheme 
of \Eqref{eq:poly_h_third}, the FRG detects nontrivial FPs also in $d=4$, for unphysical small values of $X_f$.
This holds both in the LPA and in the LPA${}^\prime$. However, the fact that the FP position and the 
critical exponents are significantly different in the two approximations was interpreted as a signal 
of the need to include further boson-fermion interactions in the truncation, in order to understand
if these FPs are physical or merely an artifact of the approximations.
This Section reports on the changes brought by the different treatment of the Yukawa sector presented in this work.

At the level of the LPA we generated three-dimensional plots similar to the ones
illustrated in Fig.~\ref{fig:spike1}, second panel, by shooting from the origin with random values of
$(v''(0),h'(0))$, for several values of $X_f<1$, and we looked for spikes signaling possible FPs, but we have not found
any of them.
We were also not able to produce any global solution studying numerically the Cauchy problem from the asymptotic region,
along the lines of Sect.~\ref{sec:d=3_numerical_FP}. 
We then re-considered the analysis at the level of polynomial truncations.
Already trying to reproduce the results of~\cite{Gies:2009hq} in other truncations with $N_v=2$ and $N_h=1$, can be 
a nontrivial test, because of the different beta-function of the Yukawa coupling, associated to different projection
rules. We have already argued that a change of the results depending on the parameterization
employed signals the presence of errors induced by the use of inconsistent truncations.
We first concentrated on the LPA, which at least for $d<4$ is able to reproduce 
the right number of nontrivial FPs. In this case, the truncation adopted in~\cite{Gies:2009hq} allows for 
non-Gau\ss ian FPs  approximately for $X_f\leq1$. For instance, at $X_f=0.4$ one can find the following FP 
\be
\kappa=0.00165\ ,\quad 
\lambda_2=27.26\ ,\quad 
g_1^2=81.13
\ee
with two relevant directions
\be
\theta_1=2.372\ ,\quad 
\theta_2=0.592\ ,\quad 
\theta_3=-2.859 \ .
\ee
We observed that in a polynomial truncation of $y(\rho)$ as in \Eqref{eq:poly_h2_SSB},
the FP position is different
\be
\kappa=0.00167\ ,\quad 
\lambda_2=54.18\ ,\quad 
y_1=494.0
\ee
as well as the critical exponents
\be
\theta_1=1.653\ ,\quad 
\theta_2=0.932\ ,\quad 
\theta_3=-3.445 \ .
\ee
Still, the changes are not dramatic.
On the other hand, we could not find any real FP within the same order of the
truncation of $h(\phi)$ given in \Eqref{eq:poly_h_SSB}. We tried to circumvent this
problem as in $d=3$, by following the FP found in one parameterization
to higher orders, and then translating back to the other parameterization.
Yet, we were not able to reveal the FP for $y(\rho)$ for bigger values of
$N_v$ and $N_h$, nor to find it by chance in different orders of the truncation of $h(\phi)$.

Hoping that the inclusion of the wave function renormalizations could
stabilize the polynomial truncations and help us in the search for FPs, we 
then considered LPA${}^\prime$, using the results of~\cite{Gies:2009hq} as
a guide for the localization of the interesting region in the space of couplings.
While the FP is present in the first order of the truncation of \Eqref{eq:poly_h_third},
we could not find it in the parameterizations considered in this paper.
Let us once more stress that this does not completely exclude that it can be found by other methods,
even if we consider this very unlikely.
Nevertheless, for the LPA${}^\prime$ we have not tried a numerical shooting at nonvanishing $X_f$ as
in the LPA.  Hence, a more careful numerical analysis is needed, to exclude with a higher level of confidence 
the presence of low-$X_f$ FPs in the theory space described by the truncation in \Eqref{eq:truncation}.
A even better test would be to consider the full next-to-leading order of the derivative expansion.

\section{Conclusions}
\label{sec:conclusions}

A proper quantitative control of the quantum dynamics of the $\mathbb{Z}_2$-symmetric Yukawa model,
beyond the domain of applicability of perturbative methods,
is important not only from a generic field-theoretical point of view,
but also for phenomenological reasons, since the latter is very useful as a toy-model of numerous condensed matter systems,
as well as of specific sectors of modern particle theory, see Sect.~\ref{sec:intro} for more details. 
The functional renormalization group (FRG) is a simple analytic nonperturbative method that can provide
a detailed description of strongly coupled systems, under approximations that are testable and improvable in several
systematic ways. Furthermore, these results can be produced, almost simultaneously, in a continuous number of 
spacetime dimensions $d$ and fermionic degrees of freedom $X_f$, thus allowing for a quick analysis of the
dependence of the dynamics on the latter parameters.

In this work we focused on the critical behavior of the $\mathbb{Z}_2$-symmetric Yukawa model at zero temperature and density.
Our principal aim was to test the impact of multi-meson-exchange,
 encoded in a Yukawa coupling which is a full function of the scalar field, on the FRG description of the latter behavior,
a question that to our knowledge has never been considered before.
Nevertheless, our analysis is relevant not only for the FRG community. For instance, in Sect.~\ref{sec:largeXf} we 
discussed the leading order of the $1/X_f$-expansion, whose results can be directly exported out of
the FRG framework in which we produced them, and recovered also by different methods.
This study illustrated how, by allowing for multi-meson-exchange, one can describe the generation
of multi-critical conformal Yukawa models as $d$ is lowered from $d=3$ towards $d=2$, across the corresponding
upper critical dimensions $d_n=2+2/n$, with $n$ a positive integer. We also showed how the large-$X_f$ limit
quantizes the corresponding critical anomalous dimension $\eta_\phi=d_n-d$.
In Sect.~\ref{sec:LPA_Xf1_genericd} we checked that this pattern of
generation of critical theories as a function of $d$ holds also at $X_f=1$, and presumably at
any other finite $X_f$. 
This would imply that in the $d\to 4$ limit only the Gau\ss ian fixed point (FP) suvives.
The latter statement, being of special relevance for particle physics, was further analyzed in Sect.~\ref{sec:d=4},
where we argued that it applies also for $X_f<1$, at least within the ansatz of~\Eqref{eq:truncation}.
Let us remark that, as far as we know, the observation of multi-critical conformal Yukawa models at finite
$X_f$ and in continuous fractal dimensions $2<d<3$ is a novel result.

Concerning the finite-$X_f$ results, they indicate that in several cases the effect of multi-meson-exchange
cannot be neglected, either quantitatively or even qualitatively. 
We argued that these higher Yukawa interactions are required by consistency
of the truncation, otherwise the solutions of the system of differential equations defining
the flow of the scalar potential $v(\phi)$ and of the Yukawa ``potential" $h(\phi)$ would depend on the chosen 
parametrization of these functions. For instance, the same FP solutions should be reproduced using any polynomial 
truncation of these functions, at least within a certain domain.
On these grounds we believe that general FRG studies of Yukawa models should at least consider the inclusion
of these interactions, and possibly check when they can actually be neglected.

On the quantitative side, in Sect.~\ref{sec:d=3_numerical_FP} we explicitly numerically constructed these global FP solutions for $d=3$ and
several values of $X_f$. These results include the Gross-Neveu universality class for $X_f>1$, and the superconformal ${\cal N}=1$ Wess-Zumino model for $X_f=1$.
At $X_f=1$, we also numerically computed the critical exponent $\nu$, and the corresponding linear perturbation around the
FP. In Sect.~\ref{sec:polynomial} we showed how the results of the global analysis can be easily reproduced by two different kinds of high-order polynomial truncations.
However, these studies were performed in the local-potential approximation (LPA), that is by neglecting the renormalization of the fields.
Taking into account the anomalous dimensions (LPA${}^\prime$) was crucial to obtain a more accurate picture, especially for $X_f>1$,
so that in Sect.~\ref{subsec:polynomial_LPAprime} we developed a LPA${}^\prime$ analysis, based on the same polynomial truncations which were proved to be trustworthy
in the LPA studies.

This allowed us to to produce estimates of the critical exponents $\nu$, $\eta_\phi$ and $\eta_\psi$, in $d=3$ and various $X_f$,
and to compare them with some of the existing literature.
We concentrated on the especially interesting cases of two and one massless Dirac ($X_f=8$ and $4$), of a Weyl ($X_f=2$), and of a Majorana spinor ($X_f=1$).
They can be found in Tab.~\ref{tab:comparison_Xf8},~\ref{tab:comparison_Xf4},~\ref{tab:comparison_Xf2},~\ref{tab:comparison_Xf1}.
Often, there still exists some significant mismatch among the available estimates, such that 
more studies by all kinds of methods, including Monte-Carlo simulations and higher-order $\epsilon$- or $1/N_f$-expansions, are welcome.
As far as the FRG is concerned, the results seem stable for $X_f\geq 4$, while for lower number of fermions there is still room for debate,
and probably larger truncations are needed. The supersymmetric case $X_f=1$ is an exception also in this sense, since it enjoys a good agreement among the results produced 
with different methods.

Larger truncations, such as a next-to-leading order of the derivative expansion, are anyway needed for a quantitative analysis of multi-critical
models in $2<d<3$, as we argued in App.~\ref{sec:largeXf} in the large-$X_f$ limit.
Still within the LPA${}^\prime$, the next natural step is to produce global numerical studies similar to the ones presented for the LPA in 
Sect.~\ref{sec:LPA_Xf1_genericd} and~\ref{sec:d=3_numerical_FP}.
Regarding the possible applications of the present analysis to different models, one possibility is to enlarge the symmetry group from $\mathbb{Z}_2$
to $O(\text{N})$. The N$=3$ three-dimensional chiral Heisenberg universality class, for instance, can be interesting for
the physics of electrons in graphene~\cite{Janssen:2014gea}.
With an enlarged symmetry, the effect of different representations would become a natural case-study and would further widen the
class of physical applications of these studies~\cite{Mesterhazy:2012ei}.
The same kind of truncation can also be used in the context of a Yukawa model interacting with gravity, along the lines of~\cite{PV}, 
to investigate first the asymptotic safety properties of the model, and then to construct global flows from the UV to the IR.
Some scenarios could be of particular interest for cosmology.

\acknowledgements
We would like to thank Alessandro Codello, Stefan Rechenberger, Michael Scherer and Ren\'e Sondenheimer 
for inspiring discussions, and Julia Borchardt, Tobias Hellwig, Benjamin Knorr and Omar Zanusso
for providing us some of their results, which can be found in Sect.~\ref{subsec:polynomial_LPAprime}, as well as for valuable 
comments. We are grateful to Holger Gies for several explanations and suggestions and for a critical reading of the manuscript.

L. Z. acknowledges support by the DFG under grant GRK1523/2.

\appendix

\section{Regulators and threshold functions}
\label{sec:regulators_and_thresholds}

We have to evaluate the r.h.s of \Eqref{flowequation}, for which
we need the $\Gamma_k^{(2)}$ matrix.  
Considering the field $\psi$ as a column and $\bar{\psi}$ 
as a row vector, let us denote by $\Phi^{\mathrm T}(q)$ the row vector 
with components $\phi(q)$, $\psi^{\mathrm T}(q)$, $\bar{\psi}(-q)$, 
and by $\Phi(p)$ the column vector given by its transposition.  
Then $\Gamma_k^{(2)}$ is obtained by the formula
\be
 \Gamma_k^{(2)}=
\frac{\overrightarrow{\delta}}{\delta \Phi^{\mathrm{T}}(-p)}
\Gamma_k
\frac{\overleftarrow{\delta}}{\delta \Phi(q)}\,.\nonumber
\ee
This inverse propagator is regularized by addition of the
following regulator
\begin{equation*}
R_k(q,p)=\delta(p-q)
\begin{pmatrix}
R_{\mathrm{B}}(p) & 0 \\
0 & R_{\mathrm{F}}(p) \\
\end{pmatrix}\ ,
\end{equation*}
where
\begin{eqnarray*}
R_{\mathrm{B}}(p)&=&Z_{\phi}p^2r_{\mathrm{B}}(p^2),\\
R_{\mathrm{F}}(p)&=&-
\begin{pmatrix}
0 & \delta^{ij}\pslash^{\mathrm{T}}\\
\delta^{ij}\pslash & 0\\
\end{pmatrix}Z_{\psi}r_{\mathrm{F}}(p^2)\ ,
\end{eqnarray*}
is a $2d_\gamma N_f \times 2d_\gamma N_f $ matrix.
In principle one can have different regulators for the scalar
(B) and for the spinors (F).
A compact way to rewrite the flow equation is
\begin{equation*}
\partial_t \Gamma_k=\frac{1}{2}\tilde \partial_t{\rm STr}\log(\Gamma_{k}^{(2)}+R_k)\,,
\end{equation*}
where
\be\label{eq:tildedt}
\tilde\partial_t \equiv \frac{\partial_t(Z_\phi r_\mathrm B)}{Z_\phi}\!\cdot\!\frac{\delta}{\delta r_\mathrm B}
+\frac{\partial_t(Z_\psi r_\mathrm F)}{Z_\psi}\!\cdot\!\frac{\delta}{\delta r_\mathrm F}\nonumber
\ee
and $\cdot$ denotes multiplication as well as integration over the common argument of the shape functions of the two factors.
Then the regularized kinetic (or squared kinetic) terms are given by
$P_{\mathrm{B/F}} (x)= x(1+r_{\mathrm{B/F}}(x))$,
and the loop momentum integrals appearing on the r.h.s. of
the flow equation give rise to corresponding
regulator dependent threshold functions.
Introducing the abbreviation $\int_p\equiv\int \!\! \frac{d^dp}{(2\pi)^d}$
these threshold functions read
\begin{eqnarray*}
l_0^{(\mathrm{B/F})d}(\omega) =&&\mkern-18mu\frac{k^{-d}}{4 v_d}\int_p \tilde\partial_t\log\left(P_{\mathrm{B}/\mathrm{F}}+\omega k^2\right)  \\
l_1^{(\mathrm{B/F})d}(\omega) =&&\mkern-18mu-\frac{k^{2-d}}{4 v_d}\int_p \tilde\partial_t\frac{1}{P_{\mathrm{B}/\mathrm{F}}+\omega k^2}  \\
l_{n_1,n_2}^{(\mathrm{FB})d}(\omega_1, \omega_2)&&\mkern-18mu =-\frac{k^{2(n_1+n_2)-d}}{4 v_d}
\int_p \tilde\partial_t \frac{1}{(P_\mathrm F + \omega_1 k^2)^{n_1}(P_\mathrm B + \omega_2 k^2)^{n_2}} \\
m_2^{(\mathrm F)d}(\omega) &=& -\frac{k^{6-d}}{4 v_d} \int_p p^2 \tilde\partial_t \left( \frac{\partial}{\partial p^2}\frac{1}{P_\mathrm F + \omega k^2} \right)^2 \\
m_4^{(\mathrm F)d}(\omega) &=& -\frac{k^{4-d}}{4 v_d}
\int_p p^4 \tilde\partial_t   \left( \frac{\partial}{\partial p^2} \frac{1+r_{\mathrm F}}{P_\mathrm F + \omega k^2} \right)^2\\
m_{4}^{(\mathrm{B})d}(\omega_1) &=& -\frac{k^{6-d}}{4 v_d} 
\int_p p^2 \tilde\partial_t \left( \frac{\tfrac{\partial}{\partial p^2}P_\mathrm B}{(P_\mathrm B + \omega_1 k^2)^2}\right)^2 \\
m_{1,2}^{(\mathrm{FB})d}(\omega_1,\omega_2)&=& -\frac{k^{4-d}}{4 v_d} 
\int_p p^2 \tilde\partial_t \left( \frac{1+r_{\mathrm F}}{P_\mathrm F +\omega_1 k^2} \frac{\tfrac{\partial}{\partial p^2}P_\mathrm B}{(P_\mathrm B + \omega_2 k^2)^2} \right) \ .
\end{eqnarray*}
In this work we adopted the linear regulator
$x r_{\mathrm{B}}(x)=(1-x)\theta(1-x)$,
where $x=q^2/k^2$. For spinors this corresponds to a shape function
$r_{\mathrm{F}}$ such that $x(1+r_{\mathrm{B}}(x))=x(1+r_{\mathrm{F}}(x))^2$.  
For it, the threshold functions can be computed analytically, and give
\begin{eqnarray*}
l_0^{(\mathrm{B})d}(\omega) &=& \frac{2}{d}\frac{1-\tfrac{\eta_\phi}{d+2} }{1+\omega} \, , \\
l_0^{(\mathrm{F})d}(\omega) &=& \frac{2}{d}\frac{1-\tfrac{\eta_\psi}{d+1}}{1+\omega} \, , \\
l_1^{(\mathrm{B/F})d}(\omega) &=&-\tfrac{\partial}{\partial \omega}l_0^{(\mathrm{B/F})d}(\omega) \, , \\
l_{n_1,n_2}^{(\mathrm{FB})d}(\omega_1, \omega_2) &=& \frac{2}{d}\left[
n_1\frac{ 1-\frac{\eta_\psi}{d+1} }{ (1+\omega_1)^{1+n_1}(1+\omega_2)^{n_2} }
+\, n_2\frac{ 1-\frac{\eta_\phi}{d+2} }{ (1+\omega_1)^{n_1}(1+\omega_2)^{1+n_2} } \right] \, , \\
m_2^{(\mathrm F)d}(\omega) &=& \frac{1}{(1 + \omega)^4} \, , \\
m_4^{(\mathrm F)d}(\omega) &=& \frac{1}{(1 + \omega)^4} + 
\frac{1-\eta_\psi}{(d-2)(1+\omega)^3}
-\left( \frac{1-\eta_\psi}{2d-4} + \frac{1}{4} \right)\frac{1}{(1 + \omega)^2} \, , \\
m_{4}^{(\mathrm{B})d}(\omega_1) &=& \frac{1}{(1+\omega_1)^4} \, , \\
m_{1,2}^{(\mathrm{FB})d}(\omega_1,\omega_2) &=& \frac{1-\tfrac{\eta_\phi}{d+1}}{(1+\omega_1)(1+\omega_2)^2} \, .
\end{eqnarray*}
%

\section{Properties of the large-$X_f$ solutions}
\label{sec:largeXf_appendix}

In both versions of the LPA, with or without $\eta_\phi=0$, and also in the LPA${}^\prime$,
\Eqref{eq:prequantizedetaphi} enables us to write the potentials in the form
\be\label{eq:largeXFPpotentials}
h(\phi)=c_h\, \phi^n\, , \quad
v(\phi)=c_v\, \phi^{dn}-\frac{4v_d}{d^2}\, {}_2F_1
\left(1,-\frac{d}{2};1-\frac{d}{2};-c_h^2 \phi^{2n}\right)\ .
\ee
The behavior of $v$ for $\phi\to\pm\infty$ is 
\be
v_{\rm asympt}(\phi)\simeq \left(\sgn(\phi)^{dn}c_v + \frac{\Gamma(-d/2)}{2^{d+1}\pi^{d/2}} |c_h|^{d}\right) |\phi|^{dn}
\ee 
and, since we are assuming $2<d<4$, the gamma function in front of $|c_h|^d$ is positive.
If $c_v\neq 0$, the scalar potential can be real only if $(-1)^{dn}$ has a real
branch, that is if
\be\label{eq:dn_rational_odd_denominator}
d=\frac{m}{n\, j}\ , \quad j\in\{1,3,5,\dots\} \ , \ m\in\mathbb{N}\ ,\ 2n j<m<4n j  \ .
\ee
Its stability further requires
\be\label{eq:stabilitycondition}
|c_h|^d\geq \frac{2^{d+1}\pi^{d/2}}{\Gamma(-d/2)}\max\{-c_v,(-1)^{1+dn}c_v\}=c_{h,\mathrm{crit}}^d
\ee
and for special values of $c_v$ and $c_h$, namely when $|c_h|=|c_{h,\mathrm{crit}}|$, 
it can become asymptotically flat (possibly only on one side) instead of growing like $\phi^{dn}$.

In order to understand the physical properties of the large-$X_f$ FP's, we need to consider
the RG flow in vicinity of the corresponding critical points. In particular we consider the
linearization of the flow, by looking at
small fluctuations of the potentials $v=v+\delta v$, $h=h+\delta h$ and 
for eigenvalue solutions
\begin{equation*}
\dot{\delta v}=-\theta \delta v\ , \quad \dot{\delta h}=-\theta \delta h\ .
\end{equation*}
These equations at large-$X_f$ are extremely simple and, for the linearized
regulator, they read
\bea
-\theta \delta v&=&-d\delta v + \frac{1}{n}\phi\delta v'+ \frac{\delta\eta_\phi}{2}\phi v'+\frac{4v_d}{d} \frac{2 h \delta h}{(1+h^2)^2}\\
-\theta \delta h&=&-\delta h + \frac{1}{n}\phi\delta h' + \frac{\delta\eta_\phi}{2}\phi h' \ .
\eea
In this appendix we want to sketch a study of the properties of these FPs as well as of 
the linearized flow around them. We believe it can be instructive to consider separately
the results obtained with or without the inclusion of the flow equation for $\eta_\phi$.
This will make evident that larger truncations, out of the reach of the present work,
are necessary to get a complete picture of the large-$X_f$ multicritical Yukawa theories.
 
\subsection{LPA}

\begin{figure}[!t]
\begin{center}
 \includegraphics[width=0.4\textwidth]{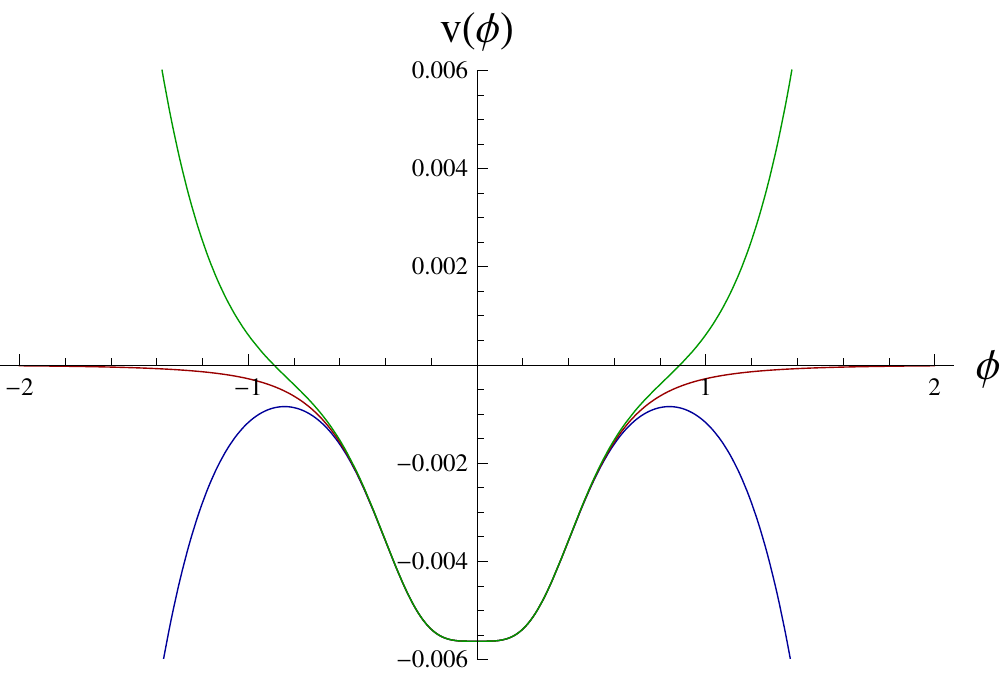}
 \includegraphics[width=0.4\textwidth]{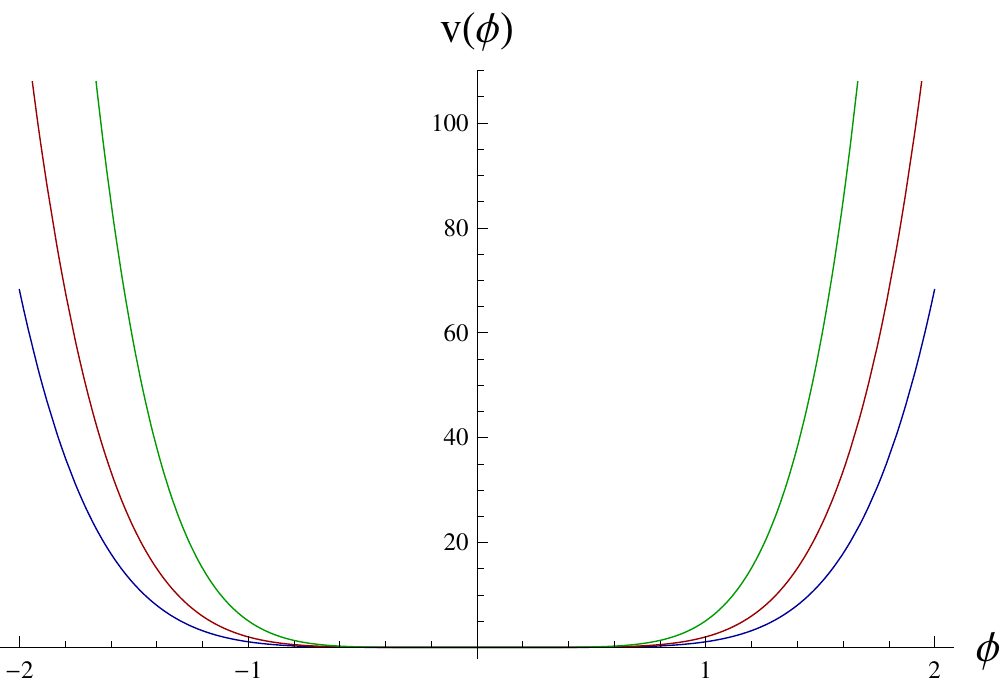}
 \caption{The $d=3$, $n=2$, FP scalar potential at nonvanishing $c_v$.
Left panel: $c_v=-1$ and $c_h \in\{c_{h,\mathrm{crit}}+10^{-3},c_{h,\mathrm{crit}},c_{h,\mathrm{crit}}-10^{-3}\}$,
 from bounded (green) to unbounded (blue).
Right panel: $c_v=1$ and $c_h \in\{c_{h,\mathrm{crit}}+2,c_{h,\mathrm{crit}},c_{h,\mathrm{crit}}-2\}$,
 from steeper (green) to broader (blue).
}
\label{fig:LXf_withcv_d3_n2_FPv}
\end{center}
\end{figure}

\begin{figure}[!t]
\begin{center}
 \includegraphics[width=0.4\textwidth]{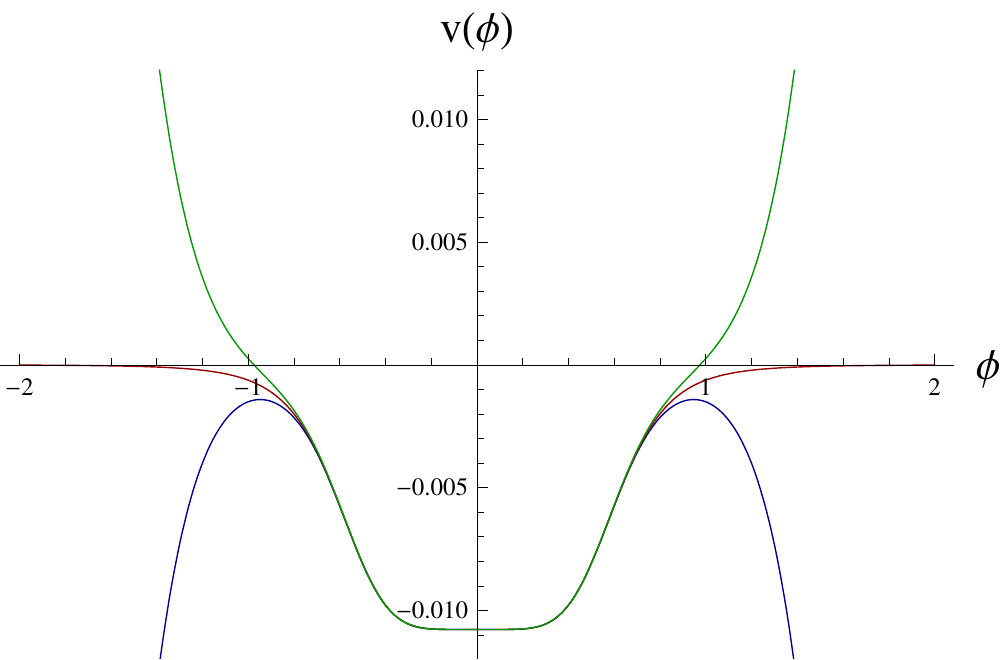}
 \includegraphics[width=0.4\textwidth]{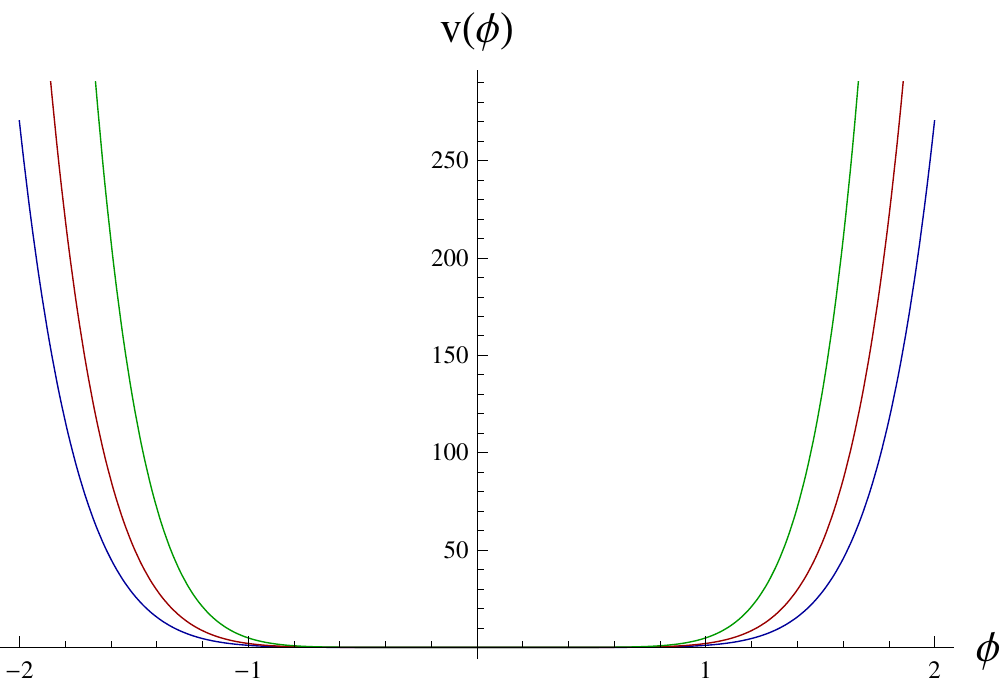}
 \caption{The $d=8/3$, $n=3$, FP scalar potential at nonvanishing $c_v$.
Left panel:  
 $c_v=-1$ and $c_h \in\{c_{h,\mathrm{crit}}+10^{-3},c_{h,\mathrm{crit}},c_{h,\mathrm{crit}}-10^{-3}\}$,
 from bounded (green) to unbounded (blue).
Right panel: $c_v=1$ and $c_h \in\{c_{h,\mathrm{crit}}+2,c_{h,\mathrm{crit}},c_{h,\mathrm{crit}}-2\}$,
 from steeper (green) to broader (blue).
}
\label{fig:LXf_withcv_d8o3_n3_FPv}
\end{center}
\end{figure}

\begin{figure}[!t]
\begin{center}
 \includegraphics[width=0.4\textwidth]{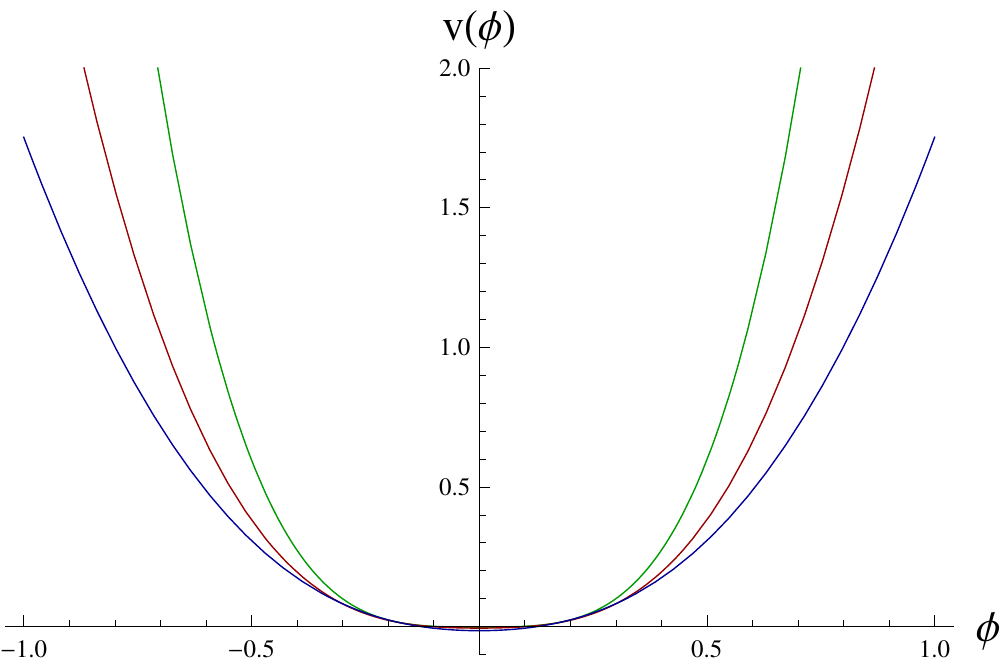}
 \includegraphics[width=0.4\textwidth]{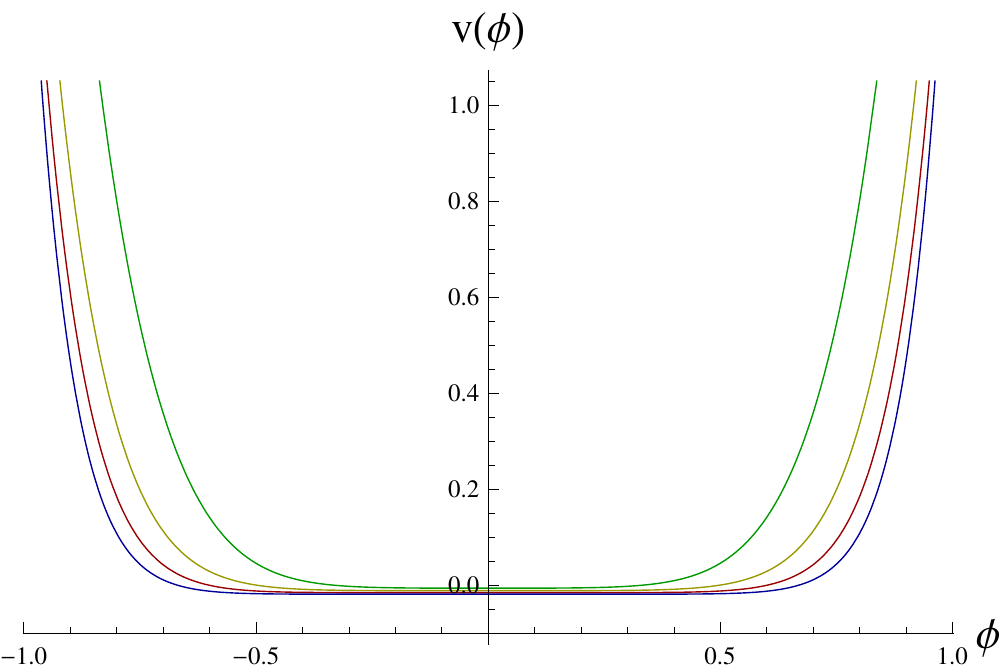}
 \caption{The even FP scalar potentials for $c_v=0$.
For illustration the value of $c_h$ has been chosen according to~\Eqref{eq:n=1fixedch},
even if this is mandatory only for $n=1$ in the LPA${}^\prime$.
Left panel: $n=1$ and $d\in\{3.5,3,2.5\}$, from steeper (green) to broader (blue).
Right panel: $n\in\{2,3,4,5\}$, in the corresponding dimension $d=2+2/n$,
 from steeper (green) to broader (blue).
}
\label{fig:LXf_nocv_FPv}
\end{center}
\end{figure}

If we set by hand $\eta_\phi=0$, regardless of $c_v$ or $c_h$
\Eqref{eq:prequantizedetaphi} leaves a discrete
set of dimensions as the only possibility, the ones in \Eqref{eq:uppercritdim}.
As a consequence $d n=2(n+1)$ and  
the scalar potential is real and even also in case $c_v\neq 0$.
The stability properties, depending on $c_h$ and $c_v$ according to 
\Eqref{eq:stabilitycondition}, are illustrated in the plots of 
Fig.~\ref{fig:LXf_withcv_d3_n2_FPv} and Fig.~\ref{fig:LXf_withcv_d8o3_n3_FPv}.
The special case $c_v=0$ is shown in Fig.~\ref{fig:LXf_nocv_FPv}.

Let us now turn to the linear perturbations of these FP's. 
By definition in the LPA one neglects a possible change of anomalous dimension.
Thus, setting $\delta\eta_\phi=0$, the solution for the perturbations reads
\bea
\delta h(\phi)&=&\delta c_h \phi^N \\
\delta v(\phi)&=&\delta c_v \phi^{(d-1)n+N}-\frac{4 v_d}{d}c_h \delta c_h \phi^{n+N}
\left[  \frac{1}{1+c_h^2 \phi^{2n}}-\frac{d}{d-2}\,  {}_2F_1\!\left(
1,1-\frac{d}{2};2-\frac{d}{2};-c_h^2 \phi^{2n}\right)\right]\nonumber
\eea
Here we restricted our analysis to the perturbations with $\delta c_h\neq0$,
and required their smoothness by setting $(1-\theta)n=N\in\mathbb N$.
For the special case $\delta c_h=0$ the solution is simply $\delta v(\phi)=\delta c_v \phi^{M}$
with critical exponent $\theta_M=d-M/n$, and will not be discussed any further.
Notice that these eigenfunctions are independent of
$c_v$, which is due to the suppression of scalar loops in the large-$X_f$ limit.
They are regular at the origin, since the leading behavior is
\be
\delta v(\phi) \mathop{\sim}_{\phi\to0}\frac{8 v_d}{d(d-2)}c_h\delta c_h \phi^{n+N}\ .
\ee
Recall that the FP potential had, as leading small field dependence, $\phi^{2n}$; as a consequence,
the relevant perturbations with $N<n$ change the behavior of the potential at the origin,
the marginal ones only change the coefficient in front of $\phi^{2n}$, and the irrelevant ones
leave the leading term unaltered.
For large value of the field
\be
\delta v(\phi) \mathop{\sim}_{\phi\to\infty}\left(\delta c_v+\sgn(\phi)^{dn}\frac{d \Gamma(-d/2)}{2^{d+1}\pi^{d/2}}|c_h|^{d-2}c_h\delta c_h\right) \phi^{dn+N-n}
\ee
while the FP potential behaves like $|\phi|^{dn}$ at infinity.
As a consequence, the irrelevant perturbations with $N>n$ completely change
the asymptotic behavior of the potential for large fields, the marginal ones with $N=n$ only
change the coefficient in front of the leading power, and the relevant ones only change
the sub-leading terms. Clearly this is not the case for those potentials,
with  special values of $c_v$, that are asymptotically flat.

Let's now discuss the symmetry properties of the perturbations.
Trivially, the symmetry of Yukawa potential under ${\mathbb Z}_2$ is
preserved or violated depending on $n$ and $N$. We now want to understand
what this entails for the scalar potential.
Recall that in the LPA $dn=2(n+1)$
and the FP $v$ is always even.
Then, the fluctuations behave as $\phi^{n+N+2}$, and whenever
$N+n$ is odd, the ${\mathbb Z}_2$ symmetry of both $h$ and $v$
at the FP is spoiled by the perturbations.
Among these symmetry breaking perturbations, the irrelevant ones, with $N>n$,
give rise to unstable potentials.
Notice that the relevant perturbations, even if spoiling symmetry,
do not directly cause instabilities (though they might induce them indirectly, i.e. beyond linearization).
The possibility to have stable theories with no definite ${\mathbb Z}_2$
symmetry emanating from symmetric FPs in the UV or IR is in any case
a question that requires a global study of the RG flow, and it is beyond
the scope of this work.

\subsection{LPA${}^\prime$}

%
\begin{figure}[!t]
\begin{center}
 \includegraphics[width=0.4\textwidth]{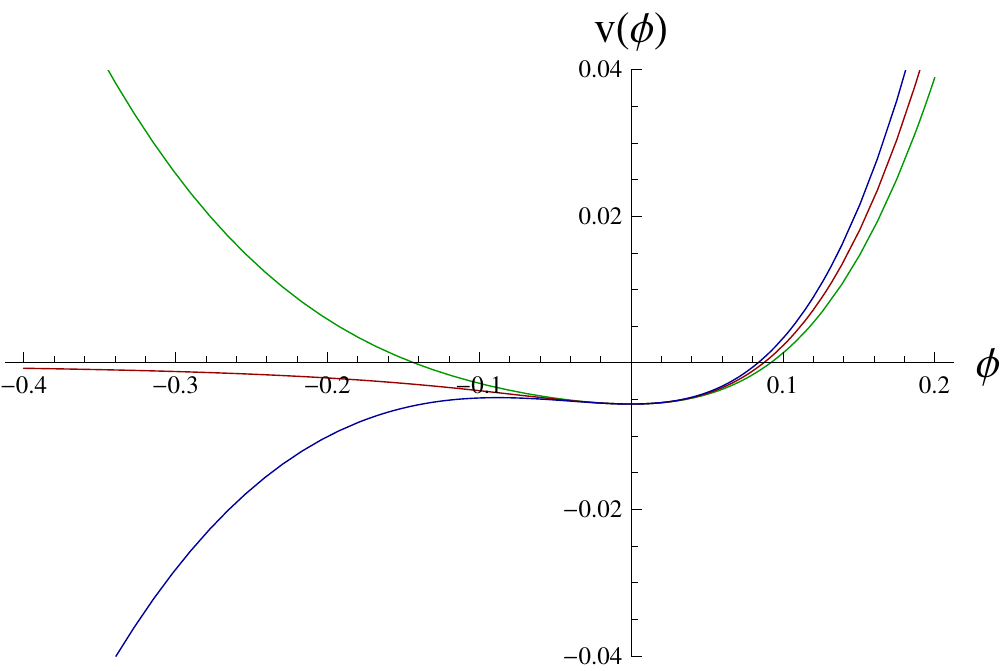}
 \includegraphics[width=0.4\textwidth]{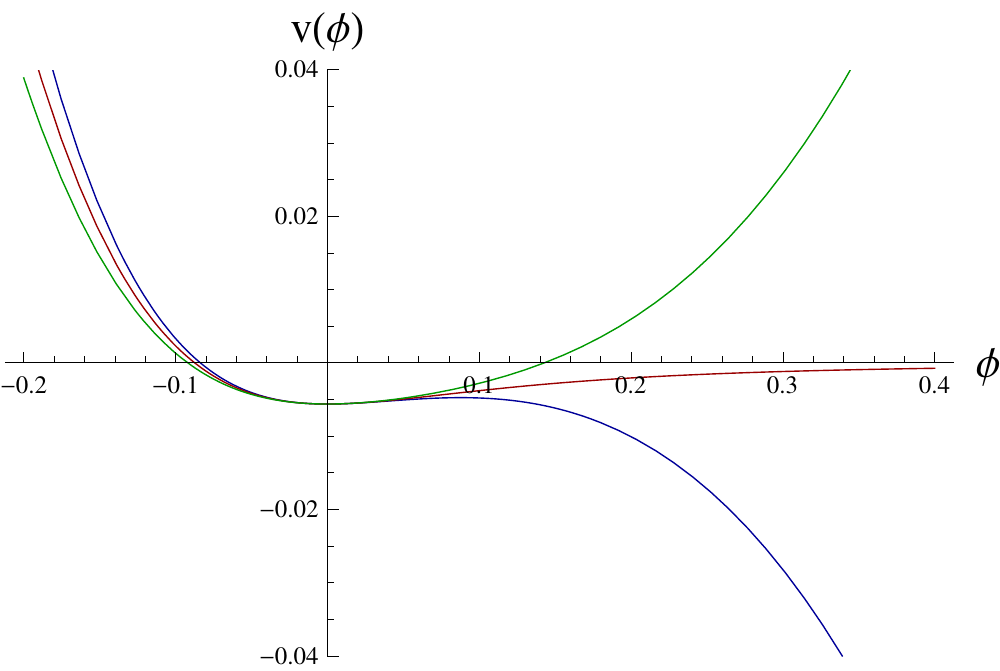}
 \caption{The $d=3$ and $n=1$ FP scalar potential at nonvanishing  $c_v$.
Left panel:   $c_v \in\{c_{v,\mathrm{crit}}-1,c_{v,\mathrm{crit}},c_{v,\mathrm{crit}}+1\}$,
 from bounded (green) to unbounded (blue).
Right panel:  $c_v \in\{-c_{v,\mathrm{crit}}-1,-c_{v,\mathrm{crit}}, -c_{v,\mathrm{crit}}+1\}$,
 from bounded (green) to unbounded (blue).
Notice that the value of the potential at the origin is arbitrary, while its behavior for large fields is not.
}
\label{fig:LXf_withcv_d3_n=1_FPv}
\end{center}
\end{figure}
\begin{figure}[!t]
\begin{center}
 \includegraphics[width=0.4\textwidth]{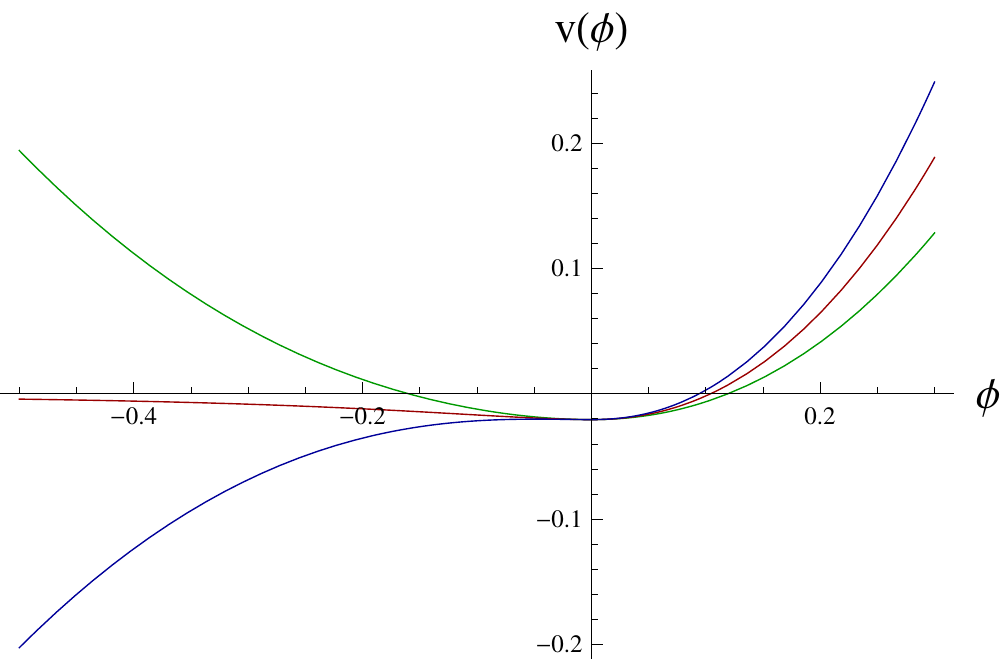}
 \includegraphics[width=0.4\textwidth]{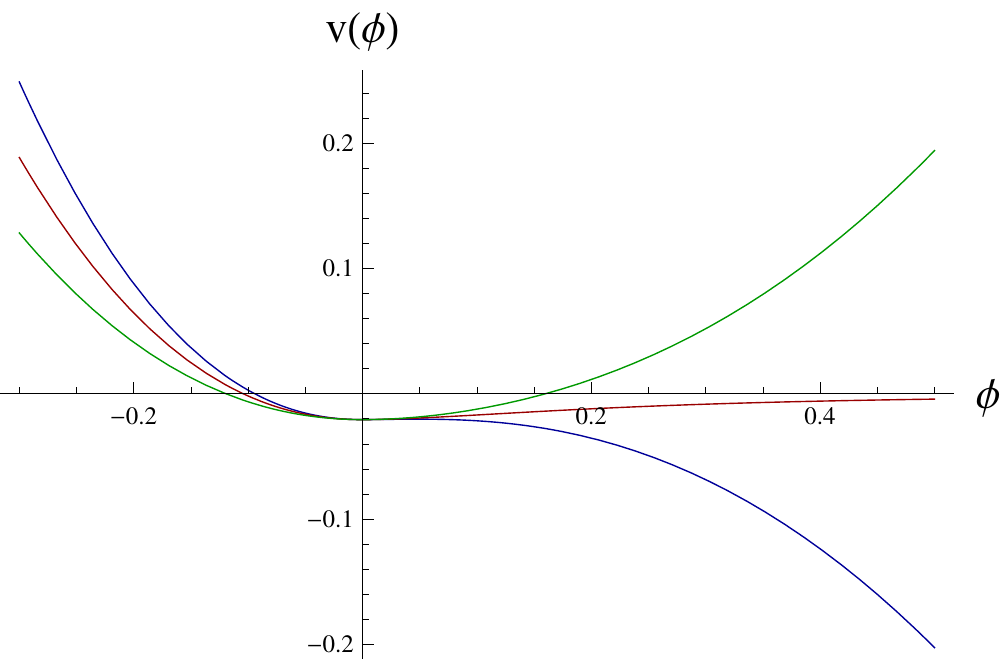}
 \caption{The $d=7/3$ and $n=1$ FP scalar potential at nonvanishing  $c_v$.
Left panel:   $c_v \in\{c_{v,\mathrm{crit}}-1,c_{v,\mathrm{crit}},c_{v,\mathrm{crit}}+1\}$,
 from bounded (green) to unbounded (blue).
Right panel:  $c_v \in\{-c_{v,\mathrm{crit}}-1,-c_{v,\mathrm{crit}}, -c_{v,\mathrm{crit}}+1\}$,
 from bounded (green) to unbounded (blue).
}
\label{fig:LXf_withcv_d7o3_n=1_FPv}
\end{center}
\end{figure}

So far we have not used the flow equation for $\eta_\phi$.
In order to do so, we first have to analyze the possible presence of a nontrivial minimum
for $v$.
The general expectation is that, since only fermion loops survive in the 
leading order of the $1/X_f$ expansion, the potential is always in the symmetric regime.
This is suggested by the expansion of the potential around the origin, based on \Eqref{eq:hyperFsmallx}.
We assume that this is always the case for the time being, 
as it is indeed for every specific example we have considered. 
Under this assumption, we need to take the $\phi\to0$ limit of the 
equation for $\eta_\phi$, which is proportional to $h'(\phi)^2$,
i.e. to $\phi^{2(n-1)}$.
Therefore, only for $n=1$ such a limit can be nonvanishing.
This shows how LPA${}^\prime$ is an improvement of LPA
only for the $n=1$ critical theory.
For the remaining values of $n$, one finds again $\eta_\phi=0$,
which artificially forces the dimension $d$ to its critical value.
We expect this condition to be lifted by more general truncations,
and a nontrivial $\eta_\phi$ should emerge for any $n$.

Let's then discuss the change brought by LPA${}^\prime$
in the description of the large-$X_f$  $n=1$ FP.
As argued in Sect.~\ref{sec:largeXf}, the nontrivial
$\eta_\phi$ allows for the existence of the non Gau\ss ian FP
in any $d<4$, as long as
\be\label{eq:largeXfetan1}
\eta_\phi=4-d\ ,\quad n=1 \, .
\ee
Actually this is the case only for the $\mathbb{Z}_2$ symmetric solution with $c_v=0$.
As soon as $c_v\neq 0$ the reality of the potential requires
\be
d=\frac{m}{j}\ , \quad j\in\{1,3,5,\dots\} \ , \ m\in\mathbb{N}\ ,\ 2j<m<4j  \ .
\ee
Regardless of $c_v$, by using \Eqref{eq:largeXfetan1} 
the flow equation for $Z_\phi$ can be solved for
$c_h$ as a function of $d$, giving~\cite{Braun:2010tt}
\be\label{eq:n=1fixedch}
c_h^2=\frac{d(4-d)(d-2)}{v_d(6d-8)}\ .
\ee
Then, the stability condition~\Eqref{eq:stabilitycondition} for the nonvanishing $c_v$ FP's
is best phrased as a bound on $c_v$
\be\label{eq:lower_bound_on_cv}
c_v\geq - c_{v,\mathrm{crit}}\ , \quad c_{v,\mathrm{crit}}=\frac{\Gamma(-d/2)}{2^{d+1}\pi^{d/2}}\left[\frac{d(4-d)(d-2)}{v_d(6d-8)}\right]^{d/2}
\ee
and additionally, only for odd $m$, $c_v\leq c_{v,\mathrm{crit}}$.
The scalar FP potential with $c_v\neq0$ is an even function if and only if $m$ is even.

Let us then turn to perturbations, and allow for a nontrivial $\delta\eta_\phi$.
We postpone for a while the task of solving the linearized equation for $\eta_\phi$,
which provides us the first correction $\delta \eta_\phi$ to the anomalous dimension,
as a function of  the FP $h$ and $\delta h$.
This is because such an equation involves the variation $\delta\phi_0$ in the
location of the minimum of the potential, which in turn can be computed
from the variation of the potential by the formula
\be\label{eq:deltaphi0}
\delta\phi_0=-\frac{\delta v'(0)}{v''(0)}
\ee
where we stuck to our assumption that the minimum of the FP potential is
always trivial. As a consequence we first solve for $\delta v$ and $\delta h$
as parametric functions of $\delta\eta_\phi$, and then plug \Eqref{eq:deltaphi0}
into the linearized equation for $\eta_\phi$, to compute the actual $\delta\eta_\phi$.
Solving for $\delta h$ is again trivial, and it immediately allows us to extract the
eigenvalues of the linearized flow.
When $\theta\neq 0$ the solution for $\delta h$ is
\be\label{eq:largeXf_deltah}
\delta h(\phi)
=\delta c_h \phi^N-\frac{\delta\eta_\phi}{2}\frac{n^2}{n-N}c_h\phi^n\ , \quad N\in\mathbb N \ , \quad N\neq n
\ee
where again we focused on $\delta c_h\neq 0$ and set  $N=(1-\theta)n\in\mathbb N$.
For $\theta=0$ instead 
\be\label{eq:largeXf_deltah_marginal}
\delta h(\phi)=\delta c_h \phi^{n}-\frac{\delta\eta_\phi}{2}n^2c_h\phi^n\log(\phi)
\ee
Notice that the second term in the last equation is simply the first order in the expansion of 
$c_h\phi^{2/(d-2+\eta_\phi)}$, which is the exactly marginal $h$, around the $n$-th FP.
As a consequence, the apparent instability that can come from 
the second term in \Eqref{eq:largeXf_deltah_marginal} is actually a fake of linearization,
 as long as $\delta\eta_\phi >-2/n$. On the other hand, a logarithmic singularity at
the origin appears even beyond linearization, and we believe this to be a pathology produced by the leading order
in $1/X_f$. The solution to this pathology will come soon, in the form of the constraint
$\delta\eta_\phi=0$ for these perturbations.

The equation for $\delta v$ is much more involved in the LPA${}^\prime$ than in the LPA,
since it now depends on the FP potential.
Yet, its solutions for generic $\delta\eta_\phi$ can be easily given analytically. 
It is not necessary to show them here.
It suffices to report that quite in general they have the property $\delta v'(0)=0$,
as it could be expected by the argument that fermion loops are generally associated
with scalar potentials with a trivial minimum~\footnote{For $\delta c_h\neq 0$ only the $n=1$, $N=0$ case
gives rise to a nonvanishing $\delta v'(0)$. For $\delta c_h= 0$ only the $M=1$ case.
In what follows we discard these cases.}. As a consequence the scalar potential
stays in the symmetric regime. Notice that this does not entail that the $\delta v(\phi)$
is also in the symmetric regime.

With this piece of information, one can work out the linearized $\delta\eta_\phi$,
by varying the r.h.s. of the flow equation for $Z_\phi$ with respect to $h$ and $v$ 
(whose fluctuations still depend parametrically on $\delta\eta_\phi$ itself) and $\eta_\phi$, 
while keeping $\phi_0$ fixed, and then taking $\phi_0\to0$.
The latter limit makes the r.h.s. vanishing unless $n=1$, in which case it reaches a
$d$-dependent constant times $c_h^2\delta\eta_\phi$.~\footnote{Such a constant
is actually infinite for the marginal perturbation, the r.h.s. inheriting a logarithmic
singularity at the origin from $\delta h$. Yet the simple way to cure this pathology and
get a self-consistent answer is to set $\delta\eta_\phi=0$.}
Hence, for general $n$ and $N$ we find $\delta\eta_\phi=0$, which
boils the analysis of the linearized perturbations down to the one
sketched in the last Section within the LPA.

\subsection{$d=4$}

The expression in \Eqref{eq:largeXFPpotentials}, cannot be used in
$d=4$ nor in $d=2$, since the hypergeometric function in $v$ has simple
poles at these values. 
The $d\to2$ case is out of the reach of the present paper.
In the $d\to4$ limit, instead, the canonical dimensional terms survive also in the LPA,
and by integrating the large-$X_f$ system of flow equations one can find the following
FP solutions
\be
h(\phi)=c_h\, \phi^n\, , \quad
v(\phi)=c_v\, \phi^{4n}+\frac{1}{64\pi^2}
\left( c_h^2 \phi^{2n} - c_h^4 \phi^{4n} \log(c_h^2+\phi^{-2n})\right)
\ee
where we already demanded the Yukawa potential to be smooth, according to 
\Eqref{eq:prequantizedetaphi}.
The crucial fact is again that the minimum of $v$ is always trivial.
This allow us to take the $\phi_0\to0$ limit of the equation for $\eta_\phi$.
For $n=1$ this leaves us with the equation $c_h^2=\eta_\phi=0$, 
thus boiling every feature of the critical theory down to the classical counting.
For $n\geq2$ we find the constraint $\eta_\phi=0$, which is inconsistent with
\Eqref{eq:prequantizedetaphi} and therefore eliminates these solutions.


\end{document}